\documentclass[prd,a4paper,superscriptaddress,aps,showpacs,floats,nofootinbib]{revtex4}
\usepackage{epsfig}
\usepackage{graphicx}
\usepackage{amsmath}
\usepackage{amssymb}
\usepackage{color}
\usepackage{soul}
\usepackage{hyperref}
\usepackage{url}
\usepackage{float}
\usepackage{booktabs}

\newcommand{\scalesum}[2]{%
	\mathop{\scalebox{#1}{$\displaystyle\sum$}}\limits_{#2}%
}
\DeclareMathAlphabet\mathbfcal{OMS}{cmsy}{b}{n}

\begin{document}

\title{Graph-based Summary Statistics for Revealing the Stochastic Gravitational Wave Background in Pulsar Timing Arrays}

\author{M. Alakhras}
\email{m\_alakhras@sbu.ac.ir}
\affiliation{Department of Physics, Shahid Beheshti University, 1983969411, Tehran, Iran}
\affiliation{Department of Physics, Faculty of Sciences, Damascus University, Damascus, Syria}

\author{S. M. S. Movahed}
\email{m.s.movahed@ipm.ir}
\affiliation{Department of Physics, Shahid Beheshti University, 1983969411, Tehran, Iran}
\affiliation{School of Astronomy, Institute for Research in Fundamental Sciences (IPM), P. O. Box 19395-5531, Tehran, Iran}
\affiliation{Department of Mathematics and Statistics, The University of Lahore,
	1-KM Defence Road Lahore-54000, Pakistan}

\date{\today}

\begin{abstract}

In this work, we propose a graph-based method implemented on the pulsar timing residuals (PTRs) for stochastic gravitational wave background (SGWB) detection within the nano-Hertz frequency regime and examining uncertainties of its parameters. We construct a correlation graph with pulsars as its nodes, and analyze the graph-based summary statistics, including structural characteristics of complex network, for identifying SGWB in the real and synthetic datasets. The effect of the number of pulsars, the observation time span, and the strength of the SGWB on the graph-based feature vector is evaluated. 
Our results demonstrate that the  Discriminative Summary Statistics for common signal detection consists of the average clustering coefficient and the edge weight fluctuation. The SGWB detection conducted after the observation of a common signal and then exclusion of non-Hellings \& Downs templates is performed by the second cumulant of edge weight for angular separation thresholds $\bar{\zeta}\gtrsim 40^{\circ}$. The lowest detectable value of SGWB strain amplitude utilizing our graph-based measures at the current PTAs sensitivity is $A_{\rm SGWB}\gtrsim 1.2\times 10^{-15}$.  Fisher forecasts confirmed that the uncertainty levels of $\log_{10} A_{\rm SGWB}$  and spectral index reach   $1.5\%$ and $19.5\%$, respectively, at $2\sigma$ confidence interval.  A weak evidence for an SGWB at $\sim 2.3\sigma$ level is obtained by applying our graph-based method to the NANOGrav 15-year dataset.

\end{abstract}

\maketitle

\section{Introduction}
\label{introduction}
Gravitational wave (GW) which is the direct prediction of Einstein's theory of general relativity \citep{1916SPAW.......688E,1918SPAW.......154E}, open a distinct window onto the exploring the fundamental nature of Universe \citep[and references therein]{1993IrAJ...21..161H,1997rggr.conf..373A,2009LRR....12....2S,2018CQGra..35p3001C,2019RPPh...82a6903C}. The LIGO collaboration's direct detection of gravitational wave from the merger of two black holes in 2015 \citep{2016PhRvL.116f1102A} presented an encouraging opportunity for assessing the fabric of the Universe through the propagating ripples in the curvature of spacetime.
A specific type of GW which may be treated by non-deterministic strain signals is known as the stochastic gravitational wave background (SGWB) \citep{2009Natur.460..990A,2018CQGra..35p3001C,2023PrPNP.12804003V}. The SGWB can be generated in many theories, including cosmological inflation \citep[and references therein]{1988PhRvD..37.2078A,2016NCimR..39..399G,2021Univ....7..398D,2021iSci...24j2860Y,2022PhRvD.105b3521T,2025GReGr..57...82S}, primordial phase transitions \citep{1985PhRvD..31.3052V,1985PhR...121..263V,1995RPPh...58..477H,2001PhRvD..64f4008D,2011PhRvD..83h3514B,Ringeval:2017eww,Blanco-Pillado:2017oxo,2023JCAP...04..045H}, modified gravity theories \citep[and references therein]{1997rggr.conf..373A,2016PhRvL.117i1102M,2019RPPh...82a6903C}, theories that go beyond the standard model of particles \citep{2000PhR...331..283M,2018CQGra..35p3001C,2019RPPh...82a6903C} and the superposition of enormous number of unresolved astrophysical GW events \citep{2012PhRvL.109q1102M}. 

Thanks to numerous observational surveys designed to probe the Universe, high-precision collections of datasets with multidimensional nature have become accessible. One might identify some of the currently operational, ongoing, and planned surveys that are either focused on or a part of their missions dedicated to the detection of GWs, such as: LISA (Laser Interferometer Space Antenna) \citep{2017arXiv170200786A}, LIGO-Virgo-KAGRA \citep{2021PhRvD.104j2001A}, Einstein Telescope (ET) \citep{Punturo_2010}, Deci-hertz Interferometer Gravitational Wave Observatory (DECIGO) \cite{2021PTEP.2021eA105K},   Cosmic Explorer (CE) \citep{2015PhRvD..91h2001D,2019BAAS...51c.141R}, the North American Nanohertz Observatory for Gravitational Waves (NANOGrav) \citep{mc13,NANOGRAV21,NANOGRAV22,2023ApJ...951L..11A}, the Parkes Pulsar Timing Array (PPTA) \citep{PPTA1,PPTA2,PPTA3,man13}, the European Pulsar Timing Array (EPTA) \citep{2013CQGra..30v4009K,2016MNRAS.458.3341D}, the Indian PTA (InPTA) \citep{2018JApA...39...51J}, and the International Pulsar Timing Array (IPTA) \citep{2010CQGra..27h4013H,2016MNRAS.458.1267V}. Looking ahead, key radio facilities that will significantly enhance PTA sensitivity include the Square Kilometre Array (SKA) \citep{cor04,laz13,5136190} and the Five-hundred-meter Aperture Spherical radio Telescope (FAST) \citep{2013MS&E...44a2022N}. Furthermore, an intensive explanation about the details of frequency band for GW detection can be found in \citep{2022Galax..10...34R}.

The development of numerous simulation suites to generate synthetic GW over different frequency ranges \citep{2012PhRvD..86l2001R,2014PhRvD..89h4046R} has encouraged researchers to employ field-level inferences and to introduce more sophisticated summary statistics for evaluating the detection pipeline, subsequently enabling cosmological inferences.
A list of some famous suites for GW mock data simulations and analysis toolkits are: LALSuite \citep{2020SoftX..1200634W}, BILBY \citep{BILBY2019}, Legacy LISA Code  \citep{2008PhRvD..77b3002P}, 
Bayesian PTA analysis with GW signal injection (ENTERPRISE) \citep{ellis_2020_4059815}, TEMPO2 \citep{2006MNRAS.369..655H}, PINT \citep{2021ApJ...911...45L}, reconstructing the spectral shape of GW background (GWBackFinder) \citep{2024JCAP...09..032D}, CLASS-GWB \citep{2022JCAP...06..030B}, Gravitational Wave Universe Toolbox \citep{2022A&A...663A.155Y} and Einstein Toolkit \citep{2012CQGra..29k5001L}.

Millisecond radio pulsars (MPs), which are emitting intensive coherent electromagnetic radio wave is particular class of pulsars. They have very stable rotational period leading to one of the suitable indicators for revealing SGWB transiting  between earth and pulsar timing arrays (PTAs). The random distribution of pulsars in the sky makes the PTA as a possible detector with astrophysical arm length size to identify very low frequency GW ($f\in [10^{-9}-10^{-7}]$ Hz) \citep{2023ApJ...951L..11A}. The operating frequency of PTA lies in the frequency range of SGWB generated by cosmological and astrophysical phenomena. The continuous waves from individual binaries  is also in the detectability range of PTA.
Both the cosmological interpretation of the SGWB signal and the default astrophysical SMBHBs background can describe the PTAs data. Neglecting the environmental effects on the evolution of binaries leads to stronger evidence for cosmological interpretation \citep{2023ApJ...951L..11A}. However, by considering the environmental effects, the astrophysical interpretation captures the frequency dependence of the PTAs signal as well as other cosmological models \citep{2024PhRvD.109b3522E}.

For detecting GW by PTA, the residual signal - the difference between measured and predicted time of arrivals (TOAs) - is calculated \citep{2016MNRAS.458.1267V}. The Pulsar Timing Residuals ($PTR$s) are very sensitive indicators to elucidate interesting physical properties of pulsars and other cosmological and astrophysical foreground processes \citep{2012hpa..book.....L,man13,Issa}. It is essential to thoroughly elucidate the different components associated with the modeling of deterministic contributions to arrival times, the stochastic astronomical and instrumental delays, and eventually the elimination of systematic errors \citep{2019BAAS...51g.195R}. It has become evident that preparing  appropriate catalog of pulsars for subsequent inferences necessitates the incorporation of sophisticated data pre-processing. Motivated by these discrepancies and inspired by advances in data analysis routines, several researchers have investigated $PTR$s to identify SGWB and estimate its physical parameters.

Searching the quasi-quadrupole spatial cross-correlation of PTA reveals the unique signature of SGWB, and it was the pioneering approach used by Hellings \& Downs \citep{1983ApJ...265L..39H, 2005ApJ...625L.123J}. Inspired by the concept of self-similarity, \cite{Issa} introduced an innovative approach for irregular fluctuation analysis, which is a suitable method for both SGWBs detection and even for discriminating between different sources of GW. Resolving bright individual SMBHBs in the presence of SGWB through future PTAs has been examined in \citep{2025arXiv250510284F}. 
Extracting the relevant physical observables from power spectrum and overlap reduction function (Hellings \& Downs curve) in PTA was carried out by \cite{2023PhRvD.107d4007B}.

Robust learning from noisy, incomplete, and relatively sparse data, especially those series that exemplify GW, can be achieved by seamlessly integrating data with physical priors, a methodology known as Physics-Informed Machine Learning \citep{2021NatRP...3..422K,2022arXiv221108064H}. A common approach to take into account the physical priors is making the proper feature vector. A pioneer research has been done by \citep{2018PhLB..778...64G}. In pursuit of quality enhancement in GW data, various approaches have been recognized, encompassing the reconstruction of GW signals, accurate assessment of statistical and systematic errors, as well as the identification, fast signal characterization, robustly have been noticed in \citep[and references therein]{2021MLS&T...2a1002C}. SGWB from  binary black hole mergers \citep{2021PhRvD.103d3020S}, Noise reduction in GW data \citep{2020PhRvR...2c3066O,2020PhRvD.101d2003V}, localization of a GW source through machine learning (ML) \citep{2022PhRvD.105j3030S,2022PhRvD.106b3032K,2022APS..APRE17006C} are just a short list of studies representing the promising of ML applications in GW physics. ML has been used as well to predict the characteristics of the SGWB signal in PTA data generated by SMBHB across various regions of parameter space \citep{2024A&A...687A..42B}. Also, emulator has been constructed using Gaussian processes and dense neural networks to map SMBHB-parameters to the resulting SGWB strain properties \citep{2025ApJ...982...55L}. Component separations (noise and signal), likelihood free parameter estimations, source identification and performing field-level inferences can reliably conducted by so-called Simulation-Based-Inference (SBI) approach \citep[and references therein]{tejero2020sbi,2016arXiv160506376P,2019MNRAS.488.4440A,2020PNAS..11730055C,2022mla..confE..24H}, and it is undoubtedly crucial and necessary for upcoming generation of both space- and ground-based facilities that are focused on GW sky \citep{2024PhRvD.109h3008A}.

Various kinds of data are recorded  through observational surveys and they can be classified into {\it time series}, {\it field}, {\it point cloud} and {\it graph} \citep{abedi2024}. Some mappings between aforementioned classes to make them appropriate for further analysis are known as \textit{recurrent plot} for multi-variable time series to field \citep{1987EL......4..973E,2007PhR...438..237M,2011SSPMA,chen2018recurrence}, {\it time delay embedding} \citep{takens1981detecting,packard1980geometry} and {\it state-space} \citep{2019PhRvE.100b2314M,yesilli2022topological,2023PhRvE.107c4303M} for transferring the time-series to point cloud, {\it visibility graph} \citep{lacasa2012time} and  {\it correlation graph} \citep{yang2009visibility}, {\it ordinal partition graph} \citep{small2013complex,2015Chaos..25e3101M}, {\it proximity graph}, {\it cycle graph}, {\it recurrence graph} and {\it transition graph} methods for conversion time series to graphs \citep[and references therein]{campanharo2011duality,silva2021time,2019PhR...787....1Z}.  In the preceptive of complex network analysis, any conceivable analysis should inherently focus on elucidation of the network topology encoded by various measures such as average path length, clustering coefficient and degree distribution \citep{2002RvMP...74...47A,2006PhRvL..97r8701Y,2013PhyA..392.3833R}. In such approach topology and geometry reveal the global and local properties of constructed complex networks, respectively.   
For our case study, i.e. PTAs, we have  multi-variable time series in $(N_{\rm PTA}+1)-$dimension. Assuming the $N_{\rm PTA}$ is the number of observed pulsars and for each we have recorded $PTR$s time series denoted by $PTR^{(a)}(t_i, \hat{p}^{(a)})$ such that $i=1,...,M^{(a)}$ and $a=1,...,N_{\rm PTA}$\footnote{In practical scenario, the timestream exhibits irregularities over the time, which implies that $\Delta t_{i}\equiv|t_{i+1}-t_{i}|$ depends on the index $i$. This irregularity introduces certain discrepancies in data analysis. For the sake of generality, we will now assume that synthetic data for all pulsars are regular, possessing the same time interval \citep{Issa}. The term of irregularity is also referred to {\it Poisson}, {\it intermittent} and {\it Gaussian} regimes as pointed out by \cite{2022Galax..10...34R}.}. The $\hat{p}^{(a)}$ is the unit vector from earth to $a$th pulsar direction. In terms of general definition of stochastic field, we instead have multi-variable time series in $(N_{\rm PTA}+1)$-dimension as \citep{kantz2003nonlinear}:
\begin{equation}
	\label{eq:PTRs}
	PTRs \equiv \left\{PTR^{(a)}(t_i, \hat{p}^{(a)})\right\}_{(i=1,a=1)}^{(M^{(a)},N_{\rm PTA})}.
\end{equation}

The complex networks based on graph theory has gained considerable attention in multidisciplinary research area that cuts across physics, biology, ecology, economics, sociology, and the humanities \citep{network_book2018,2016nesc.book.....B,2002RvMP...74...47A}.
The graph theory as an informative tool for characterizing observations in cosmology has been utilized, particularly in the context of cosmic web analysis \citep{1999ApJ...526..560U,2012NatSR...2..793K,2016arXiv160403236C,2016MNRAS.459.2690H}.  More recently, the graph-based classifier has been used  for identifying the dark matter cosmic web environments of galaxies \citep{kololgi2025learning}. This theory has also been used for the classification of MPs \citep{2024A&A...692A.187G} and in the visualization of pulsar population \citep{2022MNRAS.515.3883G}. The topology of pulse profiles of pulsars was explored by constructing a complex network  \citep{2024A&A...687A.113V}.

In this study, inspired by the approach mentioned above to reveal hidden information in multi-variable time series and to assess the higher order measures for the detection of SGWB, we would construct complex networks for both synthetic and real PTA. Subsequently, we inspect the sensitivity of several structural characteristics (topology and geomety) of {\it a priori}\footnote{We commence by presuming physics-informed graph fabrication.} the constructed graph, including average clustering coefficient, average node strength, the mean edge weight, and the standard deviation of edge weights to the presence of SGWB. We show that this approach leads to consistent constraints of associated parameters: characteristic strain amplitude ($A_{\rm SGWB}$), and spectral index ($\gamma_{\rm SGWB}$) with traditional Bayesian technique.
\begin{figure}[!t]
	\centering
	\includegraphics[width=\textwidth]{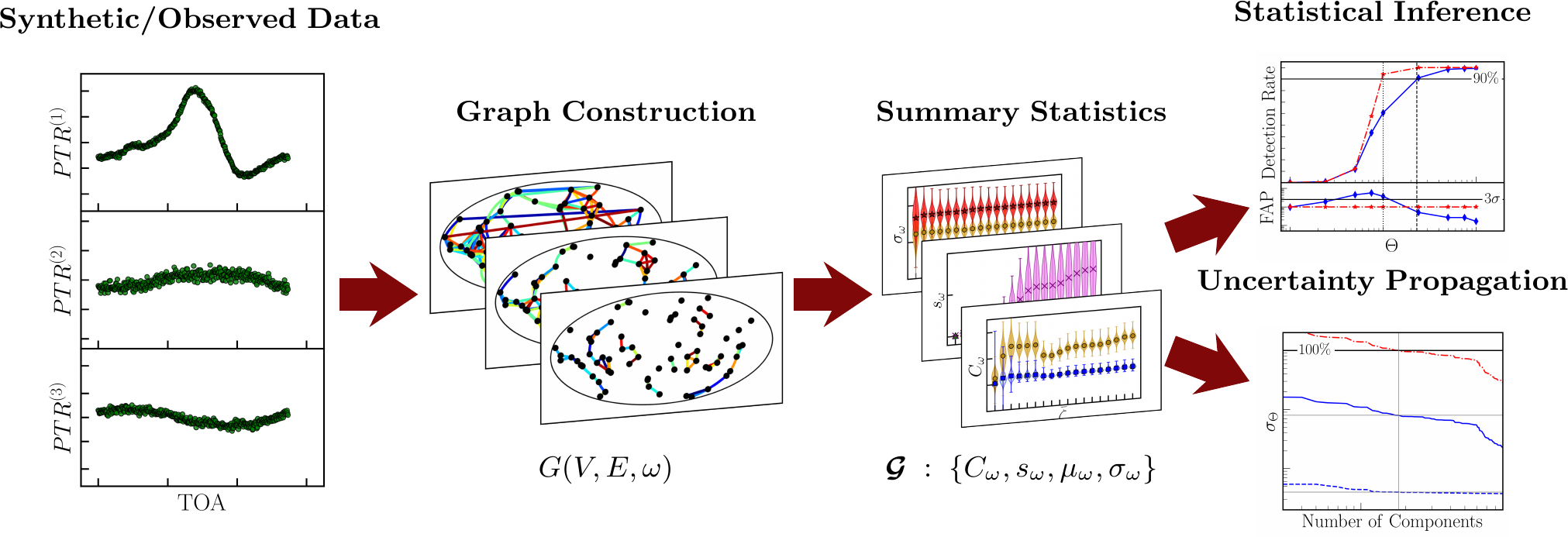}
	\caption{The proposed pipeline for PTA data analysis using graph-theory. Timing residuals, $PTR$s, are converted into complex networks, from which graph-based summary statistics are extracted. These features are used to detect the  SGWB and to evaluate uncertainties of parameters.}
	\label{fig:fig_Work_Flow}
\end{figure}
We outline the main novelties of our research as follows:\\
I) We carefully simulate $PTR$s affected by a scenario of SGWB and noise, with red noise component superimposed to construct as most as like realistic data for various values of model parameters. Relying on the graph theory, the complex network from $PTR$s of pulsars distributed in our celestial sphere, will be created.\\ 
II) Given the fact that the effect of SGWB is highly degenerated by red noise and other foreground phenomena, the Hellings \& Downs curve is investigated as the main potential indicator of SGWB footprint. \\
III) A comprehensive pipeline is proposed to ensure the input $PTR$s is adequately prepared for the implementation of our exclusive summary statistics based on graph theory (Fig.~\ref{fig:fig_Work_Flow}).\\
IV) Also, the significance of detecting the SGWB signal produced by astrophysical sources (SMBHB) captured through PTA will be assessed from the perspective of the structural characteristics of the constructed complex network by computing the $AUC$, ${\rm Welch's}$ $t$-test $p$-value and ${\rm Cohen's}$ $d$ metrics. Additionally, we will explore how the number of pulsars, total observation time span and the SGWB strength affect the capability of detection, along with the uncertainties associated with the SGWB strain amplitude and spectral index.\\
V) According to our graph-based analysis, we will assess their constraining capability on the SGWB parameters by performing the Fisher information matrix analysis.\\
VI) Finally, we implement our graph-based method on NANOGrav 15-year dataset.

The traditional inference pipeline usually focuses on summary statistics, relying on compressing observational data into low-order statistics -- most commonly the power spectrum of the two-point correlation function -- and comparing them to analytical models within the conventional inference methods. They inherently serve the explicit analytical likelihood modeling. Going beyond common summary statistics \citep{2024JCAP...09..034Y} usually leads to introduce highly non-linear measures which we are unable to theoretically predict them for a given set of cosmological parameters. Therefore, they essentially cause the absence of tractable likelihood. To mitigate such discrepancy, the likelihood-free SBI methods can be used. The SBI framework  eliminates  the necessity of explicit analytical likelihood modeling \citep{Jalali2025, 2024PhRvL.133v1006N}, and non-linear summary statistics such as the graph-based measures can be adopted to extract additional information which is essential to detect SGWB and put significant constraint on its parameters.
Given this issue, a plethora of complementary methods have been proposed and here we can clarify some of  advantages of our approach as follows:\\
I) The graph-based summary statistics provide multiple traces to extract the information from the PTA data which can in principle elucidate the SGWB imprint.\\
II) The graph-based features are inherently model-independent and therefore it is not necessary to assume particular model to compute structural characteristics of constructed graph.\\   
III) It provides suitable feature vectors for post-processing using ML algorithms like Gaussian Processes Regression (GPR) and SBI.\\
IV) It gives also a complementary to the usual Bayesian approach to cross check the results.	

The rest of the paper is organized as follows: In Sec.~\ref{sec:data}, we will give a bird eye on the SGWB imprints on the PTA data.  Data description for both synthetic and observed PTA data are also introduced in this section. We will give the graph theory foundations in Sec. \ref{sec:Graph_theory}.  We will elucidate the details of complex network construction and subsequently the derived quantities, the imprint of SGWB on the graph-based summary statistics of PTA data, alongside the significance of their distributions in  Sec.~\ref{sec:Implementation_of_Graph}. Implementation of graph theory on PTAs including  the scaling effects, error estimations of $A_{\rm SGWB}$ and $\gamma_{\rm SGWB}$ and quantifying the SGWB information content from NANOGrav 15-year data will be explained in Sec.~\ref{sec:Results}. Summary and conclusion are given in Sec.~\ref{sec:Conclusion}. 

\section{From SGWB to PTR}
\label{sec:data}

In this section, at first, we will give a brief on theoretical tenets of PTR as a tracer of SGWB. For the sake of brevity, we will collect the details in Appendix \ref{sec:SGWBs} and here we keep just the main definitions. Secondly, we will describe theoretical models for generating SGWB signals on the PTA data. We will clarify the various components of timing residuals in synthetic PTA to examine the behavior of summary statistics extracted from graph-based analysis as an indicator of the SGWB. Furthermore, we will discuss observed PTAs to assess the detection capability and compare them with the established pipeline currently used by the international PTA groups.

\subsection{PTR as a tracer of SGWB }
As mentioned in the introduction, the SGWB is the superposition of GWs produced by numerous unresolved astrophysical and/or cosmological sources with incoherent temporal phases which is statistically isotropic and Gaussian\footnote{According to the cosmological principle, we anticipate to have an isotropic background. One can consider a weak anisotropy for an astrophysical SGWB due to the non-uniform sky distribution of SMBHB which we don't treat in this paper. A non-Gaussian signature is reasonable if at least one of the connected $m$-point correlation functions (cumulants) for $m \ge 3$ is non-zero.}. The GW energy density frequency spectrum, $\Omega_{\rm GW}(f)$, can be parameterized by a scaling behavior as \citep[and references therein]{2022Galax..10...34R}:
\begin{equation}
	\label{eq:energy_power_law}
	\Omega_{\rm GW}(f)=\Omega_{\rm GW}(f_{\rm ref})\Big(\frac{f}{f_{\rm ref}}\Big)^{\alpha},
\end{equation}
in which, the $f_{\rm ref}$ and $\alpha$ are a reference frequency and the spectral index of the signal, respectively. Various theoretical models predict different values for $\alpha$ with appropriate frequency domains \citep{2022Galax..10...34R}. For coalescing black hole binaries, we have $\alpha=\frac{2}{3}$. The $\alpha=-\frac{2}{3}$ and  $\alpha=0$ are associated with cosmic strings and primordial GWs from the Big Bang. The SGWB that exists in the entire space-time leaves an intrinsic imprint on the PTRs of pulsar timing arrays.  This fingerprint is quantified with a distinctive spatial correlation pattern known as Hellings \& Downs (HD), and it follows a quasi-quadrupolar behavior with respect to the angular separation between the pulsars. The most sensitive pulsars, MSPs, can achieve high timing precision capable of detecting the faint signal of SGWB among the noise components sourced by systematic errors and various astrophysical phenomena. 
What we are observing from PTA in the presence of SGWB is a modification in the rotational frequency of the pulsar $\nu$. For a given pulsar, the relative shift in the frequency is given by \citep{2018CQGra..35p3001C}:
\begin{equation}
	\delta\nu\equiv\frac{\Delta\nu}{\nu}= -\frac{1}{2}\int_{t_{em}}^{t_{obs}}dt^\prime \dot{h}_{ij} (t^\prime, \mathbf{r} (t^\prime)) \hat{p}^i \hat{p}^j,
\end{equation}
where $t_{em}$ and $t_{obs}$ are the emission and observation time of the pulse, respectively. The $PTR$ of $a$th pulsar located at $\hat{p}^{(a)}$ due to GWs is given by: 
\begin{equation}
	PTR^{(a)}(t,\hat{p}^{(a)})=\int_0^t dt'\delta\nu(t',\hat{p}^{(a)}).
\end{equation}
The various components of $PTR$ will be  determined in the next subsection.

\begin{table}[!b]
	\centering
	\caption{Summary of synthetic and observational PTA parameters.}
	\label{tab:simulation_parameters}
	\begin{tabular}{lllll}
		\toprule
		\textbf{Component} & \textbf{Parameter} & \textbf{Mock PTA} & \textbf{Real PTA} & \textbf{Description} \\
		\midrule
		\textbf{Pulsar Array} & $N_{\rm PTA}$ & 68 & 68 & Total number of pulsars \\
		& $T_{\rm span}$ & 16.03 years & 16.03 years & Total observation duration \\
		& $\Delta t$ & 2 weeks & 1--4 weeks & Time between observations \\
		& Sky distribution & Uniform & Non-uniform & $\qquad\qquad\qquad$-- \\
		\hline
		\textbf{White Noise} & $\sigma^{(a)}$ & $[100, 1000]~\mathrm{ns}$ & $[50, 5000]~\mathrm{ns}$ & Noise strength per pulsar \\
		\hline
		\textbf{Red Noise} & $A_{\mathrm{rn}}^{(a)}$ & $[10^{-15}, 10^{-13}]$ & $[10^{-17}, 10^{-13}]$ & Strain amplitude at $f_{\text{ref}} = 1/\text{yr}$ \\
		& $\gamma_{\mathrm{rn}}^{(a)}$ & $[2, 5]$ & $[2,6]$ & Spectral index \\
		& Inclusion rate & 60\% & 50\% & Fraction of pulsars with red noise \\
		\hline
		\textbf{SGWB} & $A_{\rm SGWB}$ & $2.4 \times 10^{-15}$ & $\qquad$-- & Characteristic strain at $f_{\text{ref}} = 1/\text{yr}$ \\
		& $\gamma_{\rm SGWB}$ & $13/3$ & $\qquad$-- & Spectral index \\
		\bottomrule
	\end{tabular}
\end{table}

\subsection{Synthetic PTA Data for SGWB}
\label{PTA_Data_Simulation}

To evaluate the detection capability of our graph-based pipeline, we simulate mock PTA data incorporating the HD spatial correlation signature with simulation settings summarized in Table~\ref{tab:simulation_parameters}. Our approach requires exact spatial correlations to validate structural detection criteria. The pulsar timing residuals (PTRs) obtained after fitting a timing model (TM) to the times of arrival (TOAs) contain multiple stochastic components:
\begin{align}\label{eq:ptrcomponents}
	PTR^{(a)}(t, \hat{p}^{(a)}) = N_{\rm wn}^{(a)}(t, \hat{p}^{(a)}) + N_{\rm rn}^{(a)}(t, \hat{p}^{(a)}) + S^{(a)}_{\rm SGWB}(t, \hat{p}^{(a)}).
\end{align}
Here, $N_{\rm wn}^{(a)}(t, \hat{p}^{(a)})$ and $N_{\rm rn}^{(a)}(t, \hat{p}^{(a)})$ are the white measurement noise and intrinsic red noise, respectively, and $S^{(a)}_{\rm SGWB}(t, \hat{p}^{(a)})$ is the stochastic gravitational wave background signal, which is common to all pulsars but exhibits a spatially correlated angular structure (HD). To account for uncertainty in timing model fitting we project the simulated data using the $R$-matrix formed as \citep{2013ApJ...762...94D}:

\begin{equation}\label{eq:r-matrix}
	R = I - \mathcal{M} (\mathcal{M}^T W \mathcal{M})^{-1} \mathcal{M}^T W,
\end{equation}
where $I$ is the identity matrix, $\mathcal{M}$ is the design matrix of the pulsar which posses constant, linear and quadratic terms to account for the spin down of the pulsar. $W$ is the Noise budget matrix formed from the white and red noise components. So, the post-fit residual is computed by $PTR^{\mathrm{post-fit}}_{M \times 1} = R_{M \times M} PTR^{\mathrm{pre-fit}}_{M \times 1}$. Here after, we use the $PTR^{\mathrm{post-fit}}$, when we notice to data sets unless the otherwise mentioned. Hereafter, for convenient, we drop the ``post-fit'' superscript.

\begin{figure}[!t]
	\centering
	\includegraphics[width=\textwidth]{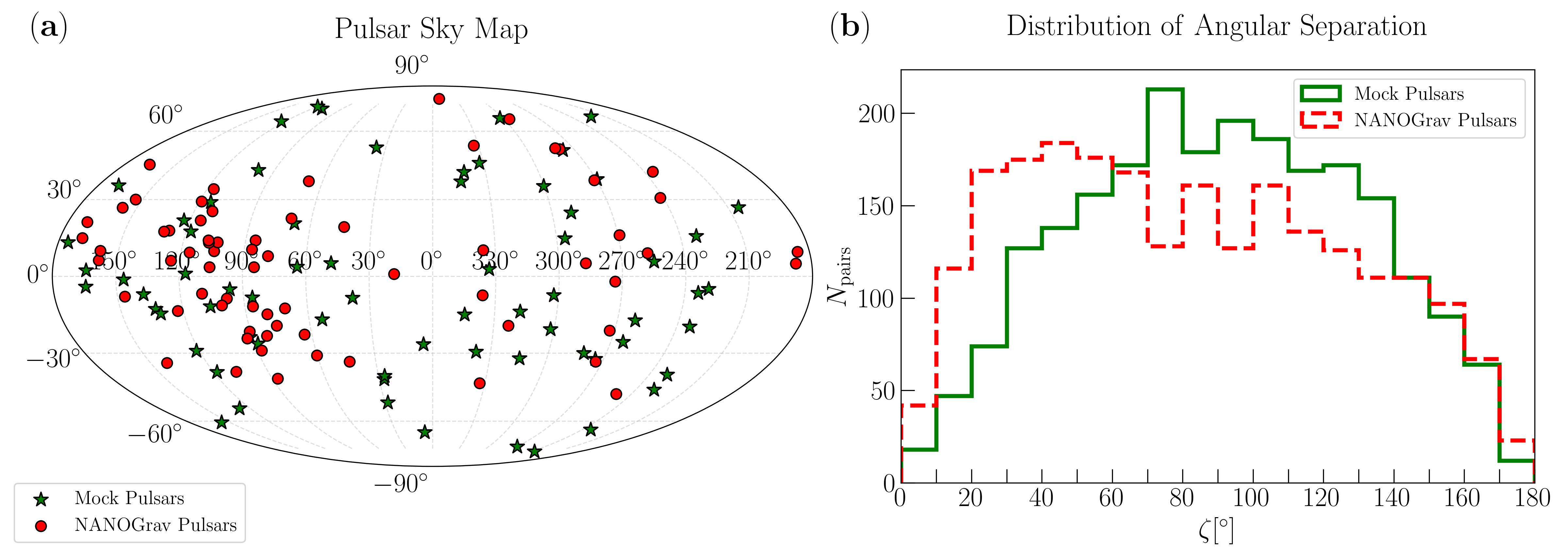}
	\caption{(a) Sky distribution of mock PTA pulsars (green stars) and NANOGrav pulsars (red circles) in celestial coordinates (Right Ascension, Declination). (b) Histogram of angular separations between all unique pulsar pairs of the mock PTA (green solid) and NANOGrav (red dashed).}
	\label{fig:pulsar_sky_map}
\end{figure}

We distribute the pulsars uniformly across the sky to avoid biases in angular correlation measurements. The left panel of Figure~\ref{fig:pulsar_sky_map} shows a realization of this uniform distribution for our mock array of 68 pulsars (green stars). The green solid line in the right panel is the distribution of angular separations of mock pulsars. The distribution is notably non-uniform where relatively fewer pulsar pairs are separated by very small (e.g., $<20^\circ$) or very large separations (e.g., $>160^\circ$), while most lie within intermediate angular ranges. As a result, the number of pulsar pairs contributing to the lowest and highest angular bins is limited, which can lead to increased statistical uncertainty in those regions.

Each pulsar's time series spans $16.03$ years of $PTR$s, sampled uniformly at a $14$-day cadence without any observational gaps or irregularities\footnote{Dealing with irregular sampled data can be accomplished through various methods. A prominent approach for creating equidistant sampling values involves convolving with a window function. This issue has been investigated in detail in \cite{Issa}.}. To construct the residuals, we first generate individual noise components, white measurement noise and intrinsic red noise, then inject the SGWB signal into the combined noises. The white noise component is modeled as uncorrelated Gaussian fluctuations with a fixed variance for each pulsar:
\begin{equation}
	\left<N_{\rm wn}^{(a)}(t, \hat{p}^{(a)})N_{\rm wn}^{(b)}(t', \hat{p}^{(b)})\right> = (\sigma^{(a)})^2 \delta (t-t') \delta_{ab},
\end{equation}
where $(\sigma^{(a)})^2$ represents the noise strength for $a$th pulsar, $\delta(t-t')$ is the Dirac delta function in time, and $\delta_{ab}$ ensures no cross-pulsar correlations. The values of $\sigma^{(a)}$ are drawn uniformly from the range $[100, 1000]~\text{ns}$ (Table~\ref{tab:simulation_parameters}). The red noise component is simulated as a stationary Gaussian process with a power-law power spectral density:
\begin{equation}
	\left<N_{\rm rn}^{(a)}(f, \hat{p}^{(a)})N_{\rm rn}^{(b)}(f', \hat{p}^{(b)})\right> = \frac{{A_{\rm rn}^{(a)}}^2}{12 \pi^2} \left(\frac{f}{f_{\mathrm{ref}}}\right)^{-\gamma_{\rm rn}^{(a)}} f_{\mathrm{ref}}^{-3} \delta (f-f') \delta_{ab},
\end{equation}
where $A_{\rm rn}^{(a)}$ and $\gamma_{\rm rn}^{(a)}$ are the strain amplitude and spectral index for $a$th pulsar, respectively, and $f_{\mathrm{ref}} = 1/\mathrm{yr}$ is the reference frequency. Red noise is included in $60\%$ of randomly selected pulsars, approximating its observed prevalence in real PTAs. Parameters $\log A_{\rm rn}^{(a)}$ and $\gamma_{\rm rn}^{(a)}$ are drawn uniformly from the ranges listed in Table~\ref{tab:simulation_parameters}.

\begin{figure}[!t]
	\centering
	\includegraphics[width=\textwidth]{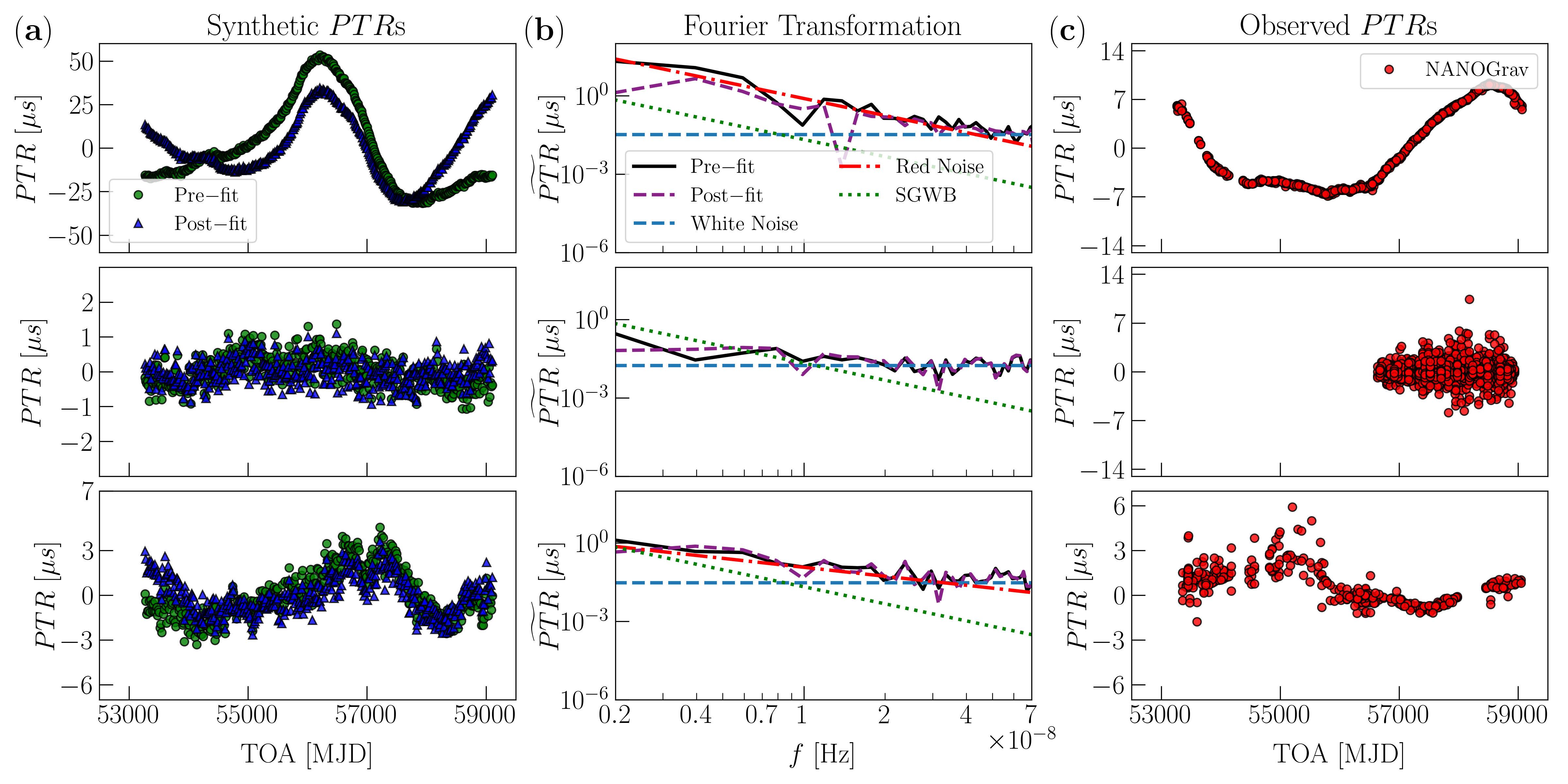}
	\caption{
		(a) Simulated post-fit residuals for three representative pulsars showing noise-only (blue triangles) and SGWB+noise (green circles) cases.
		(b) Fourier transformation of mock data, pre-fit (black solid) and post-fit (magenta dashed) with contributions from white noise (blue dashed), red noise (red dash-dotted), and SGWB signal (green dotted). The SGWB signal overlaps with red noise at low frequencies ($f \lesssim 1/\mathrm{yr}$), illustrating the challenge of detection through single pulsar analysis. 
		(c) Observed timing residuals of selected NANOGrav pulsars (red circles).
	}	
	\label{fig:sample_residuals}
\end{figure}

The isotropic, Gaussian SGWB induces a common signal in all pulsars within a PTA. While the SGWB generates timing residuals with identical power spectra across the array, its distinctive spatial correlations are encoded by the HD curve. The one-sided power spectral density of the SGWB-induced residuals is given by:
\begin{align}
	\label{eq:GW_power_spectrum}
	P_{\rm SGWB}(f) = \frac{H_0^2}{8\pi^4 f^5} \Omega_{\rm GW}(f),
\end{align}
where $H_0$ is the Hubble constant. The spectral shape of $\Omega_{\rm GW}$ depends on the physical origin of the SGWB. For astrophysical sources, such as a population of inspiraling supermassive black hole binaries (SMBHBs), the energy density spectrum is modeled as (Eq.~(\ref{eq:energy_power_law})):
\begin{align}
	\Omega_{\rm GW}^{\mathrm{SMBHB}}(f) = \frac{2 \pi^2 f^2}{3 H_0^2} A_{\rm SGWB}^2 \left(\frac{f}{f_{\mathrm{ref}}}\right)^{3-\gamma_{\rm SGWB}},
\end{align}
where $A_{\rm SGWB}$ is the characteristic strain amplitude of circular binaries evolving under gravitational radiation. The fiducial values of $A_{\rm SGWB}$ and $\gamma_{\rm SGWB}$ are listed in Table~\ref{tab:simulation_parameters}. The spatial cross-correlation function of timing residuals reads as:
\begin{eqnarray}
	\label{eq:cross_correlation}
	\mathcal{C}(\zeta, \tau)&=&\Bigg\langle \frac{1}{2T_{\mathrm{span}}}\int_{-(T_{\rm span}-\tau)}^{+(T_{\rm span}-\tau)} dt\; PTR^{(a)}(t, \hat{p}^{(a)})\;PTR^{(b)}(t+\tau, \hat{p}^{(b)}) \Bigg\rangle_{\rm pairs},
\end{eqnarray}
where $T_{\mathrm{span}}$ is the total observation duration, $\langle \rangle_{\rm pairs}$ means the average on all available pulsar pairs separated by $\cos(\zeta)\equiv|\hat{p}^{(a)}.\hat{p}^{(b)}|$.  In the presence of statistical isotropic SGWB, this quantity includes the unique footprint of SGWB which is so-called  the Hellings \& Downs signature \citep{1983ApJ...265L..39H}. The practical mathematical definition of above spatial cross-correlation is given by 
\begin{equation}
	\langle PTR^{(a)} (t, \hat{p}^{(a)}) PTR^{(b)} (t, \hat{p}^{(b)}) \rangle_t = \frac{1}{4\pi} \int d\hat{\Omega} \sum_{\diamond=+,\times} F_{\diamond}^{(a)}(\hat{\Omega}) F_{\diamond}^{(b)}(\hat{\Omega})\int^{+\infty}_{-\infty} df P_{\rm SGWB}(f)
\end{equation}
where $F_{\diamond}^{(a)} (\hat{\Omega})$ is the antenna pattern function \citep{2009PhRvD..79h4030A}. The HD Function, considering only the Earth term, is defined by:
\begin{align}\label{eq:HD_curve}
	\Gamma_{\rm HD} (\zeta) &\equiv \frac{3}{8\pi} \int d\hat{\Omega} \sum_{\diamond=+,\times} F_{\diamond}^{(a)}(\hat{\Omega}) F_{\diamond}^{(b)}(\hat{\Omega})\notag\\
	&= \frac{3}{2}\psi\ln(\psi)-\frac{\psi}{4}+\frac{1}{2}+\frac{1}{2} \delta_{ab}
\end{align}
where $\psi\equiv[1-\cos(\zeta)]/2$.

\begin{figure}[!t]
	\centering
	\includegraphics[width=\textwidth]{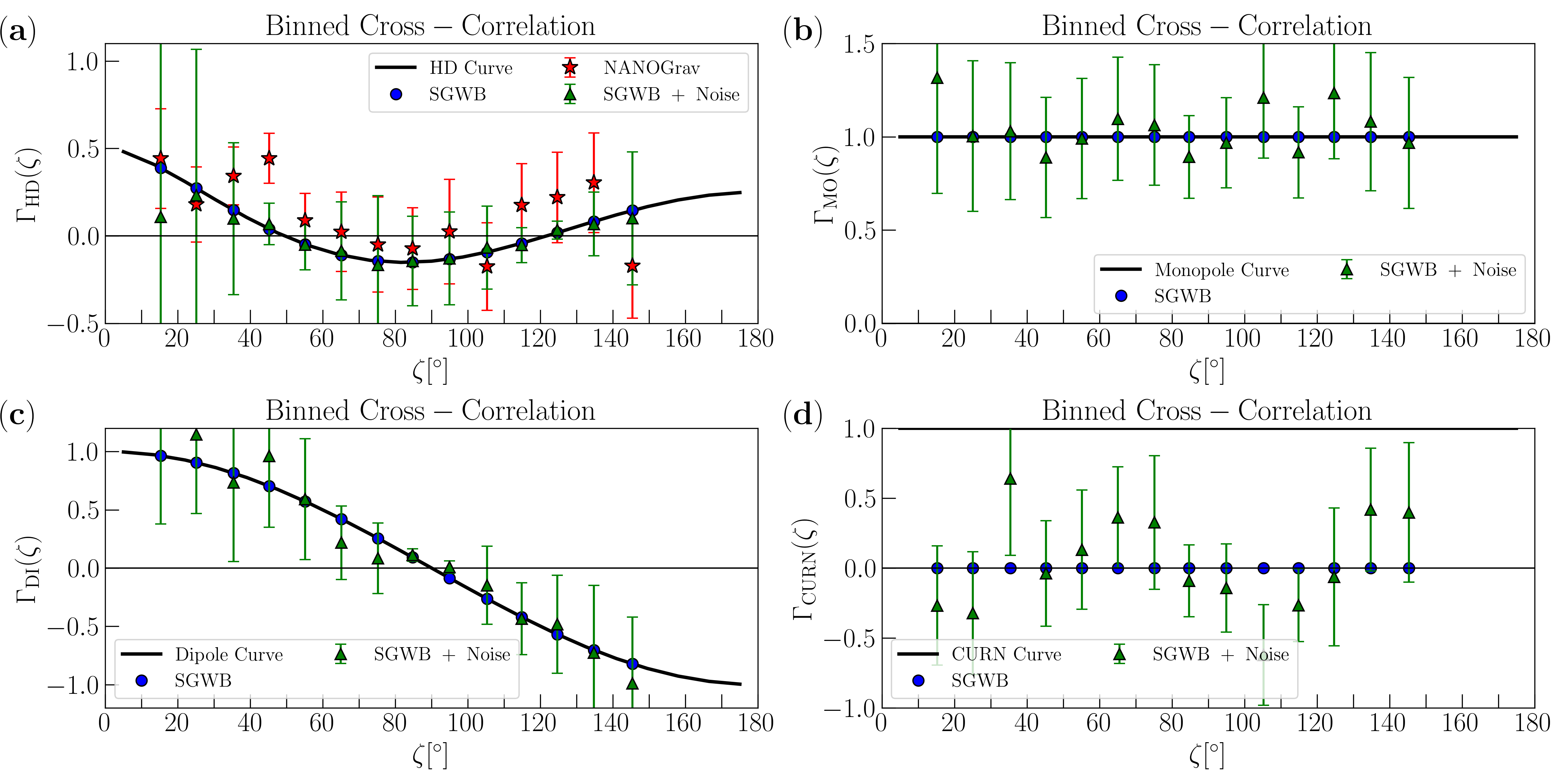}
	\caption{Recovered angular correlation patterns computed from optimal statistics for (a) Hellings \& Downs (HD), (b) Monopole, (c) Dipole, and (d) CURN models. Blue circles show pure signal realizations, green triangles signal+noise, red stars NANOGrav observations, and black lines the theoretical correlation curves. For NANOGrav data, we use the maximum-a-posteriori values of white and red noises parameters. }
	\label{fig:correlation_models}
\end{figure}

Figure~\ref{fig:sample_residuals}(a) displays simulated $PTR$s for three representative pulsars, comparing pre-fit and post-fit residuals. The Fourier transformation of pulsar timing residuals ($\widetilde{PTR}$s) is depicted in Figure~\ref{fig:sample_residuals}(b). The first pulsar (upper panel) exhibit strong red noise component which dominates the $PTR$ signal, while the second pulsar (middle panel) compromise of white noise and SGWB signal without any intrinsic red noise. This makes the slight effect of SGWB to be visible on the timing residuals at low frequency regime. The third pulsar (lower panel) has comparable SGWB and red noise. The SGWB signal is indistinguishable from intrinsic red noise in individual pulsars, emphasizing the necessity of cross-correlation analysis for detection. It turns out that the timing model projection suppresses the low-frequency power, leading to more difficulty in identifying SGWB. 

As mentioned before, the distinctive signature of an isotropic, Gaussian SGWB in PTAs is the HD spatial correlation (Eq.~(\ref{eq:HD_curve})). However, additional spatial patterns may arise from different physical processes such as system-wide clock errors resulting in constant monopole correlation pattern, $\Gamma_{\rm MO}(\zeta) = 1$. Dipole pattern indicates uncertainties in the solar system ephemeris, $\Gamma_{\rm DI}(\zeta) = \cos \zeta$. According to the NANOGrav dataset, the dipole pattern is disfavored and therefore, we ignore its contribution in our analysis \citep{2023ApJ...951L...9A,2023ApJ...951L...8A}. To complete our investigation, we also consider the case where all pulsars share a common uncorrelated red noise signal (CURN), $\Gamma_{\rm CURN}(\zeta) = 0$ in absence of the SGWB. Figure~\ref{fig:correlation_models} compares these patterns using simulated data, showing both idealized (pure signal) and realistic (signal + noise) scenarios.
The recovered angular correlation patterns are computed using optimal statistic estimator \citep{2009PhRvD..79h4030A,2015PhRvD..91d4048C} with a 10 degrees angular bin size. For the observed data, we also carry out the maximum-a-posteriori values of white and red noises parameters captured from Bayesian analysis chains of data release \citep{2023ApJ...951L...9A,2023ApJ...951L...8A}.

The SGWB signal is generated through eigenvalue decomposition of the target correlation matrix, with details provided in Appendix~\ref{app:gwb_simulation}. This method ensures exact spatial correlations with flexible switching between different correlation patterns.

\subsection{Observed PTA Data}
\label{PTA_Data_Observed}

We implement our graph-based method on the NANOGrav 15-year dataset\footnote{\texttt{https://nanograv.org/science/data}} (NG15) \citep{2023ApJ...951L...9A}. This dataset comprises high-precision timing observations from 68 millisecond pulsars monitored between 2004 and 2020. The pulsars are monitored with cadences between $1-4$ weeks using the Arecibo Observatory (AEO), the Green Bank Telescope (GBT), and the Very Large Array (VLA). The observed pulsars distribution across the sky is shown in Figure~\ref{fig:pulsar_sky_map}(a) with filled red circles and their angular separations histogram is illustrated in Figure~\ref{fig:pulsar_sky_map}(b) with dashed red line. The distribution shows some geometrical biases due to telescope sky coverage. The timing residuals contain multiple stochastic components: white noise from radiometer and jitter effects, red noise present in approximately 50\% of pulsars attributed to spin irregularities or interstellar medium variations, and potential common processes induced by the SGWB. The white noise strengths range from $\sim$50 ns for the most stable pulsars (e.g., PSR J1937$+$21) to $\sim$5 $\mu$s for more noisy pulsars (e.g., PSR J0709$+$0458). Table~\ref{tab:simulation_parameters} summarizes the observed data parameters and Figure~\ref{fig:sample_residuals}(c) shows sample PTRs of the data. Unlike our idealized simulations, the NG15 dataset suffers from some data gaps, non-uniform data acquisition, TM and ephemeris uncertainties that may produce spurious spatial correlation patterns.

\section{Graph Theory in Nutshell}
\label{sec:Graph_theory}

Graph theory is a broad, well developed mathematical framework that originated in 18th-century work by Euler \citep{euler1741}. A modern introduction to graph theory can be found in \citep{network_book2018,2016nesc.book.....B,2002RvMP...74...47A}. In this framework, a system of interacting entities (e.g., websites, proteins, or galaxies), is mapped into an ordered triple, $G:=(V, E, \omega)$, called a graph $G$, where $V\equiv\left\{v_i\right\}_{i=1}^{N}$ is the set of nodes (vertices) which represent the system's components and $N$ is known as size of graph, $E \equiv \{\epsilon_{ij} = (v_i, v_j) | v_i, v_j \in V\} \subseteq V \times V $ are the edges (links) that reflect the interactions between the system components, and $\omega$ is a real-valued weight function $\omega(\epsilon_{ij}) = \omega_{ij}$ assigned to the edges. The edges of a graph can be directed or undirected, that is, $\epsilon_{ij} \neq \epsilon_{ji}$. It is worth mentioning that the graph construction from a typical system is not unique and it generally depends on the problem that we want to solve. For example, we can map interacting individuals according to their co-working relations so we obtain the professional graph. However, The same group of persons can be mapped into a friendship graph regarding their social relations. Another mathematical representation of a graph can be achieved  through a matrix form by the adjacency matrix $\mathcal{A}$ whose elements are the weights of the edges $\mathcal{A}_{ij} = \omega_{ij}$. The adjacency matrix of an undirected graph is symmetric, $\mathcal{A}_{ij} = \mathcal{A}_{ji}$. A path is an ordered tuple of edges connecting node $v_i$ to node $v_j$,written as $\mathrm{path}_{ij} = \left( \epsilon_{ik}, \dots, \epsilon_{lj} \right)$. Such paths may not exist between all pairs of nodes and are generally not unique. Consequently, the shortest path between two nodes $v_i$ and $v_j$ is defined as the path with the minimal number of edges. The distance $d_{ij}$ between nodes is the number of edges in this shortest path. The diameter of the graph, denoted by $D$, is the maximum such distance among all node pairs: $D = \max d_{ij}$.

A directed graph is a graph whose its all edges are directed. On the contrary, if all the edges are undirected the graph is called undirected. In the weighted graph the weights of edges are different while in an unweighted graph all edge weights are equals and normalized to unity, $\omega_{ij} = 1$. Relevant properties of the system can be inferred by analyzing the structural characteristics of the mapped graph. In another word, any conceivable analysis should inherently focus on elucidation of the network topology encoded by various measures such as average path length, clustering coefficient and degree distribution \citep{2002RvMP...74...47A,2006PhRvL..97r8701Y,2013PhyA..392.3833R}. In such approach structural characteristics include topology and geometry revealing the global and local properties of constructed complex networks, respectively. The degree of a node $k_i$ is defined as the number of edges straightly connected to the node. In an undirected network, the total number of edges $L$ is given by:
\begin{align}
	L = \frac{1}{2} \sum_{i=1}^{N} k_i.
\end{align}
And the average degree $\langle k \rangle$ is given by:
\begin{align}
	\langle k \rangle = \frac{1}{N}\sum_{i=1}^{N} k_i = \frac{2L}{N}.
\end{align}
For directed graph, we can define the incoming degree $k^{\rm in}_i$ and the outgoing degree $k^{\rm out}_i$ which represent the number of edges that point to and from the node $i$, respectively. So, we have:
\begin{align}
	k_i &= k_i^{\rm in} + k_i^{\rm out}\\
	L &= \sum_{i=1}^{N} k_i^{\rm in} = \sum_{i=1}^{N} k_i^{\rm out}\\
	\langle k^{\rm in}\rangle &= \langle k^{\rm out}\rangle = \frac{L}{N}
\end{align}
The probability distribution for degree, $p(k)$, of all nodes in a graph is defined by:
\begin{equation}
	p(k)=\frac{1}{N}\sum_{i=1}^{N}\delta_{k,k_i}.
\end{equation}
One can define the $n$th cumulant of node degree, $\langle k^n \rangle$, as:
\begin{equation}
	\langle k^n \rangle = \int k^n p(k) dk,
\end{equation}
which is denoted by graph cumulants. The local clustering coefficient for node $v_i$, which quantify the tendency of a node's neighbors to connect to each other, is given by:
\begin{equation}
	C_i = \frac{2 L_i}{k_i(k_i - 1)},
\end{equation}
where $L_i$ is the number of edges between the $k_i$ neighbors which is equal to the number of triangles that node $v_i$ participates in. $C_i$ take values between $0$ and $1$. The average clustering coefficient is:
\begin{equation}
	\langle C_i \rangle = \frac{1}{N} \sum_{i=1}^{N} C_i.
\end{equation}
The correlations between node degrees are captured by the assortativity measure. Assortative graph is a graph in which nodes of similar degree tend to connect to each other. The assortativity measure in an unweighted network is given by:
\begin{equation}
	r = \frac{ \scalesum{1.3}{j > i} \Big[k_i - \langle k \rangle\Big] \Big[k_j - \langle k \rangle \Big]}{\scalesum{1.3}{i} \Big[k_i - \langle k \rangle\Big]^2}.
\end{equation}
The positive value suggests hub formation whereas the negative value indicates mixing state \citep{network_book2018}. For weighted graph, the node strength $s_i$ is defined as the sum of all edge weights connected to the $i$th node:
\begin{equation}
	s_i = \sum_{j\neq i} \omega_{ij},
\end{equation}
and average node strength becomes:
\begin{equation}
	\langle s \rangle = \frac{1}{N} \sum_{i=1}^{N} s_i.
\end{equation}
The relative importance of nodes are captured through the centrality measures. The degree centrality is defined as: $k_i / (N - 1)$. Another centrality measure is betweenness centrality, which capture how often a node participates in shortest paths between other node pairs, and is given by: $\sum_{s,t\neq i} \Big[ \mathrm{number\ of\ }\mathrm{path}_{st}\mathrm{\ that\ pass\ through\ }i / \mathrm{number\ of\ }\mathrm{path}_{st} \Big]$. Closeness centrality evaluate how close a node is, to all other nodes: $\Big[ \frac{1}{N-1} \sum_{j\neq i}d_{ij} \Big]^{-1}$.
In the next section, we will generalize the above quantities to account for weighted graph constructed for PTAs.

\section{From PTA to Graph-based Summary Statistics}
\label{sec:Implementation_of_Graph}

The unique signature of an isotropic SGWB is the HD pattern which is encoded in the spatial correlation of the pulsar timing residuals. Traditional detection approaches, whether Bayesian or frequentist, rely on pre-modeling the SGWB's power spectral density and noise components, which make them sensitive to parameter priors and simplification assumptions, (e.g., Gaussianity). In contrast to Bayesian frameworks, our detection step does not require fitting a parametric SGWB model to the data.  Meanwhile, it is motivated to introduce more complicated summary statistics that can extract non-trivial information from data.

In this work, we propose an approach where the PTA pulsars are represented as a structured graph in the framework of graph theory. Constructing a graph from a single time-series is well established and there are various methods to accomplish this, e.g.  \textit{proximity graph}, \textit{cycle graph}, \textit{visibility graph}, \textit{recurrence graph} and \textit{transition graph} (see e.g. \cite{2019PhR...787....1Z} and references therein). To establish a graph from multi-variable time series (Eq.~(\ref{eq:PTRs})), two methods were introduced, the visibility graph \citep{lacasa2012time} and correlation network \citep{2019PhR...787....1Z}.

\subsection{Complex Network Construction from PTA Data}
\label{subsec:network_construction}

To investigate the correlation pattern between the pulsars, we use the correlation network approach where the graph nodes represent the pulsars and their mutual cross-correlation determine the weight function, $\omega$. The HD correlation signature is naturally encoded in the graph and by analyzing the global and local properties of the graph we expect to trace the footprint of SGWB in the PTA data from different perspective.

\begin{figure}[!t]
	\centering
	\includegraphics[width=\textwidth]{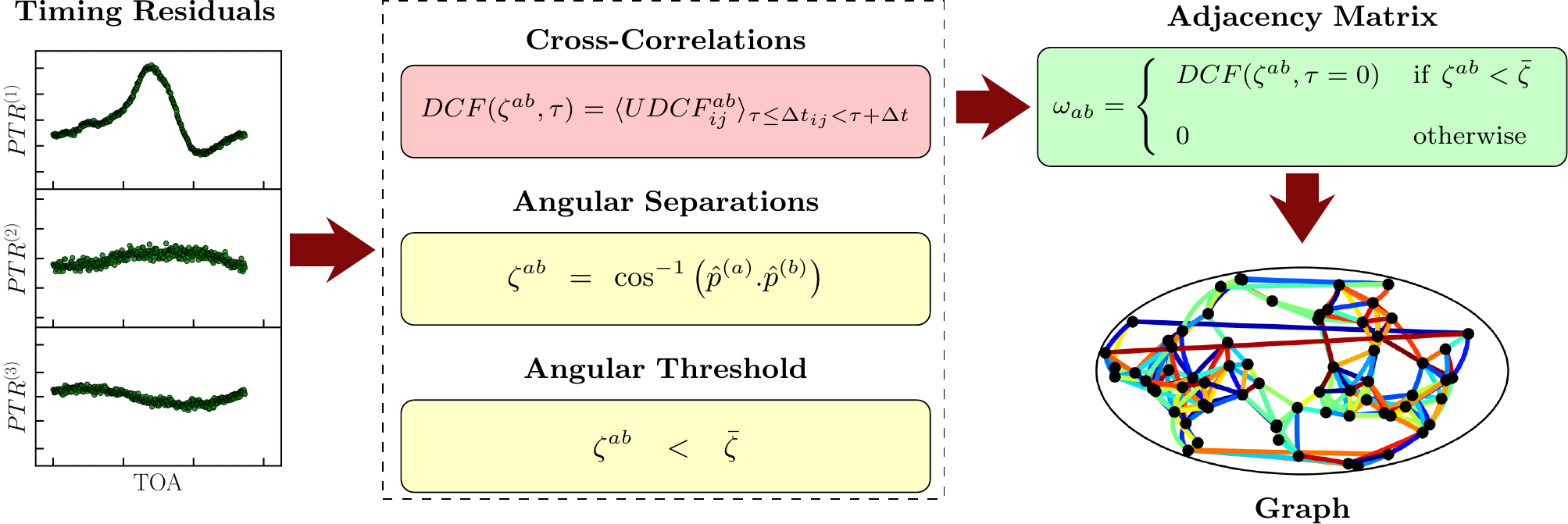}
	\caption{
		The pipeline for constructing a weighted correlation network from PTA data. Nodes represent pulsars, pulsar pairs that have angular separation below a chosen threshold, $\bar{\zeta}$, will be connected, and edge weights correspond to the discrete cross-correlations (DCF) of timing residuals, resulting in the adjacency matrix that defines the graph $G(V, E, \omega)$.
	}
	\label{fig:network_construction}
\end{figure} 

\begin{figure}
	\centering
	\includegraphics[width=\textwidth]{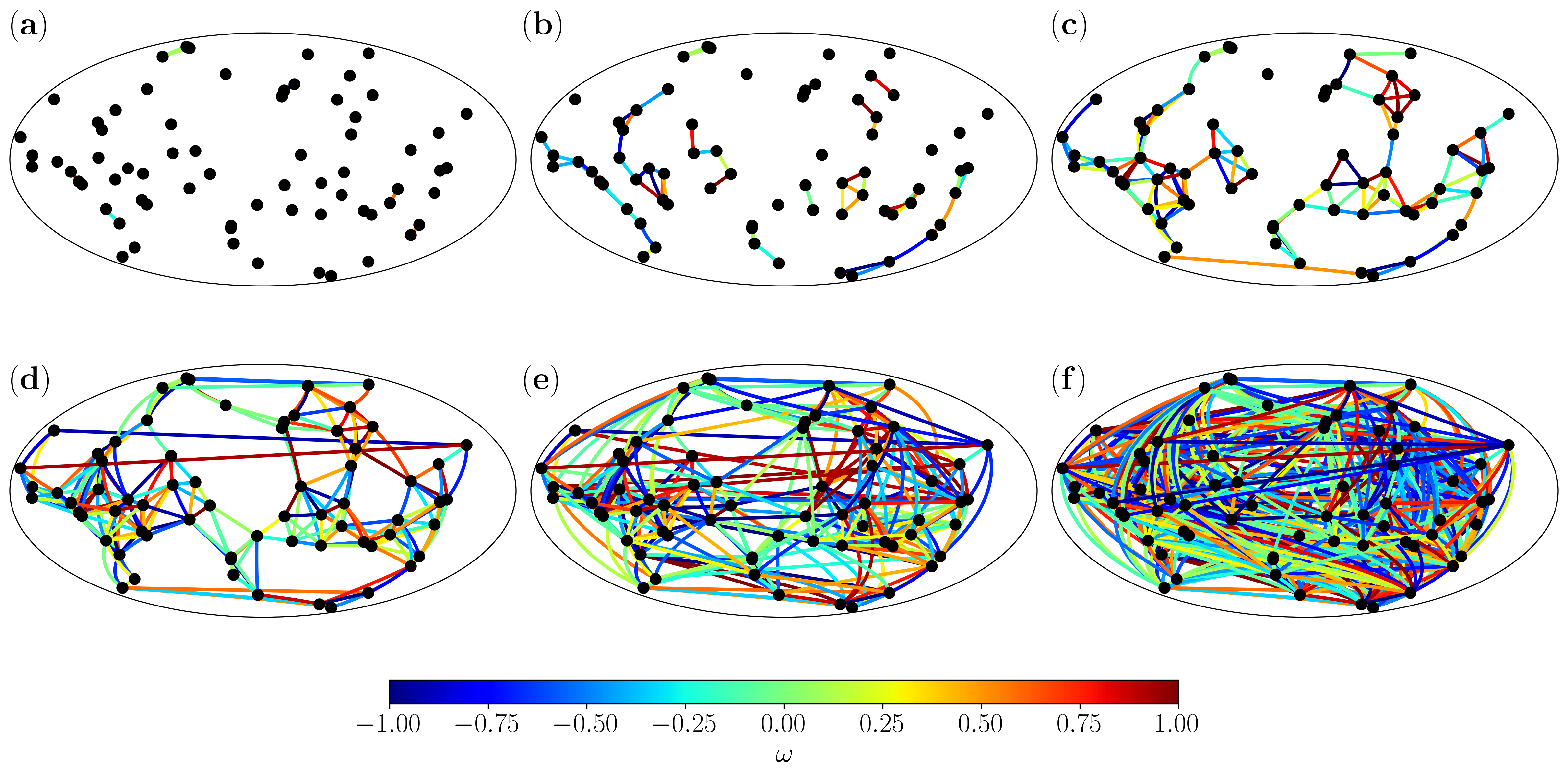}
	\caption{
		Correlation graphs constructed from mock PTA data for different angular thresholds: (a) $10^\circ$, (b) $20^\circ$, (c) $30^\circ$, (d) $50^\circ$, (e) $70^\circ$, and $90^\circ$. Nodes (filled black circles) represent pulsars.
	}
	
	\label{fig:sample_pta_graphs}
\end{figure}

Figure~\ref{fig:network_construction} illustrates the pipeline used to construct a graph from PTA data. Starting from the $PTR$s of the pulsars, we should identify the basic components of the network, namely $G=(V,E,\omega)$. The link in the graph can be constructed with various approaches. A reasonable method is to adopt the statistical correlation coefficient, which, in principle, quantifies the statistical relationship between two data sets, or, in other words, it measures how well one data set can be statistically predicted based on observations of the other set. A robust estimator for the correlation coefficient of two general time series is the discrete correlation coefficient \citep{1988ApJ...333..646E,2014MNRAS.445..437M}. Therefore, at first, we compute the unbinned discrete correlation function for a time lag of $\Delta t_{ij}\equiv t_i-t_j$ as:    
\begin{equation}
	UDCF^{(ab)}_{ij}\equiv\frac{\left[PTR^{(a)}(t_i, \hat{p}^{(a)})-\langle PTR^{(a)}(t, \hat{p}^{(a)})\rangle_{t} \right]\left[PTR^{(b)}(t_j, \hat{p}^{(b)})-\langle PTR^{(b)}(t, \hat{p}^{(b)})\rangle_{t}\right]}{\sigma_a\sigma_b},
\end{equation}
where 
\begin{eqnarray}
	\langle PTR^{(a)}(t, \hat{p}^{(a)})\rangle_{t}&=&\frac{1}{M^{(a)}}\sum_{i=1}^{M^{(a)}}PTR^{(a)}(t_i, \hat{p}^{(a)}),\\
	\sigma^2_{a}&=&\frac{1}{M^{(a)} - 1}\sum_{i=1}^{M^{(a)}}\left[PTR^{(a)}(t_i, \hat{p}^{(a)})-\langle PTR^{(a)}(t, \hat{p}^{(a)})\rangle_{t}\right]^2,
\end{eqnarray}
The discrete  cross-correlation function is estimated within independent time bins of width $\Delta t$ as:		 
\begin{eqnarray}
	DCF(\zeta^{ab},\tau)\equiv\langle UDCF^{(ab)}_{ij} \rangle\Big |_{\tau\le \Delta t_{ij}< \tau+\Delta t}.
\end{eqnarray}
To construct the weight for our graph, we take $\tau=0$ and finally we obtain the pairwise cross-correlation for a given angular separations, $\zeta^{ab}\equiv\cos^{-1}(\hat{p}^{(a)}.\hat{p}^{(b)})$. The statistical significance of the mentioned measure should be examined through a Monte-Carlo simulation. For our case, as explained in detail, we have performed the Baseline simulation and compared the derived measures for graph-based summary statistics, demonstrating their statistical significance.	
A weighted and undirected graph, $G(V, E, \omega)$, is then constructed by applying an angular separation threshold, $\bar{\zeta}$, such that, if $\zeta^{ab} < \bar{\zeta}$, an edge is added between pulsars $a$ and $b$ with weight equal to their cross-correlation; otherwise, no edge is added. However, we exclude the angular ranges between $\left[39.5^\circ-63^\circ\right]$ and $\left[104^\circ-137.5^\circ\right]$ where the absolute value HD-curve are less than $0.1$ and it resembles a noise-like fluctuations. This strategy leads to lose almost $41\%$ of total pulsar pair connections. It is worth mentioning that in our approach, the edge weight may have negative value depending on the corresponding angular separation, $\zeta^{ab}$.

To study how the spatial correlation structure evolves, we construct a family of graphs by systematically varying the angular separation threshold, $\bar{\zeta}$, between $0^\circ$ and $180^\circ$. A sample of resulting graphs is shown in Figure~\ref{fig:sample_pta_graphs}. For small thresholds (e.g., $\bar{\zeta} = 30^\circ$), the graph is sparsely connected, capturing only the correlations between nearby pulsar pairs. As $\bar{\zeta}$ increases, more distant pairs are added, leading to a denser graph, until the threshold reaches $180^\circ$, the graph contains the maximum number of allowed connections. This $\bar{\zeta}$ procedure is applied to explore the evolution of PTA correlation graph properties with the angular scales. To quantify the structural signatures of an SGWB encoded in the weight function $\omega$ of the correlation graph, we compute the weighted version of the graph characteristics introduced in sec. \ref{sec:Graph_theory} \citep{1998Natur.393..440W, RevModPhys.74.47, Onnela2005PhRvE, Newman10.1093,2020arXiv200203959G}.

{\bf I) Average clustering coefficient:} This quantity measures the tendency of nodes (pulsars) to form tightly connected triangles. A higher value indicates locally clustered groups with strong correlations, consistent with the spatial structure from an SGWB. The weighted clustering coefficient for node $a$ is
\begin{align}\label{eq:comega0}
	C_a (\bar{\zeta})= \frac{1}{k_a(k_a - 1)} \sum_{b\neq d\neq a} (\hat{\omega}_{ab}\; \hat{\omega}_{ad}\; \hat{\omega}_{bd})^{1/3},
\end{align}
where $k_a$ its degree and $\hat{\omega}_{ab}$ is the edge weight between node $a$ and $b$ normalized by the largest weight in the graph. Accordingly, the global average is:
\begin{equation}\label{eq:comega}
	C_{\omega}(\bar{\zeta})\equiv\langle C_a(\bar{\zeta}) \rangle_a = \frac{1}{|V|} \sum_a C_a(\bar{\zeta}).
\end{equation}	

{\bf II) Average node strength:} The node strength $s_a$ is the sum of all edge weights connected to the $a$th node:
\begin{equation}
	s_a (\bar{\zeta})= \sum_{b\ne a} \omega_{ab}.
\end{equation}
The mean of all individual node strengths within the graph which reflects overall pulsars correlation strength: 
\begin{align}\label{eq:somega}
	s_{\omega}(\bar{\zeta}) \equiv\langle s_a(\bar{\zeta}) \rangle_a= \frac{1}{|V|} \sum_{a=1}^{N_{\rm PTA}} \sum_{b>a} \omega_{ab}.
\end{align}
Higher values indicates pervasive correlations in the data which is useful for discriminating the signal from pure noise. 	

{\bf III) Mean edge weight:} This is the first graph cumulant derived from probability distribution of edge weights and it reads as:  
\begin{align}\label{eq:muomega}
	\mu_{\omega}(\bar{\zeta}) = \frac{1}{|E_{\bar{\zeta}}|} \sum_{a\neq b} \omega_{ab}.
\end{align}
This average pairwise cross-correlation is sensitive to signal.

{\bf IV) Standard deviation of edge weight:} The second graph cumulant of edge weights, which is the fluctuation of edge weights, or in other words, the correlation variability, reads as:
\begin{align}\label{eq:sigomega}
	\sigma_{\omega}(\bar{\zeta}) = \sqrt{ \frac{1}{|E_{\bar{\zeta}}| - 1} \scalesum{1.3}{a\ne b} \Big[\omega_{ab} - \mu_{\omega}(\bar{\zeta})\Big]^2 }.
\end{align}
Low values represent homogeneity, while higher values indicate somehow correlation anisotropies in the PTA.

\subsection{Imprint of SGWB on the Graph-based summary statistics of PTAs}
\label{subsection:summary_statistics} 

Thus far, we defined the weighted characteristics of graph, which we anticipate to be almost sensitive to SGWB fingerprint on the PTRs. However, we have also evaluated various unweighted characteristics of the graph (such as centrality measures, node degree statistics, assortativity), and our results reveal that they do not have  adequate sensitivity to identify GW. Subsequently, we introduce our graph-based summary statistics as: $\mathbfcal{G}:\{ C_{\omega}, s_{\omega}, \mu_{\omega},\sigma_{\omega}\}$ and in the following, we will look for the influence of various components in our synthetic PTA including noise as the ${\rm Baseline}$ and injected signal with various spatial correlation patterns  ($\Gamma_{\rm HD}$, $\Gamma_{\rm MO}$ and $\Gamma_{\rm CURN}$ (see subsection \ref{PTA_Data_Simulation} for more details)) on the summary statistics, $\mathbfcal{G}$. Our graph-based summary statistics have also angular separation threshold degree of freedom, hence,  $\mathbfcal{G}:\{\mathbfcal{G}_{\diamond}(\bar{\zeta}_i)\}$ and $\diamond $ can be replaced by $C,s,\mu,\sigma$. The $i$ denotes the label of each bin, which ranges from 1 to 18, corresponding to $\bar{\zeta}\in[10^{\circ},180^{\circ}]$ with a bin size $\Delta\bar{\zeta}=10^{\circ}$. Therefore $\mathbfcal{G}$ has total of 72 elements.  

\begin{figure}[!t]
	\centering
	\includegraphics[width=1.0\textwidth]{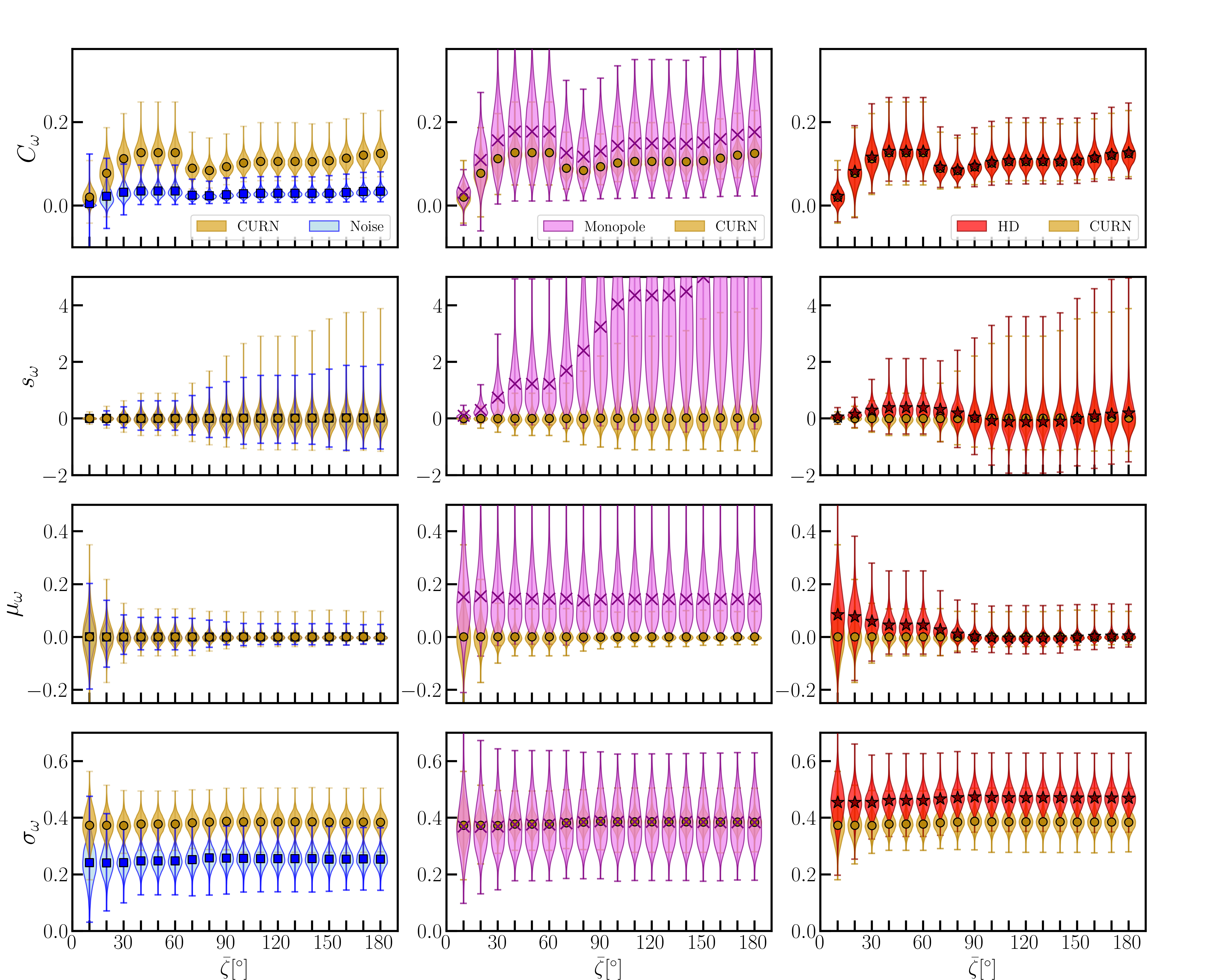}
	\caption{The empirical distributions of graph measures across angular thresholds for 10000 realizations. The graph measures are from top to bottom: I) Average clustering coefficient, II) Average node strength, III) Mean edge weight, and IV) Standard deviation of edge weight. Left column compares the ${\rm Baseline+CURN}$ (gold violins) with ${\rm Baseline}$ (blue violins) distributions. Comparison of ${\rm Baseline+Monopole}$ (violet violins) with ${\rm Baseline+CURN}$ are depicted in the middle column. Last column shows the comparison between ${\rm Baseline+HD}$ (red violins) and ${\rm Baseline+CURN}$.}
	\label{fig:network_measures}
\end{figure}

Figure~\ref{fig:network_measures} shows the empirical distributions of graph-based summary statistics computed over 10,000 realizations of synthetic data with fiducial parameter values listed in Table \ref{tab:simulation_parameters} for different spatial correlation models as a function of angular separation thresholds. Each row is assigned for one of the structural properties of constructed graph which we are going to describe them qualitatively as below and therefore we will report their significance quantitatively in the next subsection.

{\bf I) Average clustering coefficient:}
The first row of Figure~\ref{fig:network_measures} displays the first element of graph-based summary statistics,  $C_{\omega}$, as a function of $\bar{\zeta}$ (Eq. (\ref{eq:comega})).  This criterion quantifies the nature of 3-point statistics ($\overset{\triangle}{\scalebox{0.6}{abd}}$). The presence of common signal produces more triangles resulting in higher average clustering coefficient (filled circle symbols in the left panel of first row). Subsequently, it is an indicator for presence of common signal across the PTA.
The middle panel compares the $C_{\omega}$ for monopole with CURN template. This panel demonstrates that the ${\rm Baseline+Monopole}$ possesses higher clustering coefficient for wide range of angular separation thresholds (cross symbols). 
The right panel depicts that the value of $C_{\omega}$ (filled red star symbols) achieves a similar values with correspondence of ${\rm Baseline+CURN}$. Our results show that the average clustering coefficient is almost unable to distinguish the imprint of ${\rm Baseline+HD}$ and ${\rm Baseline+Monopole}$ from ${\rm Baseline+CURN}$. 

{\bf II) Average node strength:}
The second row of Figure ~\ref{fig:network_measures} reveals the sample distribution of average strength (Eq. (\ref{eq:somega})) for various templates. From the middle panel, $s_{\omega}$ is capable of discriminating between Baseline+Monopole and Baseline+CURN properly. The distribution function of $s_{\omega}$ for the ${\rm Baseline}$ and ${\rm Baseline+CURN}$ in the left panel confirms that average strength is not a proper measure for signal detection. The HD detection through the average strength is statistically weak (right panel).  

{\bf III) Mean edge weight:}
The third member of graph-based summary statistics which is associated with the mean edge weight (Eq. (\ref{eq:muomega})) is depicted as function of $\bar{\zeta}$ in the third row of Figure ~\ref{fig:network_measures}. The  mean edge weight $\mu_{\omega}$ for ${\rm Baseline+Monopole}$ becomes constant. According to middle panel, the mean edge weight is suitable for monopole detection. The ensemble average of $\mu_{\omega}$ for ${\rm Baseline}$ is clearly zero which is similar to ${\rm Baseline+CURN}$ template. The $\mu_{\omega}$ for ${\rm Baseline+HD}$ is positive at small angular scales but asymptotically approaches zero as $\bar{\zeta}$ increases, due to compensation from negative edge weights. Consequently, $\mu_{\omega}$ becomes an ineffective measure for detecting SGWB.

{\bf IV) Standard deviation of edge weight:}
Our final graph-based measure, the edge weight standard deviation $\sigma_{\omega}$, is demonstrated in the last row of Figure ~\ref{fig:network_measures}. The $\sigma_{\omega}$ is significantly higher for ${\rm Baseline+CURN}$ than for ${\rm Baseline}$. This makes it a crucial statistic for common signal detection. The ensemble average of $\sigma_{\omega}$ for ${\rm Baseline+HD}$ with characteristics reported in Table \ref{tab:simulation_parameters} gets the highest values compared to the rest of the templates. A comparison of details between ${\rm Baseline+CURN}$ and ${\rm Baseline+HD}$ confirms that a significant difference exists for all ranges of angular separation (right panel). As a result, this measure is aptly capable for both common signal detection and discriminating between ${\rm Baseline+CURN}$ and ${\rm Baseline+HD}$.

More explanations about the behavior of mentioned summary statistics are given in Appendix \ref{app:Discussion}. In the proceeding subsection, the significance of our graph-based summary statistics for SGWB detection will be quantitatively evaluated.

\subsection{Significance of Graph-based summary statistics}\label{subsection:summary_statistics1} 

\begin{figure}[!t]
	\centering
	\includegraphics[width=\textwidth]{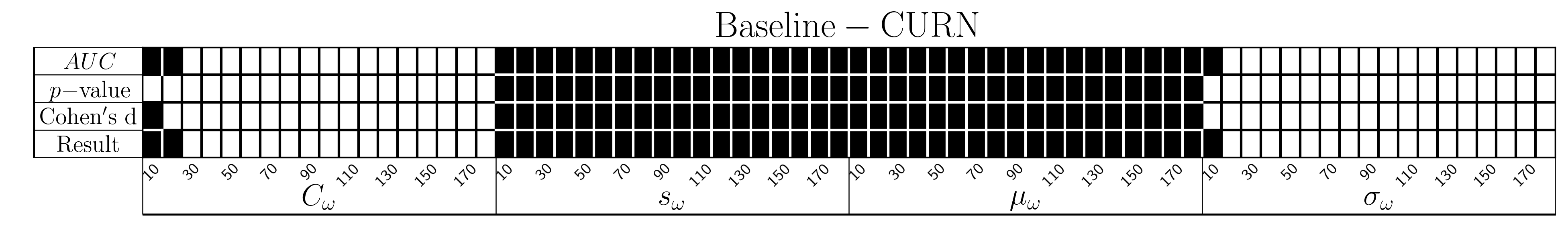}
	\includegraphics[width=\textwidth]{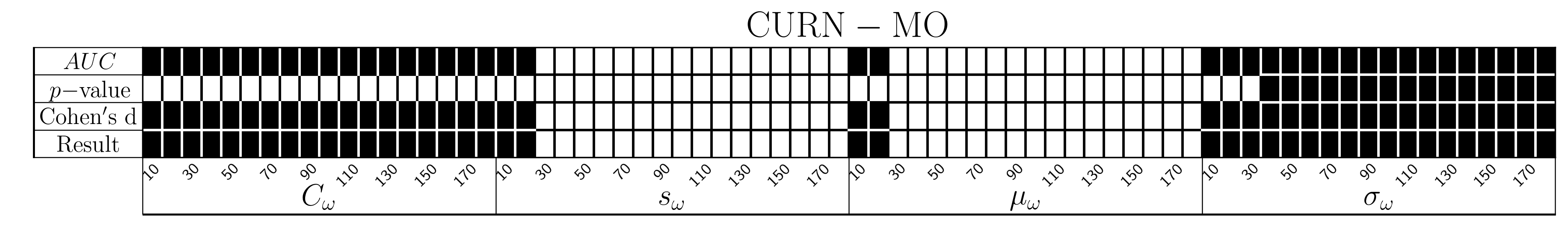}
	\includegraphics[width=\textwidth]{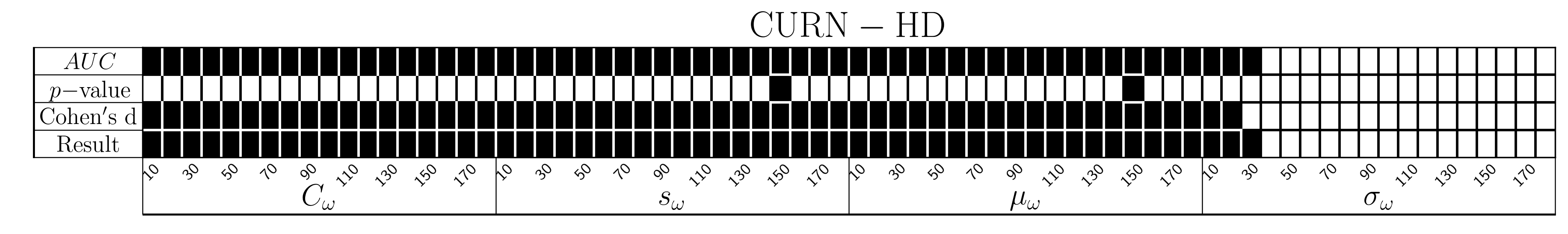}
	\caption{The binary representation of 72-component summary statistics responses to conditions assumed by evaluation metrics, such as $AUC$, $p-$value, and ${\rm Cohen's}\; d$. The white pixel refers to a component that accepts the condition associated with each metric, while the black pixel signifies a rejection of the condition. The pixels for the result row are white if all conditions imposed by the evaluation metrics are satisfied. }
	\label{fig:eligible_detection}
\end{figure}

To quantify the discriminative capability of our graph-based summary statistics, we compare the following three pairs: 
$(x_{\rm null},x_{\rm signal})\in\{({\rm Baseline},{\rm Baseline+CURN}),({\rm Baseline+CURN},{\rm Baseline+Monopole}),({\rm Baseline+CURN},{\rm Baseline+HD})\}$, through the following three metrics:

a) Area under the receiver operator characteristic curve ($AUC$) defined as the probability that a signal value ($x_{\rm signal}$) exceeds a null value ($x_{\rm null}$), $AUC = P(x_\mathrm{signal} > x_\mathrm{null})$ \citep{Hanley1982}; 

b) ${\rm Welch's}$ $t$-test $p$-value indicates a significant difference between the means of the signal and null distributions \citep{a967ba42-d0b9-3e65-976f-80cc6086b406};

c) ${\rm Cohen's}$ $d$ which quantifies the standardized effect size between distributions \citep{cohen1988spa}.

To adopt a more conservative approach, we assume that a  feature is regarded as a \emph{detection indicator} if it satisfies all the subsequent criteria:
\begin{eqnarray}\label{eq:conditions}
	AUC_{\diamond} &\ge& 0.95,\nonumber\\
	p_{\diamond}-{\rm value} &\le& 0.0027,\nonumber\\
	{\rm Cohen's}\; d_{\diamond} &\ge& 2,
\end{eqnarray}
where $\diamond$ can be replaced by element of $\mathbfcal{G}$. The aforementioned criteria ensure excellent separability, high significant ($\gtrsim 3\sigma$) and large difference between the means of signal and null distributions. Throughout this paper for convenience, we use following abbreviations for null-signal pairs: 
\begin{eqnarray}
	{\rm Baseline-CURN}& \equiv&({\rm Baseline},{\rm Baseline+CURN}),\nonumber\\
	{\rm CURN-MO}&\equiv&({\rm Baseline+CURN},{\rm Baseline+Monopole}),\nonumber\\
	{\rm CURN-HD}&\equiv&({\rm Baseline+CURN},{\rm Baseline+HD}),\nonumber
\end{eqnarray}

For common signal detection, ${\rm Baseline-CURN}$, we examine all 72-component summary statistics, $\mathbfcal{G}$. Figure \ref{fig:eligible_detection} indicates the behavior of each element of $\mathbfcal{G}$ taking into account  $AUC$, $p-$value, and ${\rm Cohen's}\; d$ evaluation metrics in the binary style. The white and black boxes are denoted to accept and reject metric conditions (Eq. (\ref{eq:conditions})), respectively. The result row is white if and only if all metrics meet the conditions. For common signal detection the $C_{\omega}$ and $\sigma_{\omega}$ are almost proper measures that can discriminate between Baseline and CURN cases. This is also consistent with the results indicated in the first and last left panels of Figure \ref{fig:network_measures}. For the Monopole scenario, the $s_{\omega}$ and $\mu_{\omega}$ are more suitable, as confirmed by the middle panels in the second and third rows of Figure \ref{fig:network_measures}. The only measure that performs well for SGWB detection is $\sigma_{\omega}$. This accomplishment is validated by the lower right panel of Figure  \ref{fig:network_measures}. 

\begin{figure}
	\centering
	\includegraphics[width=0.9\textwidth]{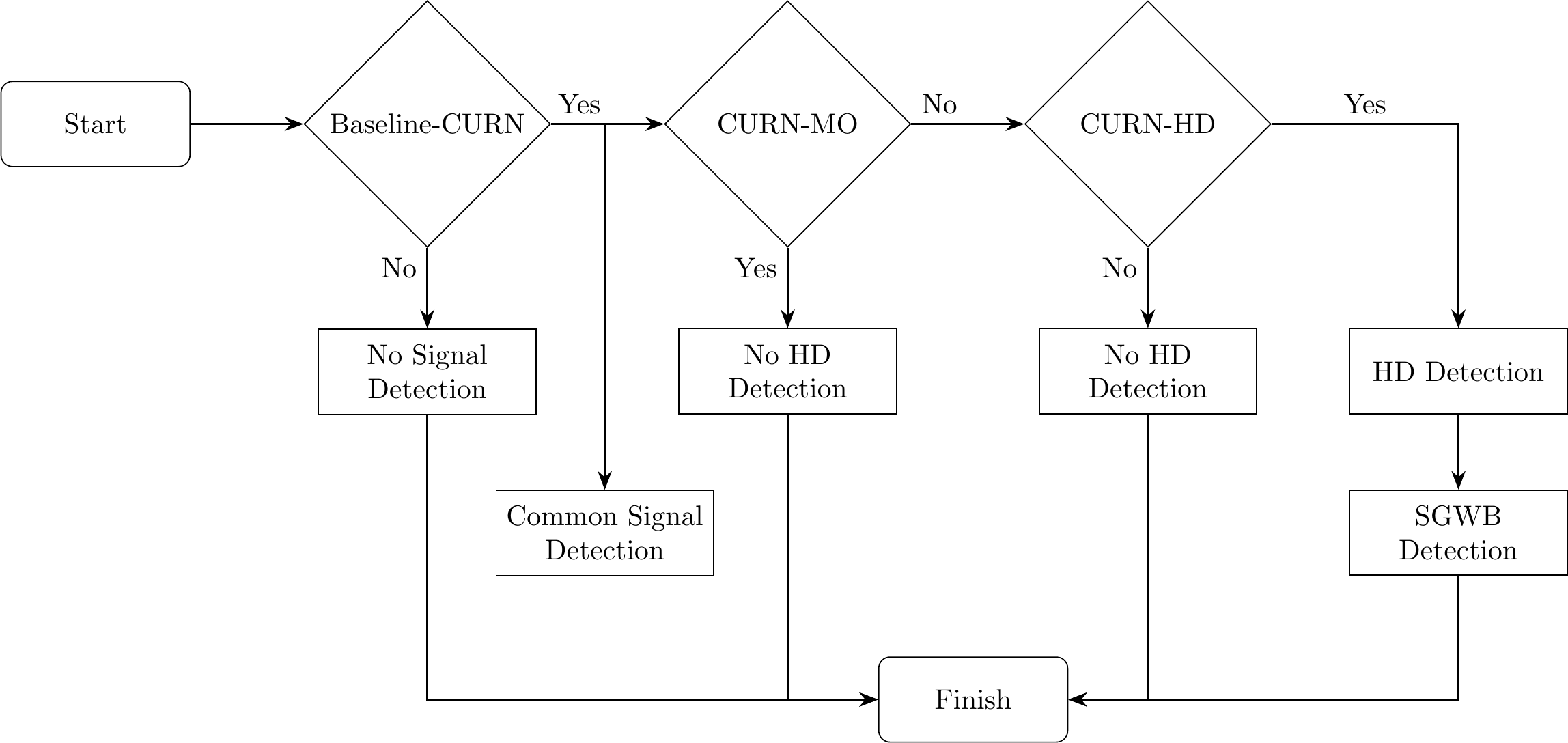}
	\caption{Decision strategy for the SGWB detection pipeline. The sole pathway for a positive detection requires a positive result in ${\rm Baseline-CURN}$, a negative result in ${\rm CURN-MO}$, and a positive result in ${\rm CURN-HD}$.}
	\label{fig:strategy}
\end{figure}

Now, according to the results from the upper panel of Figure \ref{fig:eligible_detection} (${\rm Baseline-CURN}$), we define the ``Discriminative Summary Statistics" for common signal detection as: $\mathbfcal{V}_{\rm detection}$. The $\mathbfcal{V}_{\rm detection}$ consists of a subset of members from $\mathbfcal{G}$ that fulfill all metric conditions (Eq. (\ref{eq:conditions})) to discriminate between Baseline and CURN cases. Having the $\mathbfcal{V}_{\rm detection}:\{C_{\omega},\sigma_{\omega}\}$, we turn to dealing with unknown $PTR$ data recorded in the observation for signal detection purpose and  the $\mathbfcal{V}_{\rm detection}^{\rm data}$ is computed for the observed data assuming that it is drawn from the CURN distribution. Furthermore,  the $C_{\omega}$ and $\sigma_{\omega}$ are computed for the ${\rm Baseline}$ data sets, resulting in the determination of  $\mathbfcal{V}_{\rm detection}^{\rm Baseline}$. To establish a quantitative threshold for declaring a common signal detection, we define a $z$-score for each component of the Discriminative Summary Statistics.  This score, $z\equiv(\mathbfcal{V}_{\rm detection}^{\rm data}-\mathbfcal{V}_{\rm detection}^{\rm Baseline})/\sigma_{\mathbfcal{V}_{\rm detection}^{\rm Baseline}}$ compares the Discriminative Summary Statistics values for the CURN like data set to the distribution of ${\rm Baseline}$ values. 
The components of $\mathbfcal{V}_{\rm detection}$ with a $z$-score value exceeding a specified threshold are qualified for signal detection. We implement a voting system requiring the majority of elements of $\mathbfcal{V}_{\rm detection}$ to be triggered to claim a common signal detection. This ensures that the false alarm probability (FAP) to be lower than 0.0027, which is associated with the probability of triggering below the $3 \sigma$ level of the multivariate distribution, with degree of freedom equivalent to the size of the $\mathbfcal{V}_{\rm detection}$. According to the implemented voting system, the criterion for which we are eventually allowed to announce a common signal siren, is defined by counting the number of elements in $\mathbfcal{V}_{\rm detection}$ that have been triggered. This is only permissible if this number is equal to or greater than the minimum required number of elements in $\mathbfcal{V}_{\rm detection}$, where this minimum number is determined by the predefined FAP value. The probability of triggering $m$ elements from a total of $n$ is approximately estimated by taking into account the probability of choosing the $m$ elements from $n$, multiplied by the probability of the $m$ elements being activated under the null hypothesis, with triggering threshold of $3\sigma$ allowing for faint signal detection. This condition results in approximately $\sim65\%$ of the elements in the feature vector $\mathbfcal{V}_{\rm detection}$ being activated for detection, corresponding to a creditability level of $3\sigma$.

When a positive alert for common signal detection occurs, regardless of the spatial correlation pattern, it now becomes essential to examine the nature of the spatial correlation. Considering the prevalence of monopole patterns in the $PTR$s observations, this part of the observation pipeline should be concentrated on studying the distinctions between the CURN and monopole patterns ${\rm (CURN-MO)}$. The corresponding Discriminative Summary Statistics for the Monopole strategy is $\mathbfcal{V}_{\rm Monopole}:\{s_{\omega},\mu_{\omega}\}$ which has been obtained from the middle panel of Figure \ref{fig:eligible_detection}. In analogy to the common signal detection approach, we compute  $z\equiv(\mathbfcal{V}_{\rm Monopole}^{\rm data}-\mathbfcal{V}_{\rm Monopole}^{\rm data-shuf})/\sigma_{\mathbfcal{V}_{\rm Monopole}^{\rm data-shuf}}$. Through the computation of the components of $\mathbfcal{V}_{\rm Monopole}^{\rm data-shuf}$ for the shuffled data, from which the spatial correlation pattern has been eliminated through either phase shuffling  or by randomizing the positions of pulsars in the sky. When at least $65\%$ of the components of $\mathbfcal{V}_{\rm Monopole}$ have a $z$ value exceeding $3$, the outcome of ${\rm CURN-MO}$ discrimination reveals a pattern resembling monopole in the observed data. If our criteria indicate the presence of a monopole in the data, any subsequent search for a HD pattern is meaningless. This conclusion is justified by examining the behavior of $s_{\omega}$ and $\mu_{\omega}$ (Figure \ref{fig:network_measures}) for the CURN and HD models which is demonstrating  similarity between two spatial patterns.
This, in turn, conservatively rules out the presence of a SGWB by means of our graph-based summary statistics.

If the monopole pattern is ruled out, the analysis should subsequently proceed to the HD search (${\rm CURN-HD}$). The Discriminative Summary Statistics aptly utilized for SGWB detection is denoted as $\mathbfcal{V}_{\rm HD}:\{\sigma_{\omega}\}$, as depicted in the lower panel of Figure \ref{fig:eligible_detection}. We then compute $z\equiv(\mathbfcal{V}_{\rm HD}^{\rm data}-\mathbfcal{V}_{\rm HD}^{\rm data-shuf})/\sigma_{\mathbfcal{V}_{\rm HD}^{\rm data-shuf}}$. Validation of the SGWB detection requires that at least 65\% of the components in $\mathbfcal{V}_{\rm HD}$ have a $z\ge 3$, corresponding to a $3\sigma$ credibility level.

Figure \ref{fig:strategy} outlines our proposed strategy for SGWB detection according to the proposed graph-based analysis. A conservative HD detection is only claimed following a positive ${\rm Baseline-CURN}$ alert, a rejection of Monopole hypothesis, and finally, a significant ${\rm CURN-HD}$ difference.

\section{Implementation of Graph Theory on PTAs}
\label{sec:Results}

Here, we apply the graph theory approach explained in the previous section on the synthetic and NG15 datasets. The robustness of our graph-based summary statistics with the enhancement of PTA sensitivity including the total observation time span and the number of pulsars, will be assessed. The uncertainty of SGWB parameters will be evaluated through the Fisher forecast formalism. Finally, we will implement our graph-based detection pipeline on the NG15 data.

\subsection{Scaling effects on the graph-based detection}
\label{subsectionsec:scaling}

\begin{figure}[!t]
	\centering
	\includegraphics[width=\textwidth]{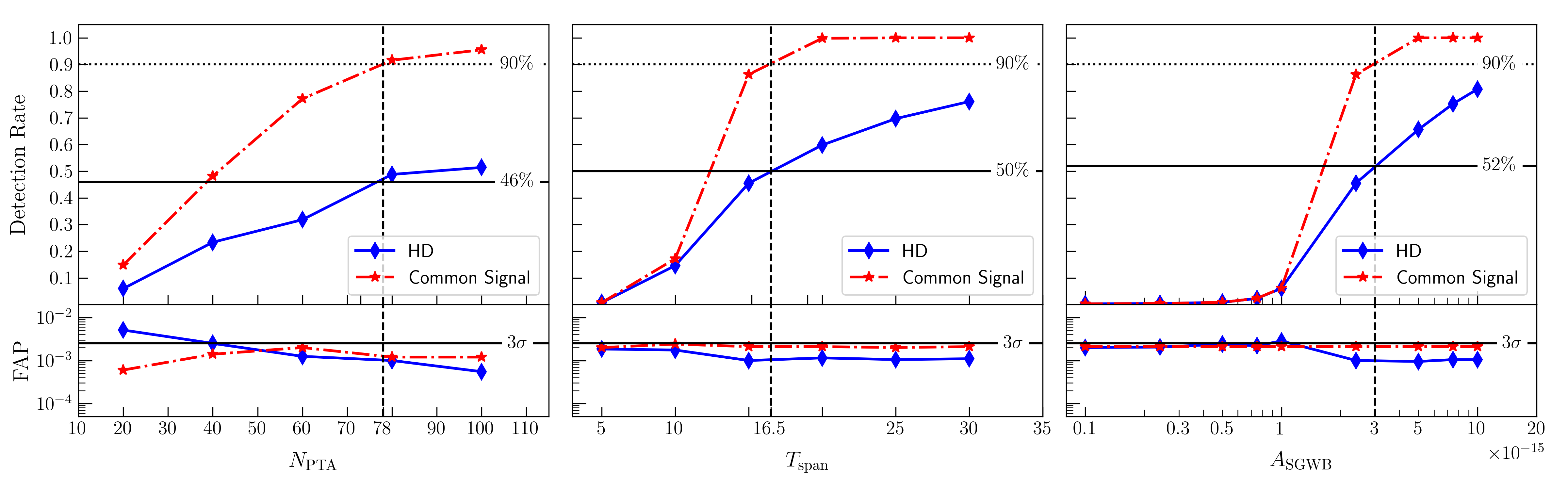}
	\caption{Detection rate and false alarm probability (FAP) of the SGWB graph-based pipeline as a function of number of pulsars $N_{\mathrm{PTA}}$ (left panel), observing time span $T_{\mathrm{span}}$ (middle panel), and SGWB strain amplitude $A_{\mathrm{SGWB}}$ (right panel). The Blue solid lines show the SGWB detection, and the red dash-dotted line depicts the common signal detection. The vertical black dashed lines mark the thresholds where the detection rate for common process exceeds $90\%$.}
	\label{fig:scaling_properties}
\end{figure}
To evaluate the robustness and performance of our graph-based detection framework under varying observational and astrophysical conditions, $\{N_{\rm PTA},T_{\rm span},A_{\rm SGWB}\}$, we conduct a systematic scalability study. Specifically, we generate 10,000 realizations for each configuration while independently varying three key parameters around their fiducial values given in Table~\ref{tab:simulation_parameters}: $N_{\mathrm{PTA}} = \{20, 40, 60, 80, 100\}$ pulsars, $T_{\mathrm{span}} = \{5, 10, 15, 20, 25, 30\}$ years, and $A_{\mathrm{SGWB}} = \{1.0\times 10^{-16}, 2.5 \times 10^{-16}, 5.0 \times 10^{-16}, 7.5 \times 10^{-16}, 1.0\times 10^{-15}, 2.4 \times 10^{-15}, 5.0 \times 10^{-15}, 7.5 \times 10^{-15}, 1.0\times 10^{-14}\}$.

Detection performance is evaluated using the feature-based strategy described in subsection~\ref{subsection:summary_statistics1}. Figure~\ref{fig:scaling_properties} shows the detection rate and the FAP of the pipeline with respect to $N_{\mathrm{PTA}}$, $T_{\mathrm{span}}$, and $A_{\mathrm{SGWB}}$. The FAP of common signal (red dash-dotted line) is defined as the ratio of ${\rm Baseline}$ cases (pure noise) falsely classified as signal detection to the total number of ${\rm Baseline}$ realizations. The FAP is controlled below the $3\sigma$ level for all PTA configurations thanks to our voting system described in subsection~\ref{subsection:summary_statistics1}. 

The detection rate of the common signal is also defined as the number of CURN realizations successfully classified as signal detected divided by the total number of realizations that include the signal. This rate increases with PTA sensitivity, approaching $\sim 90\%$ for the fiducial parameters. The minimum number of pulsars needed to have $90\%$ chance of detection while keeping other parameters fixed to their fiducial values is $78$ pulsars (left panel of Figure~\ref{fig:scaling_properties}). The minimum observation time to pierce the $90\%$ level is $16.5$ years (middle panel of Figure~\ref{fig:scaling_properties}), while shorter observation time limits the detection chance. Naturally, SGWB with higher characteristic strain amplitude has a higher chance of detection than fainter signals, and the lowest strain amplitude to get $90\%$ detection rate of common signal is $3 \times 10^{-15}$ (right panel of Figure~\ref{fig:scaling_properties}).

Upon confirmation of the detection of a common signal, further classification of the signal spatial correlation pattern is essential. We assert the detection of SGWB within $3\sigma$ credibility level only if an HD-pattern is recognized, as described in our strategy illustrated in Figure~\ref{fig:strategy}. The blue solid lines in Figure~\ref{fig:scaling_properties} show the detection rate and FAP of SGWB with respect to PTA parameters. Similar to the common signal, the FAP of SGWB is defined as the ratio of Monopole or CURN realizations that are misclassified as HD to the total number of Monopole/CURN cases.

The FAP for less sensitive PTA configurations lies higher than $3\sigma$ level, as the discrimination between the Monopole and HD cases is diminished due to the significant overlap of their feature vectors distributions, specifically the components of $\mathbfcal{V}_{\rm Monopole}:\{s_{\omega},\mu_{\omega}\}$ and $\mathbfcal{V}_{\rm HD}:\{\sigma_{\omega}\}$. By enhancing the sensitivity of PTA, specifically through the addition of more pulsars, extending the observation duration, and/or higher SGWB strain amplitude, the capability of distinguishing signals is restored, resulting in the FAP falling below the $3\sigma$ level (right panel of Figure~\ref{fig:scaling_properties}). Accordingly, the minimum strain amplitude of SGWB can be detected at $3\sigma$ confidence level is $A_{\rm SGWB}\gtrsim 1.2\times 10^{-15}$.

Now, we also define the detection rate of SGWB as the ratio of HD realizations successfully identified as HD detections to the total number of HD cases. The HD detection rate reaches high values for a more sensitive PTA configuration, and its value for our fiducial parameters is $\sim 45\%$. The detection rate of HD pattern with $78$ pulsars (corresponds to $90\%$ detection rate of common signal with other parameters fixed to their fiducial values) is $46\%$. Similarly, the detection rate reach $50\%$ for $16.5$ years of observations (without adding any new pulsar to the array), and $52\%$ with higher SGWB amplitude of $3 \times 10^{-15}$ at the fiducial PTA configurations.

\subsection{Error Estimation and Fisher Forecast}
\label{sec:fisher_forecast}

\begin{figure}[!t]
	\centering
	\includegraphics[width=\linewidth]{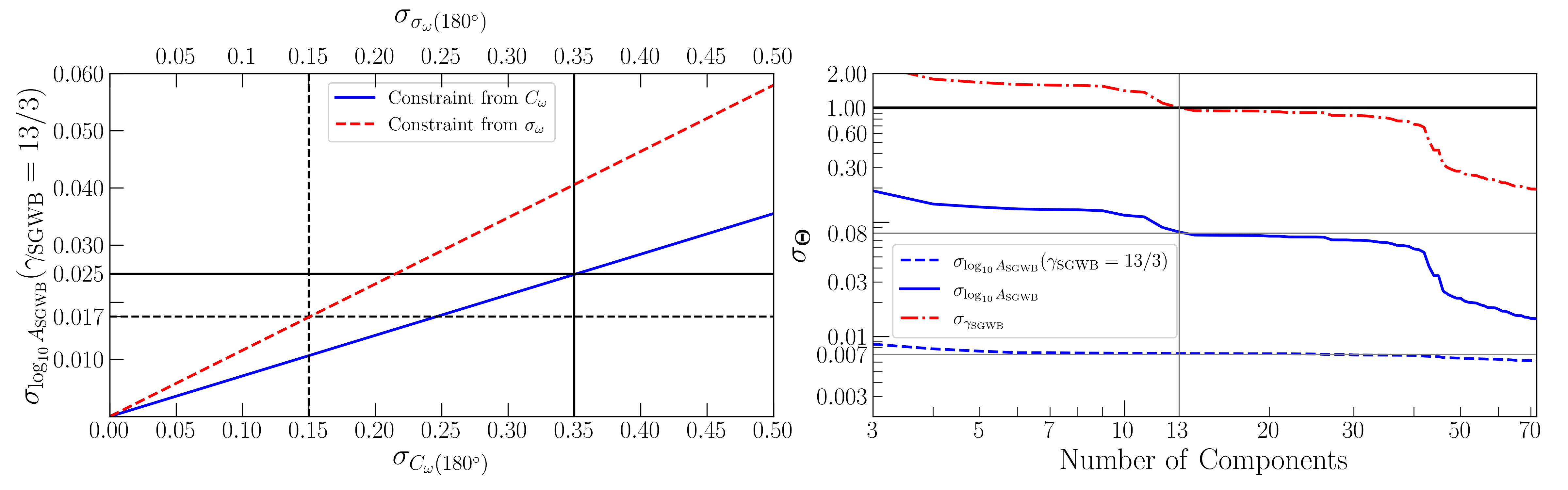}
	\caption{left panel: The relative statistical error, $\sigma_{\rm log_{10}A_{\rm SGWB}}$ for the fiducial value ${\bf \Theta}_{\rm fiducial}:\{\log_{10} A^{\rm fiducial}_{\rm SGWB}=-14.6, \gamma^{\rm fiducial}_{\rm SGWB}=13/3\}$ as a function of error associated with $C_{\omega}$ (blue solid line) and $\sigma_{\omega}$ (red dashed line). The vertical solid line corresponds to the  35\% relative error of $C_{\omega}$ leading to 2.5\% relative error on the $\log_{10}A_{\rm SGWB}$ at the fiducial point. The vertical dashed line shows the 15\% relative error of $\sigma_{\omega}$, which is associated with 1.7\% relative error on the $\log_{10}A_{\rm SGWB}$. The value of relative errors adopted for $C_{\omega}$ and $\sigma_{\omega}$ in this plot has been computed at fiducial value for $95\%$ credibility level. Here we took $\bar{\zeta}=180^{\circ}$. right panel: The blue dashed line and the blue solid line display relative statistical error of $\log_{10}A_{\rm SGWB}$ for the cases with and without (fixed) marginalization over the spectral index, respectively. While the red dash-dot line depicts the relative uncertainty of SGWB spectral index marginalized over $\log_{10}A_{\rm SGWB}$. The solid vertical line corresponds to taking into account the first 13 principal components  such that the $\sigma_{\gamma_{\rm SGWB}}=100\%$.}
	\label{fig:relative_error}
\end{figure}

To extract cosmological information and put pristine constraints on the SGWB model free parameters, ${\bf \Theta}:\{\log_{10}A_{\rm SGWB}, \gamma_{\rm SGWB}\}$, through the graph-based feature vector, $\mathbfcal{G}$, and due to its highly non-linear dependency to desired parameters,  we need to set up proper emulator for model prediction. A feasible way is training a Gaussian process emulator \citep{2014bda..book.....G} such that a machine learning procedure is adopted to predict distribution of feature vector for arbitrary parameters in training range. Similar method has been utilized in some cosmological inferences through topological data analysis e.g. \citep{2021A&A...648A..74H,2023PhDT........23G,2024Univ...10..464V}. Another robust approach to cosmological inference is SBI. This method is also known as likelihood-free inference. Particularly, for our purpose and by using the kind of our feature vector, the SBI is highly recommended for achieving reliable parameter estimation  \citep[and references therein]{tejero2020sbi,2016arXiv160506376P,2019MNRAS.488.4440A,2020PNAS..11730055C,2022mla..confE..24H}. According to either appropriate emulator or SBI, we can use NG15 dataset and by computing the data vector which is $\mathbfcal{G}$ as a new observable quantity to infer parameter estimation and their uncertainties. Alternatively, here we carry out the error propagator framework to quantify the level of uncertainty in measuring $\log_{10}A_{\rm SGWB}$ and $\gamma_{\rm SGWB}$ when the graph-based feature vector used as a new observable measures. The statistical relative error on desired parameters is given by:           
\begin{equation}
	\label{eq:uncertainty_level}
	\sigma_{\bf{\Theta}}^2 = \left(\frac{\partial \ln \mathbfcal{G}}{\partial \ln \bf{\Theta}}\right)^{-2} \sigma_{\mathbfcal{G}}^2.
\end{equation}
Here $\sigma_{\bf{\Theta}}$ and $\sigma_{\mathbfcal{G}}$ are relative statistical error of cosmological parameters and error associate with each element of our graph-based data vector, respectively. The relative uncertainty level of $\log_{10}A_{\rm SGWB}$ with fixed value of $\gamma_{\rm SGWB}$ at fiducial value  propagating from $C_{\omega} (180^\circ)$ and $\sigma_{\omega}(180^\circ)$ are shown as the blue solid line and red dashed line in the left panel of Figure~\ref{fig:relative_error}, respectively. The vertical solid line corresponds to the  35\% relative error of $C_{\omega}$ leading to 2.5\% relative error on the $\log_{10}A_{\rm SGWB}$ at the fiducial point. The vertical dashed line shows the 15\% relative error of $\sigma_{\omega}$, which is associated with 1.7\% relative error on the $\log_{10}A_{\rm SGWB}$. The value of relative errors adopted for $C_{\omega}$ and $\sigma_{\omega}$ in this plot has been computed at fiducial value for $95\%$ credibility level.  

To get linearly independent components from our feature vector $\mathbfcal{G}$, we apply principal component analysis (PCA) \citep{abdi2010principal}. Thus, we consider the principal components extracted by PCA as the feature vector,  $\mathbfcal{A}:\{PC_i\},\; i=1,...,72$, instead. Taking into account propagated uncertainties through a new set of data vectors, we compute the relative uncertainty on the $\log_{10}A_{\rm SGWB}$ given $\gamma_{\rm SGWB}=13/3$ as a function of the accumulative number of components in the right panel of Figure~\ref{fig:relative_error}. The blue dashed line represents the relative statistical error of $\log_{10}A_{\rm SGWB}$ for $\gamma_{\rm SGWB}=13/3$. The blue solid line indicates the $\sigma_{\log_{10}A_{\rm SGWB}}$ marginalized over $\gamma_{\rm SGWB}$. Whereas the red dash-dot line depicts the relative uncertainty of SGWB spectral index marginalized over $\log_{10}A_{\rm SGWB}$. The solid vertical line corresponds to taking into account the first 13 principal components where $\sigma_{\gamma_{\rm SGWB}}$ falls bellow $100\%$ level. This means that at least 13 of the largest principal components are necessary to constrain the $\gamma_{\rm SGWB}$. We also get 8\% and 0.7\% relative uncertainties of $\log_{10}A_{\rm SGWB}$ for the cases with and without (fixed) marginalization over the spectral index, respectively.

\begin{figure}[!t]
	\centering
	\includegraphics[width=0.45\linewidth]{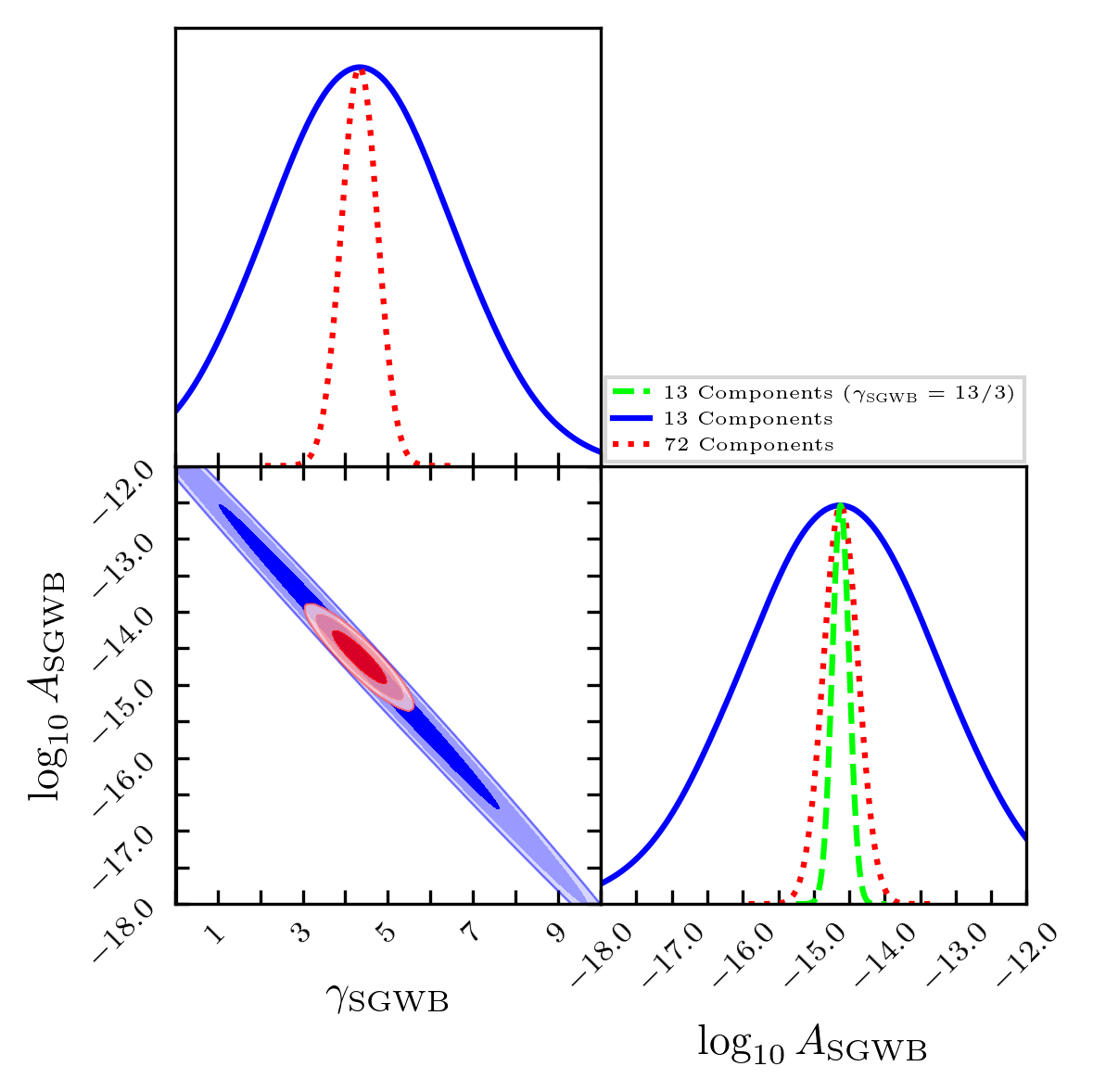}
	\caption{The constraints on $\log_{10} A_{\rm SGWB}$ and $\gamma_{\rm SGWB}$ for 13 principal components (blue contours) and 72 principal components (red contours) in the framework of Fisher forecast. The marginalized posteriors of $\log_{10} A_{\rm SGWB}$ (lower right panel) and $\gamma_{\rm SGWB}$ (upper panel) show how adding more features increase the constraining power of graph-based summary statistics. Blue lines represents the posteriors using 13 components and red dotted lines depicts the posteriors using 72 components. The posterior of $\log_{10} A_{\rm SGWB}$ for $\gamma_{\rm SGWB} = 13/3$.}
	\label{fig:constraints}
\end{figure}

Now, relying on the Fisher information matrix, we quantify the capability of graph-based measure and their combination through PCA method on constraining SGWB parameters, ${\bf\Theta}$.  The Fisher information matrix element is given by: 
\begin{equation}
	F_{\alpha\beta} =\frac{\partial \mathbfcal{A}^T}{\partial{\bf \Theta}_{\alpha}} \,
	\mathbf{C}^{-1} \,
	\frac{\partial \mathbfcal{A}}{\partial{\bf \Theta}_{\beta}} ,
\end{equation}
where $\mathbf{C}$ is the covariance matrix. To estimate the unbiased covariance matrix, we apply the standard correction on  $\mathbf{C}$ as \citep{2007A&A...464..399H}:
\begin{equation}
	\mathbf{C}^{-1}_{\mathrm{unbiased}} = \frac{N_{\mathrm{sim}} - N_{\mathrm{tot}} - 2}{N_{\mathrm{sim}} - 1} \, \mathbf{C}^{-1},
\end{equation}
where $N_{\mathrm{sim}}=10,000$ and $N_{\mathrm{tot}}$ is equal to the dimensionality of the feature vector $\mathbfcal{A}$. The minimum number of principal components to pierse  the 100\% relative error threshold on the spectral index is 13 (right panel of Figure \ref{fig:relative_error}); we begin our investigation with a data vector with the same dimension. The blue solid lines are devoted to the marginalized likelihood of SGWB parameters when the first 13 principal components are taken into account in  Figure~\ref{fig:constraints}. If we include all elements of $\mathbfcal{A}$ for Fisher forecast, the marginalized confidence contours for $\log_{10}A_{\rm SGWB}$ and $\gamma_{\rm SGWB}$ become more stringent as indicated by dark red (68\%)  and light red (95\%) contours in Figure~\ref{fig:constraints}.  For this case, the optimistic uncertainty levels of $\log_{10} A_{\rm SGWB}$  and $\gamma_{\rm SGWB}$ reach  $1.5\%$ and $19.5\%$, respectively, at $2\sigma$ confidence interval.  The posterior of $A_{\rm SGWB}$ for $\gamma_{\rm SGWB} = 13/3$ is shown via the green dashed line. It is worth noting that an increasing number of components can not change the degeneracy between SGWB strain amplitude and spectral index.

The Fisher forecast through the graph-based summary statistics can provide constraints on SGWB parameters which is in agreement with the Bayesian inference. We advocate that our  approach can be considered as a complementary method for PTA data analysis for both SGWB detection and associated parameters estimation by accounting the SBI method.

\subsection{Quantifying SGWB information content from NG15 dataset}

In the following, we will apply our method on the real PTRs which are publicly available from NANOGrav, the NG15 dataset, which is described in subsection \ref{PTA_Data_Observed} and its parameters are summarized in Table~\ref{tab:simulation_parameters}. The PTA constitutes of $68$ pulsars observed for $16.03$ years total observation time, indeed not all the pulsars have been observed for the whole time span.

\begin{figure}[!t]
	\centering
	\includegraphics[width=0.7\textwidth]{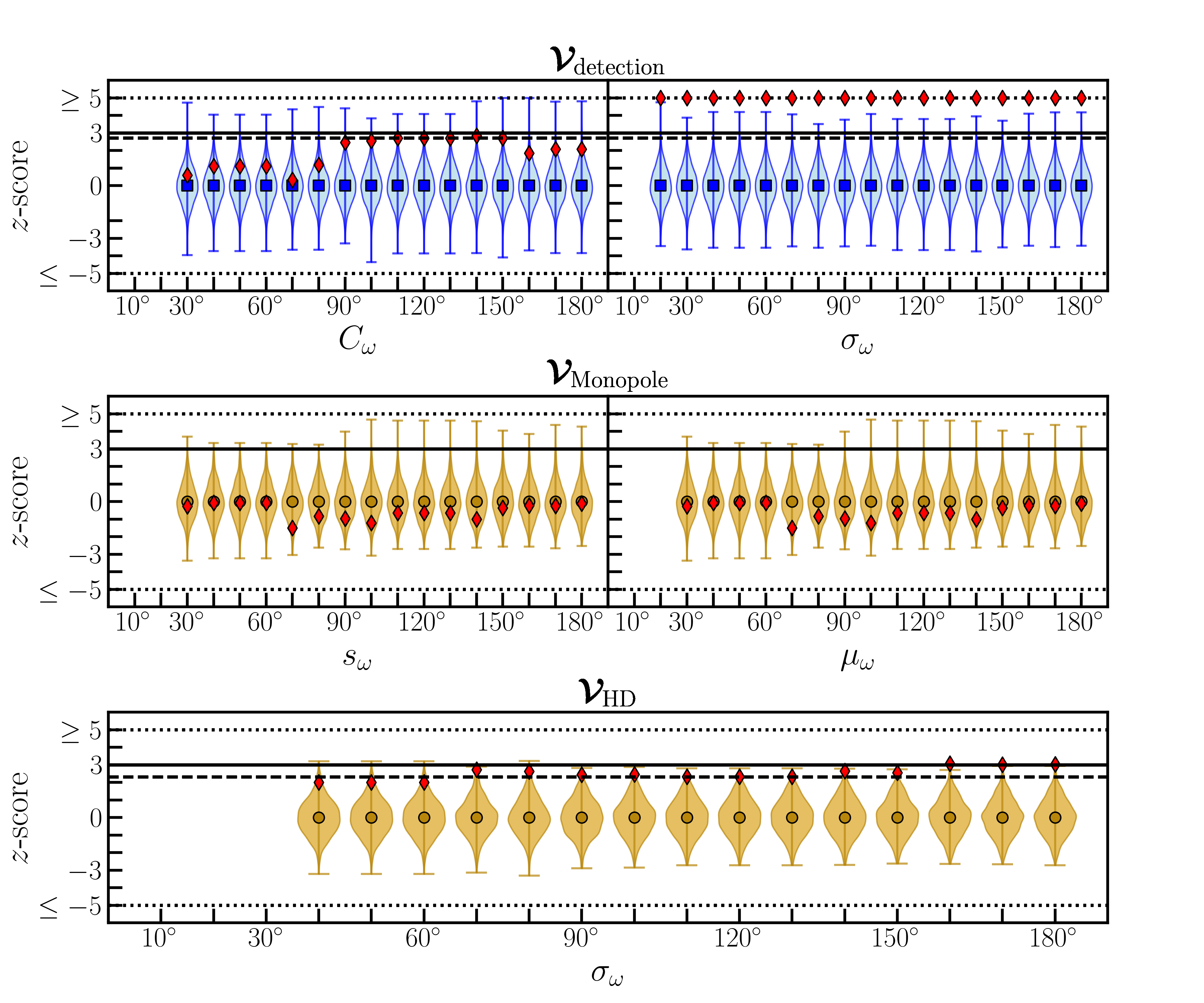}
	\caption{The Discriminative Summary Statistics for NG15 dataset (red diamond) compared to the null distributions. Upper panels shows the common signal detection feature vector, $\mathbfcal{V}_{\rm detection}$, with ${\rm Baseline}$ distributions (blue violins). The middle panels correspond to Monopole feature vector, $\mathbfcal{V}_{\rm Monopole}$, compared to CURN distributions (golden violins). The Discriminative Summary Statistics of HD detection, $\mathbfcal{V}_{\rm HD}$, is shown in the lower panel with CURN distributions overlayed. The $z$-score triggering thresholds at 3$\sigma$ are demonstrated as black solid lines. Black dashed lines correspond to the significance level that trigger more than 65\% of the features.}
	\label{fig:nano_detection}
\end{figure}

We use the reported noise parameters for individual pulsars extracted by Bayesian noise analysis performed by the NANOGrav team \citep{2023ApJ...951L..10A}. Relying on these parameters, we simulate 10,000 realizations to compute the ${\rm Baseline}$ distribution. To construct ${\rm Baseline+CURN}$ dataset, we perform random phase shuffling on the PTRs resulting in destroying any associated spatial correlation, and finally, by repeating this procedure, we obtain an ensemble of realizations which are necessary for further analysis. Figure~\ref{fig:nano_detection} shows the implementation of our detection pipeline on the NG15 dataset. The first row depicts the comparison between elements of $\mathbfcal{V}_{\rm detection}$ as a function of $\bar{\zeta}$ for ${\rm Baseline}$ and real data. The NG15 dataset shows higher clustering coefficient $C_{\omega}$ than the ${\rm Baseline}$ distributions for angular separation thresholds higher than $90^{\circ}$. However, it doesn't exceed the $3\sigma$ significance level required to claim a detection. The standard deviation of edge weights, $\sigma_\omega$, are clearly higher than the ${\rm Baseline}$ estimation. the total number of triggered features of $\mathbfcal{V}_{\rm detection}$ is $17$ out of $33$ corresponds to $52\%$ of the features. According to the adopted voting system, we can not report a high evidence ($\gtrsim3\sigma$) of common signal in the NG15 dataset within our graph-based detection pipeline. The significance level that trigger more than $65\%$ of the features is $\sim 2.7\sigma$.

To investigate the nature of the almost identified common signal, we proceed the evaluation process by computing the elements of the Discriminative Summary Statistics of Monopole pattern, $\mathbfcal{V}_{\rm Monopole}$. We compare the $\mathbfcal{V}_{\rm Monopole}$ for real data with corresponding ${\rm Baseline+CURN}$ distribution which is demonstrated in the second row of Figure~\ref{fig:nano_detection}.  The $z$-score for every component of the feature vector related to Monopole pattern is lower than the given threshold, which excludes the Monopole spatial pattern hypothesis.

To elucidate the presence of SGWB, we calculate the $\mathbfcal{V}_{\rm HD}$ for the NG15 dataset. By comparing it with ${\rm Baseline+CURN}$, our result is depicted in the last row of Figure~\ref{fig:nano_detection}. The edge weights values of real data  demonstrate a higher level of diversity than ${\rm Baseline+CURN}$ for all angular separation thresholds. Only one separation threshold, $\bar{\zeta} = 160^\circ$ has $\sigma_{\omega}$ higher than $3\sigma$ level correspond to $7\%$ of the features. The significance level to detect HD pattern with more than $65\%$ of the features correspond to $\sim 2.3\sigma$.

In conclusion, we obtain low significant evidence ($\sim 2.7\sigma$) for common signal detection and ($\sim 2.3\sigma$) for SGWB detection in NG15 using graph-based summary statistics.

\section{Summary and Conclusion}
\label{sec:Conclusion}

In this work, we developed and tested a novel pipeline for SGWB detection and characterization using graph-based summary statistics derived from PTA data (Figure~\ref{fig:fig_Work_Flow}).  A goal of this approach is to convert the PTA data into complex networks and extract its  structural characteristics by means of graph-based summary statistics to discriminate between various spatial correlation patterns, specifically monopole and HD.

Our analysis begins with the generation of synthetic PTA data to examine the proposed technique on a controlled dataset. The PTRs (Eqs.~(\ref{eq:PTRs})  and (\ref{eq:ptrcomponents})) contain white noise sourced by the systematic errors of the telescope, red noise which is caused by the pulsars' phase jitters, and various interstellar medium effects which generate per-pulsar frequency-dependent noise components, and finally the HD correlated SGWB signal (Figure \ref{fig:sample_residuals}). To account for pulsar spin down, we projected the pre-fit residuals using the $R$-matrix formalism (Eq. (\ref{eq:r-matrix})) to construct the post-fit residuals. Other correlation patterns like monopole, caused by the systematic clock errors, dipole sourced by solar system ephemeris uncertainties, and CURN, which introduces an uncorrelated common signal between the pulsars, are injected for comparison purposes (Figure~\ref{fig:correlation_models}). In addition to mock data, we have used the NANOGrav 15-year dataset (Table \ref{tab:simulation_parameters}). 

To establish the graph-based detection pipeline, we converted the PTA data into graphs by treating the pulsars as the nodes of the graphs and inter-pulsar cross-correlations as the weights of the edges (Figure~\ref{fig:network_construction}). For each graph (Figures \ref{fig:sample_pta_graphs}), we computed four graph-based measures that quantify the structural characteristics of the graph, represented as: $\mathbfcal{G} = \{C_\omega, s_\omega, \mu_\omega, \sigma_\omega\}$. 

We employed three statistical metrics to evaluate the discrimination capability of each graph measure, namely $AUC$ , ${\rm Welch's}$ $t$-test $p$-value and ${\rm Cohen's}$ $d$, over the empirical distributions of graph-based summary statistics (Figure \ref{fig:network_measures}) computed from 10,000 realizations of mock data. By adopting the Discriminative Summary Statistics, our results demonstrated that  the average clustering coefficient and the standard deviation of edge weights can significantly distinguish between the ${\rm Baseline}$ (pure noise) and ${\rm Baseline+CURN}$ cases as depicted in the upper panel of Figure~\ref{fig:eligible_detection}. The average node strengths and the edge weight mean were appropriate indicators for monopole identification (middle panel of Figure~\ref{fig:eligible_detection}). The main and sole indicator of HD was the standard deviation of edge weights (lower panel of Figure~\ref{fig:eligible_detection}).

We developed the decision strategy (Figure~\ref{fig:strategy}) in such a way that upon observing  the imprint of a common signal in the PTRs, we proceed to investigate the nature of the signal. Should the HD correlation pattern be approved after excluding other spatial patterns, we claim a detection of SGWB.

The behavior of the detection rate and false alarm probability (FAP) versus the number of pulsars, observation time span, and the $A_{\rm SGWB}$ (Figure~\ref{fig:scaling_properties}) have been investigated. Our results  exhibited the lowest detectable value of strain amplitude of SGWB utilizing our graph-based measures is $A_{\rm SGWB}\gtrsim 1.2\times 10^{-15}$.

By applying the error propagation method, we computed the statistical relative error of SGWB parameters through the uncertainty propagation of graph-based summary statistics.  Assuming the relative uncertainties of $35\%$ and $15\%$ for $C_\omega(180^\circ)$ and $\sigma_\omega(180^\circ)$,  the resulting relative errors on $\log_{10} A_{\rm SGWB}$ are $2.5\%$ and $1.7\%$, respectively (left panel of Figure~\ref{fig:relative_error}). Combining the principal components analysis with the Fisher forecast framework revealed that at least 21 of the largest principal components are necessary to constrain the $\gamma_{\rm SGWB}$, which also resulted in the relative error on $\log_{10} A_{\rm SGWB}$ approaching  $8\%$. Constraining the strain amplitude of SGWB was enhanced by 82\% when all principal components are taken into account, compared to the case where the largest 13 components are used (Figure \ref{fig:constraints}). The optimistic uncertainty levels of $\log_{10} A_{\rm SGWB}$  and $\gamma_{\rm SGWB}$ reached  $1.5\%$ and $19.5\%$, respectively, at $2\sigma$ confidence interval.

Finally, we evaluated the potential of our graph-based approach for detecting SGWB by implementing it on real PTA data, the NANOGrav 15-year (NG15) dataset. We found a weak evidence for common signal within $\sim2.7\sigma$ and for SGWB within the $\sim2.3\sigma$ credibility level using graph-based summary statistics in the NG15 dataset (Figure~\ref{fig:nano_detection}) which almost aligns with the finding of \citep{2023ApJ...951L...8A}.

In summary, our study shows that converting PTA data into a graph enables both detection and characterization of the SGWB through the graph-based summary statistics. While further refinement is needed, especially in the areas of feature selection and parameter estimation, our results motivate continued development of graph-based methods as a complementary tool in the gravitational wave data analysis techniques.

To go further, we suggest to do following tasks as the complementary subjects in the banner of  higher order summary statistics  and SGWB and will be left for the future study: while the correlation graph was a suitable choice to convert the data into a graph, but still the visibility graph method \citep{lacasa2012time}, and Topological-based data analysis \citep{zomorodian2005topology11,wasserman2018topological,dey2022computational,edelsbrunner2022computational,2024JCAP...09..034Y,2024MNRAS.535..657J,abedi2024} would be an interesting approach to be examined. Exploring the contribution of different theoretical models in elucidating the SGWB  through the Bayesian Model Averaging combined  with graph-based analysis is  intriguing for signal interpretation \citep{2022PhRvD.106h3003H,2023ApJ...951L..11A,2024MNRAS.528.1531P}.

As mentioned in subsection \ref{sec:fisher_forecast}, since the highly nonlinear relationship between graph-based summary statistics and SGWB parameters, it is essential to set up a proper emulator for parameter estimation when we are dealing with observed data. The GPR and other complementary neural density estimators to compute the posterior distribution of the model free parameters should be carried out \citep[and references therein]{2021A&A...648A..74H,2023PhDT........23G,2024Univ...10..464V, tejero2020sbi,2016arXiv160506376P,2019MNRAS.488.4440A,2020PNAS..11730055C,2022mla..confE..24H}.  One can enhance the pipeline by combining it with Machine Learning \citep{Adityan2025} to get a more sensitive, faster, and robuster detection pipeline.

\begin{acknowledgments}
	The authors are very grateful to M. H. Jalali Kanafi for his extremely useful comments on different parts of this paper. We also thank the NANOGrav team for sharing their data sets and providing extensive instructions on how to utilize the data. The graph-based calculations were performed by using \texttt{NetworkX} python package \citep{hagberg2008}. Also, the Fisher forecast plots are provided by utilizing \texttt{Getdist} python package \citep{Lewis2019xzd}
\end{acknowledgments}

\appendix

\section{A brief on theoretical foundation of SGWB}
\label{sec:SGWBs}

The mathematical description of SGWB regardless of their astrophysical and cosmological origin, is treated by metric perturbations in linearized regime as $g_{\mu\nu}=\eta_{\mu\nu}+h_{\mu\nu}$. The spatial perturbation, $h_{ij}(t, \boldsymbol{r})$, at a given position $\boldsymbol{r}$ and moment $t$, is written as:
\begin{equation}
	h_{ij}(t,\boldsymbol{r})=\int_{-\infty}^{+\infty}df\int_{\mathbb{S}^2}d\hat{n}\sum_{\diamond=+,\times}h_{\diamond}(f,\hat{n})\epsilon_{ij}^{\diamond}(\hat{n}){\rm e}^{i2\pi f(\frac{\hat{n}.\boldsymbol{r}}{c}-t)},
\end{equation}
where $f$ is the frequency of the gravitational waves, and $\epsilon_{ij}^{\diamond}(\hat{n})$ is the polarization base tensor \citep{1997PhRvD..56..545A}. The $\hat{n}$ is a unit vector that points
along the direction of travel of the waves. The energy density of GW is $\rho_{\rm GW}\equiv c^2\langle\dot{h}_{ij}\dot{h}^{ij} \rangle/ 32\pi G$, and accordingly, the GW energy density frequency spectrum integrated over solid angle, $d\hat{n}$, is defined as \citep{2019RPPh...82a6903C}: 
\begin{equation}
	\Omega_{\rm GW}(f)\equiv\frac{1}{\rho_c}\frac{d\rho_{\rm GW}(f)}{d\ln f},
\end{equation} 
with $\rho_c \equiv 3c^2H_0^2/8\pi G$.	The `one-sided' strain spectrum, $S_h (f)$, reads as:
\begin{equation}
	S_h (f) \equiv 2 \int_{\mathbb{S}^2} d\hat{n} \sum_{\diamond=+,\times} \langle h_\diamond (f, \hat{n}) h^*_\diamond (f,\hat{n}) \rangle,
\end{equation}
and the characteristic strain spectrum is $h_c (f) \equiv \sqrt{f S_h (f)}$.
Thus the energy density can be written in terms of strain spectrum as:
\begin{equation}
	\Omega_{\rm GW}(f) = \frac{\pi c^2 f^3}{4\rho_c G} S_h(f).
\end{equation}

Also, in terms of characteristic strain spectrum, we have:
\begin{equation}
	\Omega_{\rm GW}(f) = \frac{\pi c^2 f^2}{4\rho_c G} h_c^2(f).
\end{equation}

\section{SGWB Signal Generation via Eigenvalue Decomposition}
\label{app:gwb_simulation}

To simulate timing residuals with specified spatial correlations, we employ eigenvalue decomposition of the target correlation matrix $\mathcal{C}$:
\begin{align}
	\mathcal{C} = Q \Lambda Q^{T},
\end{align}
where $Q$ contains eigenvectors and $\Lambda$ is the diagonal eigenvalue matrix. 

We generate $N_{\mathrm{PTA}}$ independent Gaussian time series $Z$ with the SGWB power spectrum (Eq.~\ref{eq:GW_power_spectrum}). The correlated residuals are constructed as:
\begin{align}
	S_{\rm SGWB}(t) = Q \Lambda^{1/2} Z.
\end{align}

To ensure positive definiteness, we introduce a small perturbation $\epsilon = |\min\{0, \min(\Lambda)\}|$ and modify the eigenvalues:
\begin{align}
	\tilde{\Lambda} = \Lambda + \mathrm{diag}(\epsilon).
\end{align}

A normalization factor $\rho = \mathrm{diag}(1/(1 + \epsilon))$ compensates for the enhanced auto-correlation:
\begin{align}
	S_{\rm SGWB}(t) = \rho Q \tilde{\Lambda}^{1/2} Z.
\end{align}

\section{Complementary Notes on Graph-based Summary Statistics}
\label{app:Discussion}

\begin{figure}[!b]
	\centering
	\includegraphics[width=\textwidth]{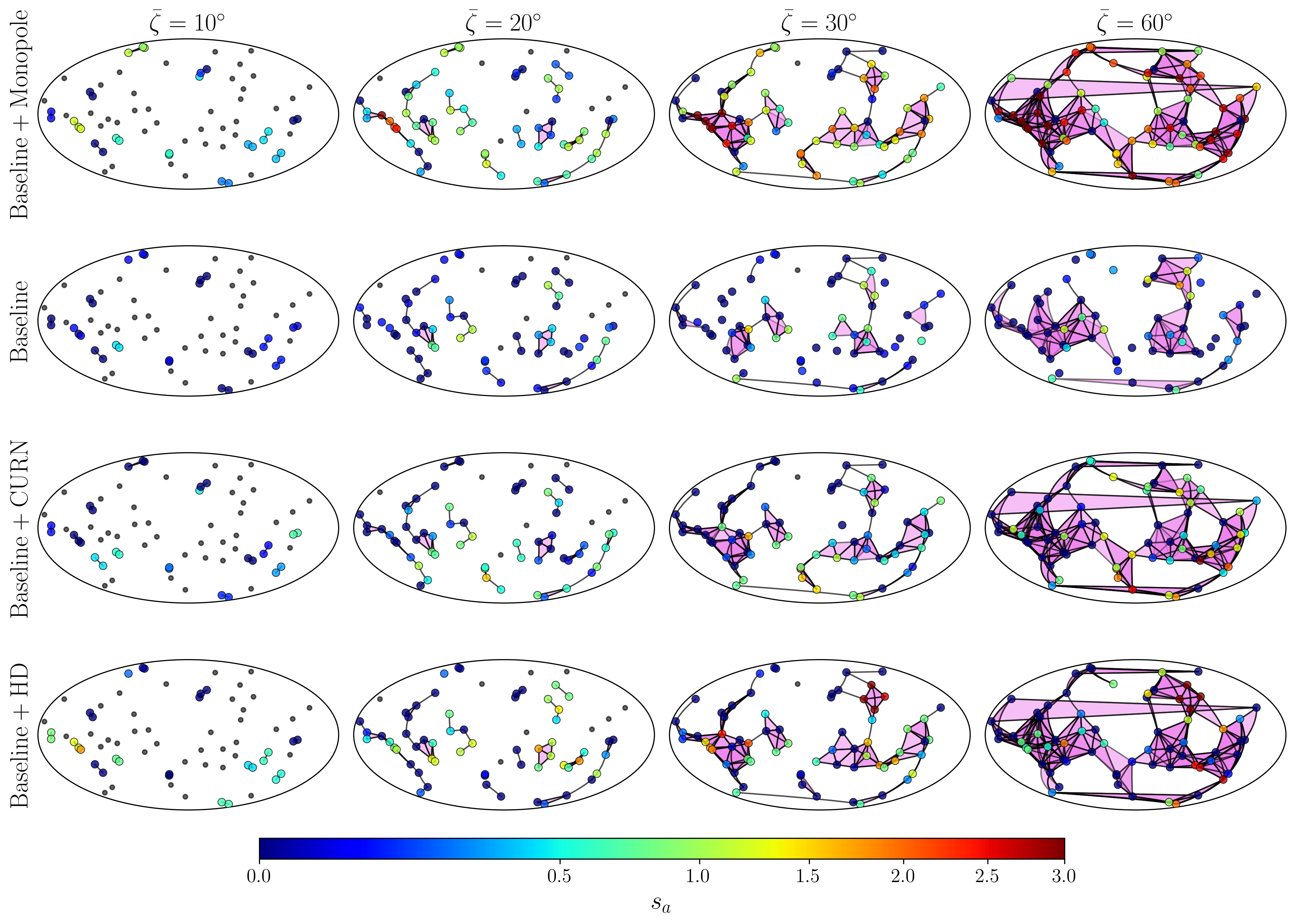}
	\caption{The constructed graph from PTA data for a range of angular separation values ($\bar{\zeta}$) from left to right.  Each row corresponds to a specific spatial correlation template identified as ${\rm Baseline+Monopole}$, ${\rm Baseline}$, ${\rm Baseline+CURN}$, and ${\rm Baseline+HD}$. The points are associated with graph nodes (pulsars), with different node colors reflecting their strength values ($s_a$). Two nodes have an edge if and only if their spatial angular separation satisfies the condition $\zeta_{ab}\le\bar{\zeta}$. The abundance of triangles, which is almost quantified by the average clustering coefficient up to a geometrical mean value (Eq. (\ref{eq:comega})), is evidently insufficient
		for the noise template.}
	\label{fig:sample_adjacency2}
\end{figure}

To give a more complete view concerning the graph-based summary statistics, we depict the constructed graph in Figure \ref{fig:sample_adjacency2}. Assuming a specific spatial correlation template, the overall properties of the graph vary.  Each row corresponds to a specific spatial correlation template identified as ${\rm Baseline+Monopole}$, ${\rm Baseline}$, ${\rm Baseline+CURN}$, and ${\rm Baseline+HD}$. The points are associated with graph nodes (pulsars), with different nodes' colors reflecting their strength values ($s_a$).  Two nodes have an edge if and only if their spatial angular separation satisfies the condition $\zeta_{ab}\le\bar{\zeta}$. An obvious change for different templates, which is visually captured, is the abundance of triangles. This statistic is quantified by the average clustering coefficient up to the geometrical mean value (Eqs. (\ref{eq:comega0}) and (\ref{eq:comega})). This criterion quantifies the nature of 3-point statistics ($\overset{\triangle}{\scalebox{0.6}{abd}}$). As expected from the clustering feature, the number of triangles for the ${\rm Baseline}$ is less than the other spatial templates. The presence of common signal, produces more triangles resulting in higher average clustering coefficient. It is worth noting that the triangles, $\overset{\triangle}{\scalebox{0.6}{abd}}$, with a geometrical mean, $(\omega_{ab}\omega_{ad}\omega_{bd})^{1/3}$, less than $0.1$ have been removed for better visualization. Figure \ref{fig:sample_adjacency2} also shows that the values of $s_a$ for the monopole template are statistically higher than the corresponding values for the ${\rm Baseline}$, ${\rm Baseline+CURN}$, and ${\rm Baseline+HD}$.

\begin{figure}[!t]
	\centering
	\includegraphics[width=\textwidth]{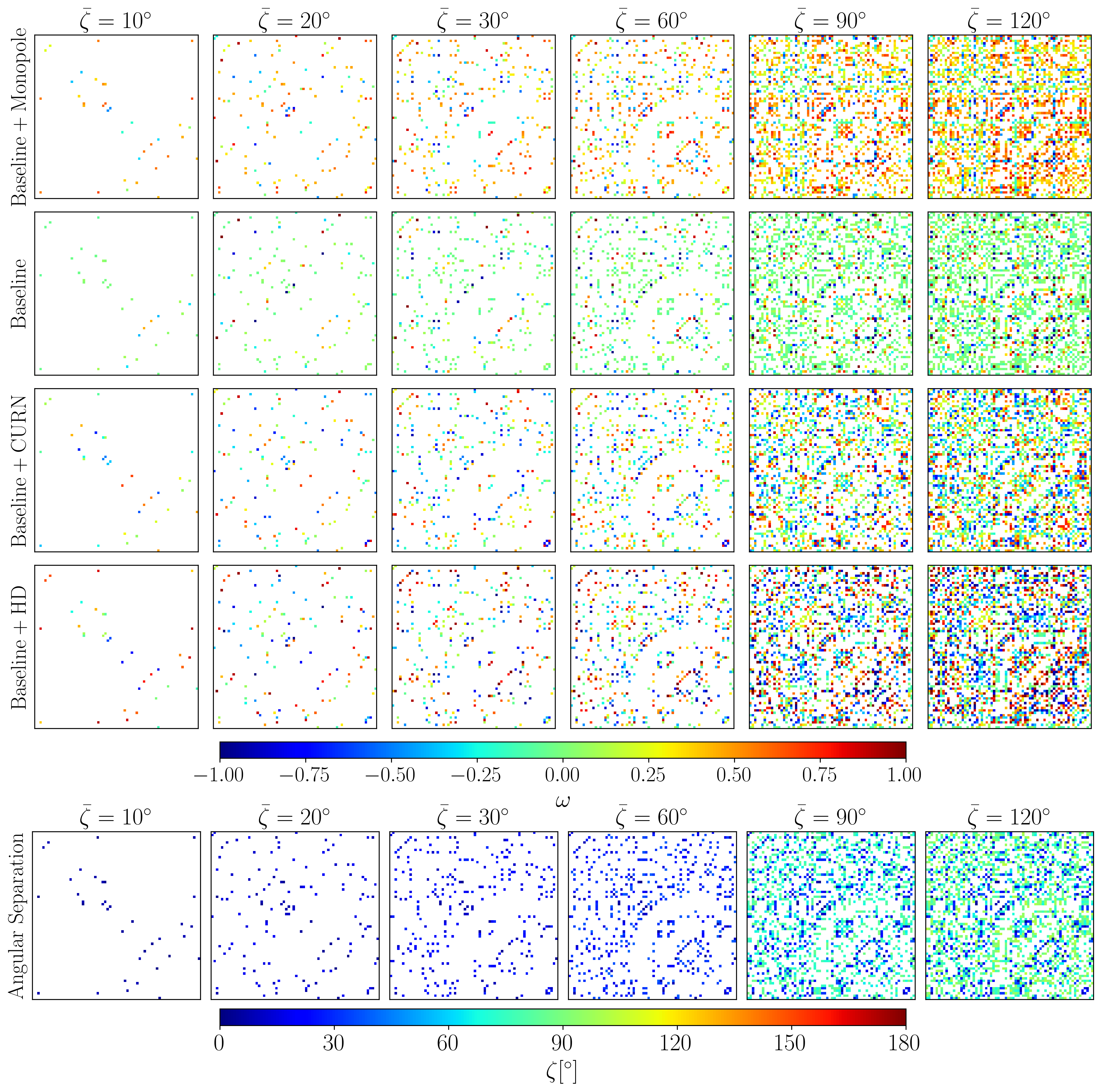}
	\caption{The density plots for distinct components employed in the construction of graphs are shown for a range of angular separation values ($\bar{\zeta}$) from left to right. Each panel is a matrix of size $68\time68$. The first four rows depict the sample averaged of adjacency matrices, with elements denoted as $\omega_{ab}$ for various spatial correlation templates applied to the PTA network.  The last row reveals the matrix whose elements consist of the values of $\zeta_{ab}$s for a particular threshold, $\bar{\zeta}$.  The white pixels in the matrices are devoted to pairs that we have excluded and whose angular separations belong to the ranges between $\left[39.5^\circ-63^\circ\right]$ and $\left[104^\circ-137.5^\circ\right]$ (see subsection \ref{subsec:network_construction} for more details).
	}
	\label{fig:sample_adjacency}
\end{figure}

To go further, we show the adjacency matrix for $N_{\rm PTA}=68$  of  size  $68 \times 68$  for various values of $\bar{\zeta}$ in the upper part of Figure \ref{fig:sample_adjacency}.  The associated overall amount of edges for constructed graph is also indicated in the last row of Figure \ref{fig:sample_adjacency}.  The first four rows in the Figure \ref{fig:sample_adjacency} correspond to a realization of adjacency matrices for ${\rm Baseline+Monopole}$, ${\rm Baseline}$, ${\rm Baseline+CURN}$ and ${\rm Baseline+HD}$ templates exerted on the PTA, respectively. The last row depicts that the higher value of $\bar{\zeta}$, the more dense graph is. The white pixels in these plots figure out the pairs that we have masked to reduce the spurious impact of noise (see subsection \ref{subsec:network_construction}). The angular separation values for masked pairs are in the ranges between $\left[39.5^\circ-63^\circ\right]$ and $\left[104^\circ-137.5^\circ\right]$. This achievement is consistent with the graphs illustrated in Figure \ref{fig:sample_pta_graphs}. Figure \ref{fig:sample_adjacency} also shows that the mean $\omega_{ab}$ values for the ${\rm Baseline+Monopole}$ template are higher than those for the ${\rm Baseline}$, ${\rm Baseline+CURN}$, and ${\rm Baseline+HD}$ templates. The diversity in $\omega_{ab}$, quantified by $\sigma_{\omega}$, is realized in the first four rows of Figure \ref{fig:sample_adjacency}. It is therefore evident when comparing the third and fourth rows, that $\sigma_{\omega}^{\rm Baseline+HD} \gtrsim \sigma_{\omega}^{\rm Baseline+CURN}$ for all values of $\bar{\zeta}$.



\begin{thebibliography}{157}
\expandafter\ifx\csname natexlab\endcsname\relax\def\natexlab#1{#1}\fi
\expandafter\ifx\csname bibnamefont\endcsname\relax
  \def\bibnamefont#1{#1}\fi
\expandafter\ifx\csname bibfnamefont\endcsname\relax
  \def\bibfnamefont#1{#1}\fi
\expandafter\ifx\csname citenamefont\endcsname\relax
  \def\citenamefont#1{#1}\fi
\expandafter\ifx\csname url\endcsname\relax
  \def\url#1{\texttt{#1}}\fi
\expandafter\ifx\csname urlprefix\endcsname\relax\def\urlprefix{URL }\fi
\providecommand{\bibinfo}[2]{#2}
\providecommand{\eprint}[2][]{\url{#2}}

\bibitem[{\citenamefont{{Einstein}}(1916)}]{1916SPAW.......688E}
\bibinfo{author}{\bibfnamefont{A.}~\bibnamefont{{Einstein}}},
  \bibinfo{journal}{Sitzungsberichte der K\&ouml;niglich Preussischen Akademie
  der Wissenschaften} pp. \bibinfo{pages}{688--696} (\bibinfo{year}{1916}).

\bibitem[{\citenamefont{{Einstein}}(1918)}]{1918SPAW.......154E}
\bibinfo{author}{\bibfnamefont{A.}~\bibnamefont{{Einstein}}},
  \bibinfo{journal}{Sitzungsberichte der K\&ouml;niglich Preussischen Akademie
  der Wissenschaften} pp. \bibinfo{pages}{154--167} (\bibinfo{year}{1918}).

\bibitem[{\citenamefont{{Hawking} et~al.}(1993)\citenamefont{{Hawking},
  {Israel}, and {Dolan}}}]{1993IrAJ...21..161H}
\bibinfo{author}{\bibfnamefont{S.~W.} \bibnamefont{{Hawking}}},
  \bibinfo{author}{\bibfnamefont{W.}~\bibnamefont{{Israel}}}, \bibnamefont{and}
  \bibinfo{author}{\bibfnamefont{B.}~\bibnamefont{{Dolan}}},
  \bibinfo{journal}{Irish Astronomical Journal} \textbf{\bibinfo{volume}{21}},
  \bibinfo{pages}{161} (\bibinfo{year}{1993}).

\bibitem[{\citenamefont{{Allen}}(1997)}]{1997rggr.conf..373A}
\bibinfo{author}{\bibfnamefont{B.}~\bibnamefont{{Allen}}}, in
  \emph{\bibinfo{booktitle}{Relativistic Gravitation and Gravitational
  Radiation}}, edited by \bibinfo{editor}{\bibfnamefont{J.-A.}
  \bibnamefont{{Marck}}} \bibnamefont{and}
  \bibinfo{editor}{\bibfnamefont{J.-P.} \bibnamefont{{Lasota}}}
  (\bibinfo{year}{1997}), pp. \bibinfo{pages}{373--417},
  \eprint{gr-qc/9604033}.

\bibitem[{\citenamefont{{Sathyaprakash} and
  {Schutz}}(2009)}]{2009LRR....12....2S}
\bibinfo{author}{\bibfnamefont{B.~S.} \bibnamefont{{Sathyaprakash}}}
  \bibnamefont{and} \bibinfo{author}{\bibfnamefont{B.~F.}
  \bibnamefont{{Schutz}}}, \bibinfo{journal}{Living Reviews in Relativity}
  \textbf{\bibinfo{volume}{12}}, \bibinfo{eid}{2} (\bibinfo{year}{2009}),
  \eprint{0903.0338}.

\bibitem[{\citenamefont{{Caprini} and {Figueroa}}(2018)}]{2018CQGra..35p3001C}
\bibinfo{author}{\bibfnamefont{C.}~\bibnamefont{{Caprini}}} \bibnamefont{and}
  \bibinfo{author}{\bibfnamefont{D.~G.} \bibnamefont{{Figueroa}}},
  \bibinfo{journal}{Classical and Quantum Gravity}
  \textbf{\bibinfo{volume}{35}}, \bibinfo{eid}{163001} (\bibinfo{year}{2018}),
  \eprint{1801.04268}.

\bibitem[{\citenamefont{{Christensen}}(2019)}]{2019RPPh...82a6903C}
\bibinfo{author}{\bibfnamefont{N.}~\bibnamefont{{Christensen}}},
  \bibinfo{journal}{Reports on Progress in Physics}
  \textbf{\bibinfo{volume}{82}}, \bibinfo{eid}{016903} (\bibinfo{year}{2019}),
  \eprint{1811.08797}.

\bibitem[{\citenamefont{{Abbott} et~al.}(2016)\citenamefont{{Abbott}, {Abbott},
  {Abbott}, {Abernathy}, {Acernese}, {Ackley}, {Adams}, {Adams}, {Addesso},
  {Adhikari} et~al.}}]{2016PhRvL.116f1102A}
\bibinfo{author}{\bibfnamefont{B.~P.} \bibnamefont{{Abbott}}},
  \bibinfo{author}{\bibfnamefont{R.}~\bibnamefont{{Abbott}}},
  \bibinfo{author}{\bibfnamefont{T.~D.} \bibnamefont{{Abbott}}},
  \bibinfo{author}{\bibfnamefont{M.~R.} \bibnamefont{{Abernathy}}},
  \bibinfo{author}{\bibfnamefont{F.}~\bibnamefont{{Acernese}}},
  \bibinfo{author}{\bibfnamefont{K.}~\bibnamefont{{Ackley}}},
  \bibinfo{author}{\bibfnamefont{C.}~\bibnamefont{{Adams}}},
  \bibinfo{author}{\bibfnamefont{T.}~\bibnamefont{{Adams}}},
  \bibinfo{author}{\bibfnamefont{P.}~\bibnamefont{{Addesso}}},
  \bibinfo{author}{\bibfnamefont{R.~X.} \bibnamefont{{Adhikari}}},
  \bibnamefont{et~al.}, \bibinfo{journal}{\prl} \textbf{\bibinfo{volume}{116}},
  \bibinfo{eid}{061102} (\bibinfo{year}{2016}), \eprint{1602.03837}.

\bibitem[{\citenamefont{{Abbott} et~al.}(2009)\citenamefont{{Abbott}, {Abbott},
  {Acernese}, {Adhikari}, {Ajith}, {Allen}, {Allen}, {Alshourbagy}, {Amin},
  {Anderson} et~al.}}]{2009Natur.460..990A}
\bibinfo{author}{\bibfnamefont{B.~P.} \bibnamefont{{Abbott}}},
  \bibinfo{author}{\bibfnamefont{R.}~\bibnamefont{{Abbott}}},
  \bibinfo{author}{\bibfnamefont{F.}~\bibnamefont{{Acernese}}},
  \bibinfo{author}{\bibfnamefont{R.}~\bibnamefont{{Adhikari}}},
  \bibinfo{author}{\bibfnamefont{P.}~\bibnamefont{{Ajith}}},
  \bibinfo{author}{\bibfnamefont{B.}~\bibnamefont{{Allen}}},
  \bibinfo{author}{\bibfnamefont{G.}~\bibnamefont{{Allen}}},
  \bibinfo{author}{\bibfnamefont{M.}~\bibnamefont{{Alshourbagy}}},
  \bibinfo{author}{\bibfnamefont{R.~S.} \bibnamefont{{Amin}}},
  \bibinfo{author}{\bibfnamefont{S.~B.} \bibnamefont{{Anderson}}},
  \bibnamefont{et~al.}, \bibinfo{journal}{\nat} \textbf{\bibinfo{volume}{460}},
  \bibinfo{pages}{990} (\bibinfo{year}{2009}), \eprint{0910.5772}.

\bibitem[{\citenamefont{{van Remortel} et~al.}(2023)\citenamefont{{van
  Remortel}, {Janssens}, and {Turbang}}}]{2023PrPNP.12804003V}
\bibinfo{author}{\bibfnamefont{N.}~\bibnamefont{{van Remortel}}},
  \bibinfo{author}{\bibfnamefont{K.}~\bibnamefont{{Janssens}}},
  \bibnamefont{and}
  \bibinfo{author}{\bibfnamefont{K.}~\bibnamefont{{Turbang}}},
  \bibinfo{journal}{Progress in Particle and Nuclear Physics}
  \textbf{\bibinfo{volume}{128}}, \bibinfo{eid}{104003} (\bibinfo{year}{2023}),
  \eprint{2210.00761}.

\bibitem[{\citenamefont{{Allen}}(1988)}]{1988PhRvD..37.2078A}
\bibinfo{author}{\bibfnamefont{B.}~\bibnamefont{{Allen}}},
  \bibinfo{journal}{\prd} \textbf{\bibinfo{volume}{37}}, \bibinfo{pages}{2078}
  (\bibinfo{year}{1988}).

\bibitem[{\citenamefont{{Guzzetti} et~al.}(2016)\citenamefont{{Guzzetti},
  {Bartolo}, {Liguori}, and {Matarrese}}}]{2016NCimR..39..399G}
\bibinfo{author}{\bibfnamefont{M.~C.} \bibnamefont{{Guzzetti}}},
  \bibinfo{author}{\bibfnamefont{N.}~\bibnamefont{{Bartolo}}},
  \bibinfo{author}{\bibfnamefont{M.}~\bibnamefont{{Liguori}}},
  \bibnamefont{and}
  \bibinfo{author}{\bibfnamefont{S.}~\bibnamefont{{Matarrese}}},
  \bibinfo{journal}{Nuovo Cimento Rivista Serie} \textbf{\bibinfo{volume}{39}},
  \bibinfo{pages}{399} (\bibinfo{year}{2016}).

\bibitem[{\citenamefont{{Dom{\`e}nech}}(2021)}]{2021Univ....7..398D}
\bibinfo{author}{\bibfnamefont{G.}~\bibnamefont{{Dom{\`e}nech}}},
  \bibinfo{journal}{Universe} \textbf{\bibinfo{volume}{7}}, \bibinfo{eid}{398}
  (\bibinfo{year}{2021}), \eprint{2109.01398}.

\bibitem[{\citenamefont{{Yuan} and {Huang}}(2021)}]{2021iSci...24j2860Y}
\bibinfo{author}{\bibfnamefont{C.}~\bibnamefont{{Yuan}}} \bibnamefont{and}
  \bibinfo{author}{\bibfnamefont{Q.-G.} \bibnamefont{{Huang}}},
  \bibinfo{journal}{iScience} \textbf{\bibinfo{volume}{24}},
  \bibinfo{pages}{102860} (\bibinfo{year}{2021}), \eprint{2103.04739}.

\bibitem[{\citenamefont{{Tasinato}}(2022)}]{2022PhRvD.105b3521T}
\bibinfo{author}{\bibfnamefont{G.}~\bibnamefont{{Tasinato}}},
  \bibinfo{journal}{\prd} \textbf{\bibinfo{volume}{105}}, \bibinfo{eid}{023521}
  (\bibinfo{year}{2022}), \eprint{2201.10333}.

\bibitem[{\citenamefont{{Sasaki}}(2025)}]{2025GReGr..57...82S}
\bibinfo{author}{\bibfnamefont{M.}~\bibnamefont{{Sasaki}}},
  \bibinfo{journal}{General Relativity and Gravitation}
  \textbf{\bibinfo{volume}{57}}, \bibinfo{eid}{82} (\bibinfo{year}{2025}).

\bibitem[{\citenamefont{{Vachaspati} and
  {Vilenkin}}(1985)}]{1985PhRvD..31.3052V}
\bibinfo{author}{\bibfnamefont{T.}~\bibnamefont{{Vachaspati}}}
  \bibnamefont{and}
  \bibinfo{author}{\bibfnamefont{A.}~\bibnamefont{{Vilenkin}}},
  \bibinfo{journal}{\prd} \textbf{\bibinfo{volume}{31}}, \bibinfo{pages}{3052}
  (\bibinfo{year}{1985}).

\bibitem[{\citenamefont{{Vilenkin}}(1985)}]{1985PhR...121..263V}
\bibinfo{author}{\bibfnamefont{A.}~\bibnamefont{{Vilenkin}}},
  \bibinfo{journal}{Physics Reports} \textbf{\bibinfo{volume}{121}},
  \bibinfo{pages}{263} (\bibinfo{year}{1985}).

\bibitem[{\citenamefont{{Hindmarsh} and {Kibble}}(1995)}]{1995RPPh...58..477H}
\bibinfo{author}{\bibfnamefont{M.~B.} \bibnamefont{{Hindmarsh}}}
  \bibnamefont{and} \bibinfo{author}{\bibfnamefont{T.~W.~B.}
  \bibnamefont{{Kibble}}}, \bibinfo{journal}{Reports on Progress in Physics}
  \textbf{\bibinfo{volume}{58}}, \bibinfo{pages}{477} (\bibinfo{year}{1995}),
  \eprint{hep-ph/9411342}.

\bibitem[{\citenamefont{{Damour} and {Vilenkin}}(2001)}]{2001PhRvD..64f4008D}
\bibinfo{author}{\bibfnamefont{T.}~\bibnamefont{{Damour}}} \bibnamefont{and}
  \bibinfo{author}{\bibfnamefont{A.}~\bibnamefont{{Vilenkin}}},
  \bibinfo{journal}{\prd} \textbf{\bibinfo{volume}{64}}, \bibinfo{eid}{064008}
  (\bibinfo{year}{2001}), \eprint{gr-qc/0104026}.

\bibitem[{\citenamefont{{Blanco-Pillado}
  et~al.}(2011)\citenamefont{{Blanco-Pillado}, {Olum}, and
  {Shlaer}}}]{2011PhRvD..83h3514B}
\bibinfo{author}{\bibfnamefont{J.~J.} \bibnamefont{{Blanco-Pillado}}},
  \bibinfo{author}{\bibfnamefont{K.~D.} \bibnamefont{{Olum}}},
  \bibnamefont{and} \bibinfo{author}{\bibfnamefont{B.}~\bibnamefont{{Shlaer}}},
  \bibinfo{journal}{\prd} \textbf{\bibinfo{volume}{83}}, \bibinfo{eid}{083514}
  (\bibinfo{year}{2011}), \eprint{1101.5173}.

\bibitem[{\citenamefont{{Ringeval} and {Suyama}}(2017)}]{Ringeval:2017eww}
\bibinfo{author}{\bibfnamefont{C.}~\bibnamefont{{Ringeval}}} \bibnamefont{and}
  \bibinfo{author}{\bibfnamefont{T.}~\bibnamefont{{Suyama}}},
  \bibinfo{journal}{Journal of Cosmology and Astroparticle Physics}
  \textbf{\bibinfo{volume}{2017}}, \bibinfo{eid}{027} (\bibinfo{year}{2017}),
  \eprint{1709.03845}.

\bibitem[{\citenamefont{{Blanco-Pillado} and
  {Olum}}(2017)}]{Blanco-Pillado:2017oxo}
\bibinfo{author}{\bibfnamefont{J.~J.} \bibnamefont{{Blanco-Pillado}}}
  \bibnamefont{and} \bibinfo{author}{\bibfnamefont{K.~D.}
  \bibnamefont{{Olum}}}, \bibinfo{journal}{\prd} \textbf{\bibinfo{volume}{96}},
  \bibinfo{eid}{104046} (\bibinfo{year}{2017}), \eprint{1709.02693}.

\bibitem[{\citenamefont{{Hindmarsh} and {Kume}}(2023)}]{2023JCAP...04..045H}
\bibinfo{author}{\bibfnamefont{M.}~\bibnamefont{{Hindmarsh}}} \bibnamefont{and}
  \bibinfo{author}{\bibfnamefont{J.}~\bibnamefont{{Kume}}},
  \bibinfo{journal}{Journal of Cosmology and Astroparticle Physics}
  \textbf{\bibinfo{volume}{2023}}, \bibinfo{eid}{045} (\bibinfo{year}{2023}),
  \eprint{2210.06178}.

\bibitem[{\citenamefont{{Maselli} et~al.}(2016)\citenamefont{{Maselli},
  {Marassi}, {Ferrari}, {Kokkotas}, and {Schneider}}}]{2016PhRvL.117i1102M}
\bibinfo{author}{\bibfnamefont{A.}~\bibnamefont{{Maselli}}},
  \bibinfo{author}{\bibfnamefont{S.}~\bibnamefont{{Marassi}}},
  \bibinfo{author}{\bibfnamefont{V.}~\bibnamefont{{Ferrari}}},
  \bibinfo{author}{\bibfnamefont{K.}~\bibnamefont{{Kokkotas}}},
  \bibnamefont{and}
  \bibinfo{author}{\bibfnamefont{R.}~\bibnamefont{{Schneider}}},
  \bibinfo{journal}{\prl} \textbf{\bibinfo{volume}{117}}, \bibinfo{eid}{091102}
  (\bibinfo{year}{2016}), \eprint{1606.04996}.

\bibitem[{\citenamefont{{Maggiore}}(2000)}]{2000PhR...331..283M}
\bibinfo{author}{\bibfnamefont{M.}~\bibnamefont{{Maggiore}}},
  \bibinfo{journal}{Physics Reports} \textbf{\bibinfo{volume}{331}},
  \bibinfo{pages}{283} (\bibinfo{year}{2000}), \eprint{gr-qc/9909001}.

\bibitem[{\citenamefont{{Mandic} et~al.}(2012)\citenamefont{{Mandic}, {Thrane},
  {Giampanis}, and {Regimbau}}}]{2012PhRvL.109q1102M}
\bibinfo{author}{\bibfnamefont{V.}~\bibnamefont{{Mandic}}},
  \bibinfo{author}{\bibfnamefont{E.}~\bibnamefont{{Thrane}}},
  \bibinfo{author}{\bibfnamefont{S.}~\bibnamefont{{Giampanis}}},
  \bibnamefont{and}
  \bibinfo{author}{\bibfnamefont{T.}~\bibnamefont{{Regimbau}}},
  \bibinfo{journal}{\prl} \textbf{\bibinfo{volume}{109}}, \bibinfo{eid}{171102}
  (\bibinfo{year}{2012}), \eprint{1209.3847}.

\bibitem[{\citenamefont{{Amaro-Seoane}
  et~al.}(2017)\citenamefont{{Amaro-Seoane}, {Audley}, {Babak}, {Baker},
  {Barausse}, {Bender}, {Berti}, {Binetruy}, {Born}, {Bortoluzzi}
  et~al.}}]{2017arXiv170200786A}
\bibinfo{author}{\bibfnamefont{P.}~\bibnamefont{{Amaro-Seoane}}},
  \bibinfo{author}{\bibfnamefont{H.}~\bibnamefont{{Audley}}},
  \bibinfo{author}{\bibfnamefont{S.}~\bibnamefont{{Babak}}},
  \bibinfo{author}{\bibfnamefont{J.}~\bibnamefont{{Baker}}},
  \bibinfo{author}{\bibfnamefont{E.}~\bibnamefont{{Barausse}}},
  \bibinfo{author}{\bibfnamefont{P.}~\bibnamefont{{Bender}}},
  \bibinfo{author}{\bibfnamefont{E.}~\bibnamefont{{Berti}}},
  \bibinfo{author}{\bibfnamefont{P.}~\bibnamefont{{Binetruy}}},
  \bibinfo{author}{\bibfnamefont{M.}~\bibnamefont{{Born}}},
  \bibinfo{author}{\bibfnamefont{D.}~\bibnamefont{{Bortoluzzi}}},
  \bibnamefont{et~al.}, \bibinfo{journal}{arXiv e-prints}
  \bibinfo{eid}{arXiv:1702.00786} (\bibinfo{year}{2017}), \eprint{1702.00786}.

\bibitem[{\citenamefont{{Abbott} et~al.}(2021)\citenamefont{{Abbott}, {Abbott},
  {Acernese}, {Ackley}, {Adams}, {Adhikari}, {Adhikari}, {Adya}, {Affeldt},
  {Agarwal} et~al.}}]{2021PhRvD.104j2001A}
\bibinfo{author}{\bibfnamefont{R.}~\bibnamefont{{Abbott}}},
  \bibinfo{author}{\bibfnamefont{T.~D.} \bibnamefont{{Abbott}}},
  \bibinfo{author}{\bibfnamefont{F.}~\bibnamefont{{Acernese}}},
  \bibinfo{author}{\bibfnamefont{K.}~\bibnamefont{{Ackley}}},
  \bibinfo{author}{\bibfnamefont{C.}~\bibnamefont{{Adams}}},
  \bibinfo{author}{\bibfnamefont{N.}~\bibnamefont{{Adhikari}}},
  \bibinfo{author}{\bibfnamefont{R.~X.} \bibnamefont{{Adhikari}}},
  \bibinfo{author}{\bibfnamefont{V.~B.} \bibnamefont{{Adya}}},
  \bibinfo{author}{\bibfnamefont{C.}~\bibnamefont{{Affeldt}}},
  \bibinfo{author}{\bibfnamefont{D.}~\bibnamefont{{Agarwal}}},
  \bibnamefont{et~al.}, \bibinfo{journal}{\prd} \textbf{\bibinfo{volume}{104}},
  \bibinfo{eid}{102001} (\bibinfo{year}{2021}), \eprint{2107.13796}.

\bibitem[{\citenamefont{{Punturo} et~al.}(2010)\citenamefont{{Punturo},
  {Abernathy}, {Acernese}, {Allen}, {Andersson}, {Arun}, {Barone}, {Barr},
  {Barsuglia}, {Beker} et~al.}}]{Punturo_2010}
\bibinfo{author}{\bibfnamefont{M.}~\bibnamefont{{Punturo}}},
  \bibinfo{author}{\bibfnamefont{M.}~\bibnamefont{{Abernathy}}},
  \bibinfo{author}{\bibfnamefont{F.}~\bibnamefont{{Acernese}}},
  \bibinfo{author}{\bibfnamefont{B.}~\bibnamefont{{Allen}}},
  \bibinfo{author}{\bibfnamefont{N.}~\bibnamefont{{Andersson}}},
  \bibinfo{author}{\bibfnamefont{K.}~\bibnamefont{{Arun}}},
  \bibinfo{author}{\bibfnamefont{F.}~\bibnamefont{{Barone}}},
  \bibinfo{author}{\bibfnamefont{B.}~\bibnamefont{{Barr}}},
  \bibinfo{author}{\bibfnamefont{M.}~\bibnamefont{{Barsuglia}}},
  \bibinfo{author}{\bibfnamefont{M.}~\bibnamefont{{Beker}}},
  \bibnamefont{et~al.}, \bibinfo{journal}{Classical and Quantum Gravity}
  \textbf{\bibinfo{volume}{27}}, \bibinfo{eid}{194002} (\bibinfo{year}{2010}).

\bibitem[{\citenamefont{{Kawamura} et~al.}(2021)\citenamefont{{Kawamura},
  {Ando}, {Seto}, {Sato}, {Musha}, {Kawano}, {Yokoyama}, {Tanaka}, {Ioka},
  {Akutsu} et~al.}}]{2021PTEP.2021eA105K}
\bibinfo{author}{\bibfnamefont{S.}~\bibnamefont{{Kawamura}}},
  \bibinfo{author}{\bibfnamefont{M.}~\bibnamefont{{Ando}}},
  \bibinfo{author}{\bibfnamefont{N.}~\bibnamefont{{Seto}}},
  \bibinfo{author}{\bibfnamefont{S.}~\bibnamefont{{Sato}}},
  \bibinfo{author}{\bibfnamefont{M.}~\bibnamefont{{Musha}}},
  \bibinfo{author}{\bibfnamefont{I.}~\bibnamefont{{Kawano}}},
  \bibinfo{author}{\bibfnamefont{J.}~\bibnamefont{{Yokoyama}}},
  \bibinfo{author}{\bibfnamefont{T.}~\bibnamefont{{Tanaka}}},
  \bibinfo{author}{\bibfnamefont{K.}~\bibnamefont{{Ioka}}},
  \bibinfo{author}{\bibfnamefont{T.}~\bibnamefont{{Akutsu}}},
  \bibnamefont{et~al.}, \bibinfo{journal}{Progress of Theoretical and
  Experimental Physics} \textbf{\bibinfo{volume}{2021}}, \bibinfo{eid}{05A105}
  (\bibinfo{year}{2021}), \eprint{2006.13545}.

\bibitem[{\citenamefont{{Dwyer} et~al.}(2015)\citenamefont{{Dwyer}, {Sigg},
  {Ballmer}, {Barsotti}, {Mavalvala}, and {Evans}}}]{2015PhRvD..91h2001D}
\bibinfo{author}{\bibfnamefont{S.}~\bibnamefont{{Dwyer}}},
  \bibinfo{author}{\bibfnamefont{D.}~\bibnamefont{{Sigg}}},
  \bibinfo{author}{\bibfnamefont{S.~W.} \bibnamefont{{Ballmer}}},
  \bibinfo{author}{\bibfnamefont{L.}~\bibnamefont{{Barsotti}}},
  \bibinfo{author}{\bibfnamefont{N.}~\bibnamefont{{Mavalvala}}},
  \bibnamefont{and} \bibinfo{author}{\bibfnamefont{M.}~\bibnamefont{{Evans}}},
  \bibinfo{journal}{\prd} \textbf{\bibinfo{volume}{91}}, \bibinfo{eid}{082001}
  (\bibinfo{year}{2015}), \eprint{1410.0612}.

\bibitem[{\citenamefont{{Reitze} et~al.}(2019)\citenamefont{{Reitze}, {LIGO
  Laboratory: California Institute of Technology}, {LIGO Laboratory:
  Massachusetts Institute of Technology}, {LIGO Hanford Observatory}, and {LIGO
  Livingston Observatory}}}]{2019BAAS...51c.141R}
\bibinfo{author}{\bibfnamefont{D.}~\bibnamefont{{Reitze}}},
  \bibinfo{author}{\bibnamefont{{LIGO Laboratory: California Institute of
  Technology}}}, \bibinfo{author}{\bibnamefont{{LIGO Laboratory: Massachusetts
  Institute of Technology}}}, \bibinfo{author}{\bibnamefont{{LIGO Hanford
  Observatory}}}, \bibnamefont{and} \bibinfo{author}{\bibnamefont{{LIGO
  Livingston Observatory}}}, \bibinfo{journal}{Bulletin of the AAS}
  \textbf{\bibinfo{volume}{51}}, \bibinfo{eid}{141} (\bibinfo{year}{2019}),
  \eprint{1903.04615}.

\bibitem[{\citenamefont{{McLaughlin}}(2013)}]{mc13}
\bibinfo{author}{\bibfnamefont{M.~A.} \bibnamefont{{McLaughlin}}},
  \bibinfo{journal}{Classical and Quantum Gravity}
  \textbf{\bibinfo{volume}{30}}, \bibinfo{eid}{224008} (\bibinfo{year}{2013}),
  \eprint{1310.0758}.

\bibitem[{\citenamefont{{Demorest} and {Nanograv
  Collaboration}}(2021)}]{NANOGRAV21}
\bibinfo{author}{\bibfnamefont{P.}~\bibnamefont{{Demorest}}} \bibnamefont{and}
  \bibinfo{author}{\bibnamefont{{Nanograv Collaboration}}}, in
  \emph{\bibinfo{booktitle}{American Astronomical Society Meeting Abstracts}}
  (\bibinfo{year}{2021}), vol. \bibinfo{volume}{237} of
  \emph{\bibinfo{series}{American Astronomical Society Meeting Abstracts}}, p.
  \bibinfo{pages}{101.01}.

\bibitem[{\citenamefont{{Dolch} et~al.}(2022)\citenamefont{{Dolch}, {Crawford},
  {Wahl}, {Baker}, {Christy}, {DeCesar}, {Gironda}, {Kaplan}, {Key}, {Lam}
  et~al.}}]{NANOGRAV22}
\bibinfo{author}{\bibfnamefont{T.}~\bibnamefont{{Dolch}}},
  \bibinfo{author}{\bibfnamefont{F.}~\bibnamefont{{Crawford}}},
  \bibinfo{author}{\bibfnamefont{H.}~\bibnamefont{{Wahl}}},
  \bibinfo{author}{\bibfnamefont{P.}~\bibnamefont{{Baker}}},
  \bibinfo{author}{\bibfnamefont{B.}~\bibnamefont{{Christy}}},
  \bibinfo{author}{\bibfnamefont{M.}~\bibnamefont{{DeCesar}}},
  \bibinfo{author}{\bibfnamefont{M.}~\bibnamefont{{Gironda}}},
  \bibinfo{author}{\bibfnamefont{D.}~\bibnamefont{{Kaplan}}},
  \bibinfo{author}{\bibfnamefont{J.}~\bibnamefont{{Key}}},
  \bibinfo{author}{\bibfnamefont{M.}~\bibnamefont{{Lam}}},
  \bibnamefont{et~al.}, in \emph{\bibinfo{booktitle}{American Astronomical
  Society Meeting \#240}} (\bibinfo{year}{2022}), vol. \bibinfo{volume}{240} of
  \emph{\bibinfo{series}{American Astronomical Society Meeting Abstracts}}, p.
  \bibinfo{pages}{315.01}.

\bibitem[{\citenamefont{{Afzal} et~al.}(2023)\citenamefont{{Afzal}, {Agazie},
  {Anumarlapudi}, {Archibald}, {Arzoumanian}, {Baker}, {B{\'e}csy},
  {Blanco-Pillado}, {Blecha}, {Boddy} et~al.}}]{2023ApJ...951L..11A}
\bibinfo{author}{\bibfnamefont{A.}~\bibnamefont{{Afzal}}},
  \bibinfo{author}{\bibfnamefont{G.}~\bibnamefont{{Agazie}}},
  \bibinfo{author}{\bibfnamefont{A.}~\bibnamefont{{Anumarlapudi}}},
  \bibinfo{author}{\bibfnamefont{A.~M.} \bibnamefont{{Archibald}}},
  \bibinfo{author}{\bibfnamefont{Z.}~\bibnamefont{{Arzoumanian}}},
  \bibinfo{author}{\bibfnamefont{P.~T.} \bibnamefont{{Baker}}},
  \bibinfo{author}{\bibfnamefont{B.}~\bibnamefont{{B{\'e}csy}}},
  \bibinfo{author}{\bibfnamefont{J.~J.} \bibnamefont{{Blanco-Pillado}}},
  \bibinfo{author}{\bibfnamefont{L.}~\bibnamefont{{Blecha}}},
  \bibinfo{author}{\bibfnamefont{K.~K.} \bibnamefont{{Boddy}}},
  \bibnamefont{et~al.}, \bibinfo{journal}{Astrophysical Journal Letters}
  \textbf{\bibinfo{volume}{951}}, \bibinfo{eid}{L11} (\bibinfo{year}{2023}),
  \eprint{2306.16219}.

\bibitem[{\citenamefont{{Hobbs}}(2015)}]{PPTA1}
\bibinfo{author}{\bibfnamefont{G.}~\bibnamefont{{Hobbs}}},
  \bibinfo{journal}{Publication of Korean Astronomical Society}
  \textbf{\bibinfo{volume}{30}}, \bibinfo{pages}{577} (\bibinfo{year}{2015}).

\bibitem[{\citenamefont{{Kerr} et~al.}(2020)\citenamefont{{Kerr}, {Reardon},
  {Hobbs}, {Shannon}, {Manchester}, {Dai}, {Russell}, {Zhang}, {van Straten},
  {Os{\l}owski} et~al.}}]{PPTA2}
\bibinfo{author}{\bibfnamefont{M.}~\bibnamefont{{Kerr}}},
  \bibinfo{author}{\bibfnamefont{D.~J.} \bibnamefont{{Reardon}}},
  \bibinfo{author}{\bibfnamefont{G.}~\bibnamefont{{Hobbs}}},
  \bibinfo{author}{\bibfnamefont{R.~M.} \bibnamefont{{Shannon}}},
  \bibinfo{author}{\bibfnamefont{R.~N.} \bibnamefont{{Manchester}}},
  \bibinfo{author}{\bibfnamefont{S.}~\bibnamefont{{Dai}}},
  \bibinfo{author}{\bibfnamefont{C.~J.} \bibnamefont{{Russell}}},
  \bibinfo{author}{\bibfnamefont{S.}~\bibnamefont{{Zhang}}},
  \bibinfo{author}{\bibfnamefont{W.}~\bibnamefont{{van Straten}}},
  \bibinfo{author}{\bibfnamefont{S.}~\bibnamefont{{Os{\l}owski}}},
  \bibnamefont{et~al.}, \bibinfo{journal}{Publications of the Astron. Soc. of
  Australia} \textbf{\bibinfo{volume}{37}}, \bibinfo{eid}{e020}
  (\bibinfo{year}{2020}), \eprint{2003.09780}.

\bibitem[{\citenamefont{{Zic} et~al.}(2023)\citenamefont{{Zic}, {Reardon},
  {Kapur}, {Hobbs}, {Mandow}, {Cury{\l}o}, {Shannon}, {Askew}, {Bailes}, {Bhat}
  et~al.}}]{PPTA3}
\bibinfo{author}{\bibfnamefont{A.}~\bibnamefont{{Zic}}},
  \bibinfo{author}{\bibfnamefont{D.~J.} \bibnamefont{{Reardon}}},
  \bibinfo{author}{\bibfnamefont{A.}~\bibnamefont{{Kapur}}},
  \bibinfo{author}{\bibfnamefont{G.}~\bibnamefont{{Hobbs}}},
  \bibinfo{author}{\bibfnamefont{R.}~\bibnamefont{{Mandow}}},
  \bibinfo{author}{\bibfnamefont{M.}~\bibnamefont{{Cury{\l}o}}},
  \bibinfo{author}{\bibfnamefont{R.~M.} \bibnamefont{{Shannon}}},
  \bibinfo{author}{\bibfnamefont{J.}~\bibnamefont{{Askew}}},
  \bibinfo{author}{\bibfnamefont{M.}~\bibnamefont{{Bailes}}},
  \bibinfo{author}{\bibfnamefont{N.~D.~R.} \bibnamefont{{Bhat}}},
  \bibnamefont{et~al.}, \bibinfo{journal}{Publications of the Astron. Soc. of
  Australia} \textbf{\bibinfo{volume}{40}}, \bibinfo{eid}{e049}
  (\bibinfo{year}{2023}), \eprint{2306.16230}.

\bibitem[{\citenamefont{{Manchester} et~al.}(2013)\citenamefont{{Manchester},
  {Hobbs}, {Bailes}, {Coles}, {van Straten}, {Keith}, {Shannon}, {Bhat},
  {Brown}, {Burke-Spolaor} et~al.}}]{man13}
\bibinfo{author}{\bibfnamefont{R.~N.} \bibnamefont{{Manchester}}},
  \bibinfo{author}{\bibfnamefont{G.}~\bibnamefont{{Hobbs}}},
  \bibinfo{author}{\bibfnamefont{M.}~\bibnamefont{{Bailes}}},
  \bibinfo{author}{\bibfnamefont{W.~A.} \bibnamefont{{Coles}}},
  \bibinfo{author}{\bibfnamefont{W.}~\bibnamefont{{van Straten}}},
  \bibinfo{author}{\bibfnamefont{M.~J.} \bibnamefont{{Keith}}},
  \bibinfo{author}{\bibfnamefont{R.~M.} \bibnamefont{{Shannon}}},
  \bibinfo{author}{\bibfnamefont{N.~D.~R.} \bibnamefont{{Bhat}}},
  \bibinfo{author}{\bibfnamefont{A.}~\bibnamefont{{Brown}}},
  \bibinfo{author}{\bibfnamefont{S.~G.} \bibnamefont{{Burke-Spolaor}}},
  \bibnamefont{et~al.}, \bibinfo{journal}{Publications of the Astron. Soc. of
  Australia} \textbf{\bibinfo{volume}{30}}, \bibinfo{eid}{e017}
  (\bibinfo{year}{2013}), \eprint{1210.6130}.

\bibitem[{\citenamefont{{Kramer} and {Champion}}(2013)}]{2013CQGra..30v4009K}
\bibinfo{author}{\bibfnamefont{M.}~\bibnamefont{{Kramer}}} \bibnamefont{and}
  \bibinfo{author}{\bibfnamefont{D.~J.} \bibnamefont{{Champion}}},
  \bibinfo{journal}{Classical and Quantum Gravity}
  \textbf{\bibinfo{volume}{30}}, \bibinfo{eid}{224009} (\bibinfo{year}{2013}).

\bibitem[{\citenamefont{{Desvignes} et~al.}(2016)\citenamefont{{Desvignes},
  {Caballero}, {Lentati}, {Verbiest}, {Champion}, {Stappers}, {Janssen},
  {Lazarus}, {Os{\l}owski}, {Babak} et~al.}}]{2016MNRAS.458.3341D}
\bibinfo{author}{\bibfnamefont{G.}~\bibnamefont{{Desvignes}}},
  \bibinfo{author}{\bibfnamefont{R.~N.} \bibnamefont{{Caballero}}},
  \bibinfo{author}{\bibfnamefont{L.}~\bibnamefont{{Lentati}}},
  \bibinfo{author}{\bibfnamefont{J.~P.~W.} \bibnamefont{{Verbiest}}},
  \bibinfo{author}{\bibfnamefont{D.~J.} \bibnamefont{{Champion}}},
  \bibinfo{author}{\bibfnamefont{B.~W.} \bibnamefont{{Stappers}}},
  \bibinfo{author}{\bibfnamefont{G.~H.} \bibnamefont{{Janssen}}},
  \bibinfo{author}{\bibfnamefont{P.}~\bibnamefont{{Lazarus}}},
  \bibinfo{author}{\bibfnamefont{S.}~\bibnamefont{{Os{\l}owski}}},
  \bibinfo{author}{\bibfnamefont{S.}~\bibnamefont{{Babak}}},
  \bibnamefont{et~al.}, \bibinfo{journal}{Monthly Notices of the RAS}
  \textbf{\bibinfo{volume}{458}}, \bibinfo{pages}{3341} (\bibinfo{year}{2016}),
  \eprint{1602.08511}.

\bibitem[{\citenamefont{{Joshi} et~al.}(2018)\citenamefont{{Joshi},
  {Arumugasamy}, {Bagchi}, {Bandyopadhyay}, {Basu}, {Dhanda Batra},
  {Bethapudi}, {Choudhary}, {De}, {Dey} et~al.}}]{2018JApA...39...51J}
\bibinfo{author}{\bibfnamefont{B.~C.} \bibnamefont{{Joshi}}},
  \bibinfo{author}{\bibfnamefont{P.}~\bibnamefont{{Arumugasamy}}},
  \bibinfo{author}{\bibfnamefont{M.}~\bibnamefont{{Bagchi}}},
  \bibinfo{author}{\bibfnamefont{D.}~\bibnamefont{{Bandyopadhyay}}},
  \bibinfo{author}{\bibfnamefont{A.}~\bibnamefont{{Basu}}},
  \bibinfo{author}{\bibfnamefont{N.}~\bibnamefont{{Dhanda Batra}}},
  \bibinfo{author}{\bibfnamefont{S.}~\bibnamefont{{Bethapudi}}},
  \bibinfo{author}{\bibfnamefont{A.}~\bibnamefont{{Choudhary}}},
  \bibinfo{author}{\bibfnamefont{K.}~\bibnamefont{{De}}},
  \bibinfo{author}{\bibfnamefont{L.}~\bibnamefont{{Dey}}},
  \bibnamefont{et~al.}, \bibinfo{journal}{Journal of Astrophysics and
  Astronomy} \textbf{\bibinfo{volume}{39}}, \bibinfo{eid}{51}
  (\bibinfo{year}{2018}).

\bibitem[{\citenamefont{{Hobbs} et~al.}(2010)\citenamefont{{Hobbs},
  {Archibald}, {Arzoumanian}, {Backer}, {Bailes}, {Bhat}, {Burgay},
  {Burke-Spolaor}, {Champion}, {Cognard} et~al.}}]{2010CQGra..27h4013H}
\bibinfo{author}{\bibfnamefont{G.}~\bibnamefont{{Hobbs}}},
  \bibinfo{author}{\bibfnamefont{A.}~\bibnamefont{{Archibald}}},
  \bibinfo{author}{\bibfnamefont{Z.}~\bibnamefont{{Arzoumanian}}},
  \bibinfo{author}{\bibfnamefont{D.}~\bibnamefont{{Backer}}},
  \bibinfo{author}{\bibfnamefont{M.}~\bibnamefont{{Bailes}}},
  \bibinfo{author}{\bibfnamefont{N.~D.~R.} \bibnamefont{{Bhat}}},
  \bibinfo{author}{\bibfnamefont{M.}~\bibnamefont{{Burgay}}},
  \bibinfo{author}{\bibfnamefont{S.}~\bibnamefont{{Burke-Spolaor}}},
  \bibinfo{author}{\bibfnamefont{D.}~\bibnamefont{{Champion}}},
  \bibinfo{author}{\bibfnamefont{I.}~\bibnamefont{{Cognard}}},
  \bibnamefont{et~al.}, \bibinfo{journal}{Classical and Quantum Gravity}
  \textbf{\bibinfo{volume}{27}}, \bibinfo{eid}{084013} (\bibinfo{year}{2010}),
  \eprint{0911.5206}.

\bibitem[{\citenamefont{{Verbiest} et~al.}(2016)\citenamefont{{Verbiest},
  {Lentati}, {Hobbs}, {van Haasteren}, {Demorest}, {Janssen}, {Wang},
  {Desvignes}, {Caballero}, {Keith} et~al.}}]{2016MNRAS.458.1267V}
\bibinfo{author}{\bibfnamefont{J.~P.~W.} \bibnamefont{{Verbiest}}},
  \bibinfo{author}{\bibfnamefont{L.}~\bibnamefont{{Lentati}}},
  \bibinfo{author}{\bibfnamefont{G.}~\bibnamefont{{Hobbs}}},
  \bibinfo{author}{\bibfnamefont{R.}~\bibnamefont{{van Haasteren}}},
  \bibinfo{author}{\bibfnamefont{P.~B.} \bibnamefont{{Demorest}}},
  \bibinfo{author}{\bibfnamefont{G.~H.} \bibnamefont{{Janssen}}},
  \bibinfo{author}{\bibfnamefont{J.~B.} \bibnamefont{{Wang}}},
  \bibinfo{author}{\bibfnamefont{G.}~\bibnamefont{{Desvignes}}},
  \bibinfo{author}{\bibfnamefont{R.~N.} \bibnamefont{{Caballero}}},
  \bibinfo{author}{\bibfnamefont{M.~J.} \bibnamefont{{Keith}}},
  \bibnamefont{et~al.}, \bibinfo{journal}{Monthly Notices of the RAS}
  \textbf{\bibinfo{volume}{458}}, \bibinfo{pages}{1267} (\bibinfo{year}{2016}),
  \eprint{1602.03640}.

\bibitem[{\citenamefont{{Cordes} et~al.}(2004)\citenamefont{{Cordes}, {Kramer},
  {Lazio}, {Stappers}, {Backer}, and {Johnston}}}]{cor04}
\bibinfo{author}{\bibfnamefont{J.~M.} \bibnamefont{{Cordes}}},
  \bibinfo{author}{\bibfnamefont{M.}~\bibnamefont{{Kramer}}},
  \bibinfo{author}{\bibfnamefont{T.~J.~W.} \bibnamefont{{Lazio}}},
  \bibinfo{author}{\bibfnamefont{B.~W.} \bibnamefont{{Stappers}}},
  \bibinfo{author}{\bibfnamefont{D.~C.} \bibnamefont{{Backer}}},
  \bibnamefont{and}
  \bibinfo{author}{\bibfnamefont{S.}~\bibnamefont{{Johnston}}},
  \bibinfo{journal}{New Astronomy Review} \textbf{\bibinfo{volume}{48}},
  \bibinfo{pages}{1413} (\bibinfo{year}{2004}), \eprint{astro-ph/0505555}.

\bibitem[{\citenamefont{{Lazio}}(2013)}]{laz13}
\bibinfo{author}{\bibfnamefont{T.~J.~W.} \bibnamefont{{Lazio}}},
  \bibinfo{journal}{Classical and Quantum Gravity}
  \textbf{\bibinfo{volume}{30}}, \bibinfo{eid}{224011} (\bibinfo{year}{2013}).

\bibitem[{\citenamefont{{Dewdney} et~al.}(2009)\citenamefont{{Dewdney}, {Hall},
  {Schilizzi}, and {Lazio}}}]{5136190}
\bibinfo{author}{\bibfnamefont{P.~E.} \bibnamefont{{Dewdney}}},
  \bibinfo{author}{\bibfnamefont{P.~J.} \bibnamefont{{Hall}}},
  \bibinfo{author}{\bibfnamefont{R.~T.} \bibnamefont{{Schilizzi}}},
  \bibnamefont{and} \bibinfo{author}{\bibfnamefont{T.~J.~L.~W.}
  \bibnamefont{{Lazio}}}, \bibinfo{journal}{IEEE Proceedings}
  \textbf{\bibinfo{volume}{97}}, \bibinfo{pages}{1482} (\bibinfo{year}{2009}).

\bibitem[{\citenamefont{{Nan} and {Li}}(2013)}]{2013MS&E...44a2022N}
\bibinfo{author}{\bibfnamefont{R.}~\bibnamefont{{Nan}}} \bibnamefont{and}
  \bibinfo{author}{\bibfnamefont{D.}~\bibnamefont{{Li}}}, in
  \emph{\bibinfo{booktitle}{Materials Science and Engineering Conference
  Series}} (\bibinfo{publisher}{IOP}, \bibinfo{year}{2013}),
  vol.~\bibinfo{volume}{44} of \emph{\bibinfo{series}{Materials Science and
  Engineering Conference Series}}, p. \bibinfo{pages}{012022}.

\bibitem[{\citenamefont{{Renzini} et~al.}(2022)\citenamefont{{Renzini},
  {Goncharov}, {Jenkins}, and {Meyers}}}]{2022Galax..10...34R}
\bibinfo{author}{\bibfnamefont{A.~I.} \bibnamefont{{Renzini}}},
  \bibinfo{author}{\bibfnamefont{B.}~\bibnamefont{{Goncharov}}},
  \bibinfo{author}{\bibfnamefont{A.~C.} \bibnamefont{{Jenkins}}},
  \bibnamefont{and} \bibinfo{author}{\bibfnamefont{P.~M.}
  \bibnamefont{{Meyers}}}, \bibinfo{journal}{Galaxies}
  \textbf{\bibinfo{volume}{10}}, \bibinfo{eid}{34} (\bibinfo{year}{2022}),
  \eprint{2202.00178}.

\bibitem[{\citenamefont{{Regimbau} et~al.}(2012)\citenamefont{{Regimbau},
  {Dent}, {Del Pozzo}, {Giampanis}, {Li}, {Robinson}, {Van Den Broeck},
  {Meacher}, {Rodriguez}, {Sathyaprakash} et~al.}}]{2012PhRvD..86l2001R}
\bibinfo{author}{\bibfnamefont{T.}~\bibnamefont{{Regimbau}}},
  \bibinfo{author}{\bibfnamefont{T.}~\bibnamefont{{Dent}}},
  \bibinfo{author}{\bibfnamefont{W.}~\bibnamefont{{Del Pozzo}}},
  \bibinfo{author}{\bibfnamefont{S.}~\bibnamefont{{Giampanis}}},
  \bibinfo{author}{\bibfnamefont{T.~G.~F.} \bibnamefont{{Li}}},
  \bibinfo{author}{\bibfnamefont{C.}~\bibnamefont{{Robinson}}},
  \bibinfo{author}{\bibfnamefont{C.}~\bibnamefont{{Van Den Broeck}}},
  \bibinfo{author}{\bibfnamefont{D.}~\bibnamefont{{Meacher}}},
  \bibinfo{author}{\bibfnamefont{C.}~\bibnamefont{{Rodriguez}}},
  \bibinfo{author}{\bibfnamefont{B.~S.} \bibnamefont{{Sathyaprakash}}},
  \bibnamefont{et~al.}, \bibinfo{journal}{\prd} \textbf{\bibinfo{volume}{86}},
  \bibinfo{eid}{122001} (\bibinfo{year}{2012}), \eprint{1201.3563}.

\bibitem[{\citenamefont{{Regimbau} et~al.}(2014)\citenamefont{{Regimbau},
  {Meacher}, and {Coughlin}}}]{2014PhRvD..89h4046R}
\bibinfo{author}{\bibfnamefont{T.}~\bibnamefont{{Regimbau}}},
  \bibinfo{author}{\bibfnamefont{D.}~\bibnamefont{{Meacher}}},
  \bibnamefont{and}
  \bibinfo{author}{\bibfnamefont{M.}~\bibnamefont{{Coughlin}}},
  \bibinfo{journal}{\prd} \textbf{\bibinfo{volume}{89}}, \bibinfo{eid}{084046}
  (\bibinfo{year}{2014}), \eprint{1404.1134}.

\bibitem[{\citenamefont{{Wette}}(2020)}]{2020SoftX..1200634W}
\bibinfo{author}{\bibfnamefont{K.}~\bibnamefont{{Wette}}},
  \bibinfo{journal}{SoftwareX} \textbf{\bibinfo{volume}{12}},
  \bibinfo{eid}{100634} (\bibinfo{year}{2020}), \eprint{2012.09552}.

\bibitem[{\citenamefont{Ashton et~al.}(2019)\citenamefont{Ashton, Huebner,
  Lasky, Talbot, Ackley, Biscoveanu, Chu, Divarkala, Easter, Goncharov
  et~al.}}]{BILBY2019}
\bibinfo{author}{\bibfnamefont{G.}~\bibnamefont{Ashton}},
  \bibinfo{author}{\bibfnamefont{M.}~\bibnamefont{Huebner}},
  \bibinfo{author}{\bibfnamefont{P.~D.} \bibnamefont{Lasky}},
  \bibinfo{author}{\bibfnamefont{C.}~\bibnamefont{Talbot}},
  \bibinfo{author}{\bibfnamefont{K.}~\bibnamefont{Ackley}},
  \bibinfo{author}{\bibfnamefont{S.}~\bibnamefont{Biscoveanu}},
  \bibinfo{author}{\bibfnamefont{Q.}~\bibnamefont{Chu}},
  \bibinfo{author}{\bibfnamefont{A.}~\bibnamefont{Divarkala}},
  \bibinfo{author}{\bibfnamefont{P.~J.} \bibnamefont{Easter}},
  \bibinfo{author}{\bibfnamefont{B.}~\bibnamefont{Goncharov}},
  \bibnamefont{et~al.}, \bibinfo{journal}{The Astrophysical Journal Supplement
  Series} \textbf{\bibinfo{volume}{241}}, \bibinfo{pages}{27}
  (\bibinfo{year}{2019}), \urlprefix\url{https://arxiv.org/abs/1811.02042}.

\bibitem[{\citenamefont{{Petiteau} et~al.}(2008)\citenamefont{{Petiteau},
  {Auger}, {Halloin}, {Jeannin}, {Plagnol}, {Pireaux}, {Regimbau}, and
  {Vinet}}}]{2008PhRvD..77b3002P}
\bibinfo{author}{\bibfnamefont{A.}~\bibnamefont{{Petiteau}}},
  \bibinfo{author}{\bibfnamefont{G.}~\bibnamefont{{Auger}}},
  \bibinfo{author}{\bibfnamefont{H.}~\bibnamefont{{Halloin}}},
  \bibinfo{author}{\bibfnamefont{O.}~\bibnamefont{{Jeannin}}},
  \bibinfo{author}{\bibfnamefont{E.}~\bibnamefont{{Plagnol}}},
  \bibinfo{author}{\bibfnamefont{S.}~\bibnamefont{{Pireaux}}},
  \bibinfo{author}{\bibfnamefont{T.}~\bibnamefont{{Regimbau}}},
  \bibnamefont{and} \bibinfo{author}{\bibfnamefont{J.-Y.}
  \bibnamefont{{Vinet}}}, \bibinfo{journal}{\prd}
  \textbf{\bibinfo{volume}{77}}, \bibinfo{eid}{023002} (\bibinfo{year}{2008}),
  \eprint{0802.2023}.

\bibitem[{\citenamefont{Ellis et~al.}(2020)\citenamefont{Ellis, Vallisneri,
  Taylor, and Baker}}]{ellis_2020_4059815}
\bibinfo{author}{\bibfnamefont{J.~A.} \bibnamefont{Ellis}},
  \bibinfo{author}{\bibfnamefont{M.}~\bibnamefont{Vallisneri}},
  \bibinfo{author}{\bibfnamefont{S.~R.} \bibnamefont{Taylor}},
  \bibnamefont{and} \bibinfo{author}{\bibfnamefont{P.~T.} \bibnamefont{Baker}},
  \emph{\bibinfo{title}{Enterprise: Enhanced numerical toolbox enabling a
  robust pulsar inference suite}}, \bibinfo{howpublished}{Zenodo}
  (\bibinfo{year}{2020}),
  \urlprefix\url{https://doi.org/10.5281/zenodo.4059815}.

\bibitem[{\citenamefont{{Hobbs} et~al.}(2006)\citenamefont{{Hobbs}, {Edwards},
  and {Manchester}}}]{2006MNRAS.369..655H}
\bibinfo{author}{\bibfnamefont{G.~B.} \bibnamefont{{Hobbs}}},
  \bibinfo{author}{\bibfnamefont{R.~T.} \bibnamefont{{Edwards}}},
  \bibnamefont{and} \bibinfo{author}{\bibfnamefont{R.~N.}
  \bibnamefont{{Manchester}}}, \bibinfo{journal}{Monthly Notices of the RAS}
  \textbf{\bibinfo{volume}{369}}, \bibinfo{pages}{655} (\bibinfo{year}{2006}),
  \eprint{astro-ph/0603381}.

\bibitem[{\citenamefont{{Luo} et~al.}(2021)\citenamefont{{Luo}, {Ransom},
  {Demorest}, {Ray}, {Archibald}, {Kerr}, {Jennings}, {Bachetti}, {van
  Haasteren}, {Champagne} et~al.}}]{2021ApJ...911...45L}
\bibinfo{author}{\bibfnamefont{J.}~\bibnamefont{{Luo}}},
  \bibinfo{author}{\bibfnamefont{S.}~\bibnamefont{{Ransom}}},
  \bibinfo{author}{\bibfnamefont{P.}~\bibnamefont{{Demorest}}},
  \bibinfo{author}{\bibfnamefont{P.~S.} \bibnamefont{{Ray}}},
  \bibinfo{author}{\bibfnamefont{A.}~\bibnamefont{{Archibald}}},
  \bibinfo{author}{\bibfnamefont{M.}~\bibnamefont{{Kerr}}},
  \bibinfo{author}{\bibfnamefont{R.~J.} \bibnamefont{{Jennings}}},
  \bibinfo{author}{\bibfnamefont{M.}~\bibnamefont{{Bachetti}}},
  \bibinfo{author}{\bibfnamefont{R.}~\bibnamefont{{van Haasteren}}},
  \bibinfo{author}{\bibfnamefont{C.~A.} \bibnamefont{{Champagne}}},
  \bibnamefont{et~al.}, \bibinfo{journal}{\apj} \textbf{\bibinfo{volume}{911}},
  \bibinfo{eid}{45} (\bibinfo{year}{2021}), \eprint{2012.00074}.

\bibitem[{\citenamefont{{Dimitriou} et~al.}(2024)\citenamefont{{Dimitriou},
  {Figueroa}, and {Zald{\'\i}var}}}]{2024JCAP...09..032D}
\bibinfo{author}{\bibfnamefont{A.}~\bibnamefont{{Dimitriou}}},
  \bibinfo{author}{\bibfnamefont{D.~G.} \bibnamefont{{Figueroa}}},
  \bibnamefont{and}
  \bibinfo{author}{\bibfnamefont{B.}~\bibnamefont{{Zald{\'\i}var}}},
  \bibinfo{journal}{Journal of Cosmology and Astroparticle Physics}
  \textbf{\bibinfo{volume}{2024}}, \bibinfo{eid}{032} (\bibinfo{year}{2024}),
  \eprint{2309.08430}.

\bibitem[{\citenamefont{{Bellomo} et~al.}(2022)\citenamefont{{Bellomo},
  {Bertacca}, {Jenkins}, {Matarrese}, {Raccanelli}, {Regimbau}, {Ricciardone},
  and {Sakellariadou}}}]{2022JCAP...06..030B}
\bibinfo{author}{\bibfnamefont{N.}~\bibnamefont{{Bellomo}}},
  \bibinfo{author}{\bibfnamefont{D.}~\bibnamefont{{Bertacca}}},
  \bibinfo{author}{\bibfnamefont{A.~C.} \bibnamefont{{Jenkins}}},
  \bibinfo{author}{\bibfnamefont{S.}~\bibnamefont{{Matarrese}}},
  \bibinfo{author}{\bibfnamefont{A.}~\bibnamefont{{Raccanelli}}},
  \bibinfo{author}{\bibfnamefont{T.}~\bibnamefont{{Regimbau}}},
  \bibinfo{author}{\bibfnamefont{A.}~\bibnamefont{{Ricciardone}}},
  \bibnamefont{and}
  \bibinfo{author}{\bibfnamefont{M.}~\bibnamefont{{Sakellariadou}}},
  \bibinfo{journal}{Journal of Cosmology and Astroparticle Physics}
  \textbf{\bibinfo{volume}{2022}}, \bibinfo{eid}{030} (\bibinfo{year}{2022}),
  \eprint{2110.15059}.

\bibitem[{\citenamefont{{Yi} et~al.}(2022)\citenamefont{{Yi}, {Nelemans},
  {Brinkerink}, {Kostrzewa-Rutkowska}, {Timmer}, {Stoppa}, {Rossi}, and
  {Portegies Zwart}}}]{2022A&A...663A.155Y}
\bibinfo{author}{\bibfnamefont{S.-X.} \bibnamefont{{Yi}}},
  \bibinfo{author}{\bibfnamefont{G.}~\bibnamefont{{Nelemans}}},
  \bibinfo{author}{\bibfnamefont{C.}~\bibnamefont{{Brinkerink}}},
  \bibinfo{author}{\bibfnamefont{Z.}~\bibnamefont{{Kostrzewa-Rutkowska}}},
  \bibinfo{author}{\bibfnamefont{S.~T.} \bibnamefont{{Timmer}}},
  \bibinfo{author}{\bibfnamefont{F.}~\bibnamefont{{Stoppa}}},
  \bibinfo{author}{\bibfnamefont{E.~M.} \bibnamefont{{Rossi}}},
  \bibnamefont{and} \bibinfo{author}{\bibfnamefont{S.~F.}
  \bibnamefont{{Portegies Zwart}}}, \bibinfo{journal}{Astronomy and
  Astrophysics} \textbf{\bibinfo{volume}{663}}, \bibinfo{eid}{A155}
  (\bibinfo{year}{2022}), \eprint{2106.13662}.

\bibitem[{\citenamefont{{L{\"o}ffler} et~al.}(2012)\citenamefont{{L{\"o}ffler},
  {Faber}, {Bentivegna}, {Bode}, {Diener}, {Haas}, {Hinder}, {Mundim}, {Ott},
  {Schnetter} et~al.}}]{2012CQGra..29k5001L}
\bibinfo{author}{\bibfnamefont{F.}~\bibnamefont{{L{\"o}ffler}}},
  \bibinfo{author}{\bibfnamefont{J.}~\bibnamefont{{Faber}}},
  \bibinfo{author}{\bibfnamefont{E.}~\bibnamefont{{Bentivegna}}},
  \bibinfo{author}{\bibfnamefont{T.}~\bibnamefont{{Bode}}},
  \bibinfo{author}{\bibfnamefont{P.}~\bibnamefont{{Diener}}},
  \bibinfo{author}{\bibfnamefont{R.}~\bibnamefont{{Haas}}},
  \bibinfo{author}{\bibfnamefont{I.}~\bibnamefont{{Hinder}}},
  \bibinfo{author}{\bibfnamefont{B.~C.} \bibnamefont{{Mundim}}},
  \bibinfo{author}{\bibfnamefont{C.~D.} \bibnamefont{{Ott}}},
  \bibinfo{author}{\bibfnamefont{E.}~\bibnamefont{{Schnetter}}},
  \bibnamefont{et~al.}, \bibinfo{journal}{Classical and Quantum Gravity}
  \textbf{\bibinfo{volume}{29}}, \bibinfo{eid}{115001} (\bibinfo{year}{2012}),
  \eprint{1111.3344}.

\bibitem[{\citenamefont{{Ellis} et~al.}(2024)\citenamefont{{Ellis},
  {Fairbairn}, {Franciolini}, {H{\"u}tsi}, {Iovino}, {Lewicki}, {Raidal},
  {Urrutia}, {Vaskonen}, and {Veerm{\"a}e}}}]{2024PhRvD.109b3522E}
\bibinfo{author}{\bibfnamefont{J.}~\bibnamefont{{Ellis}}},
  \bibinfo{author}{\bibfnamefont{M.}~\bibnamefont{{Fairbairn}}},
  \bibinfo{author}{\bibfnamefont{G.}~\bibnamefont{{Franciolini}}},
  \bibinfo{author}{\bibfnamefont{G.}~\bibnamefont{{H{\"u}tsi}}},
  \bibinfo{author}{\bibfnamefont{A.}~\bibnamefont{{Iovino}}},
  \bibinfo{author}{\bibfnamefont{M.}~\bibnamefont{{Lewicki}}},
  \bibinfo{author}{\bibfnamefont{M.}~\bibnamefont{{Raidal}}},
  \bibinfo{author}{\bibfnamefont{J.}~\bibnamefont{{Urrutia}}},
  \bibinfo{author}{\bibfnamefont{V.}~\bibnamefont{{Vaskonen}}},
  \bibnamefont{and}
  \bibinfo{author}{\bibfnamefont{H.}~\bibnamefont{{Veerm{\"a}e}}},
  \bibinfo{journal}{\prd} \textbf{\bibinfo{volume}{109}}, \bibinfo{eid}{023522}
  (\bibinfo{year}{2024}), \eprint{2308.08546}.

\bibitem[{\citenamefont{{Lorimer} and {Kramer}}(2012)}]{2012hpa..book.....L}
\bibinfo{author}{\bibfnamefont{D.~R.} \bibnamefont{{Lorimer}}}
  \bibnamefont{and} \bibinfo{author}{\bibfnamefont{M.}~\bibnamefont{{Kramer}}},
  \emph{\bibinfo{title}{{Handbook of Pulsar Astronomy}}}
  (\bibinfo{publisher}{Cambridge University Press}, \bibinfo{year}{2012}).

\bibitem[{\citenamefont{{Eghdami} et~al.}(2018)\citenamefont{{Eghdami},
  {Panahi}, and {Movahed}}}]{Issa}
\bibinfo{author}{\bibfnamefont{I.}~\bibnamefont{{Eghdami}}},
  \bibinfo{author}{\bibfnamefont{H.}~\bibnamefont{{Panahi}}}, \bibnamefont{and}
  \bibinfo{author}{\bibfnamefont{S.~M.~S.} \bibnamefont{{Movahed}}},
  \bibinfo{journal}{\apj} \textbf{\bibinfo{volume}{864}}, \bibinfo{eid}{162}
  (\bibinfo{year}{2018}), \eprint{1704.08599}.

\bibitem[{\citenamefont{{Ransom} et~al.}(2019)\citenamefont{{Ransom},
  {Brazier}, {Chatterjee}, {Cohen}, {Cordes}, {DeCesar}, {Demorest}, {Hazboun},
  {Lam}, {Lynch} et~al.}}]{2019BAAS...51g.195R}
\bibinfo{author}{\bibfnamefont{S.}~\bibnamefont{{Ransom}}},
  \bibinfo{author}{\bibfnamefont{A.}~\bibnamefont{{Brazier}}},
  \bibinfo{author}{\bibfnamefont{S.}~\bibnamefont{{Chatterjee}}},
  \bibinfo{author}{\bibfnamefont{T.}~\bibnamefont{{Cohen}}},
  \bibinfo{author}{\bibfnamefont{J.~M.} \bibnamefont{{Cordes}}},
  \bibinfo{author}{\bibfnamefont{M.~E.} \bibnamefont{{DeCesar}}},
  \bibinfo{author}{\bibfnamefont{P.~B.} \bibnamefont{{Demorest}}},
  \bibinfo{author}{\bibfnamefont{J.~S.} \bibnamefont{{Hazboun}}},
  \bibinfo{author}{\bibfnamefont{M.~T.} \bibnamefont{{Lam}}},
  \bibinfo{author}{\bibfnamefont{R.~S.} \bibnamefont{{Lynch}}},
  \bibnamefont{et~al.}, in \emph{\bibinfo{booktitle}{Bulletin of the American
  Astronomical Society}} (\bibinfo{year}{2019}), vol.~\bibinfo{volume}{51}, p.
  \bibinfo{pages}{195}, \eprint{1908.05356}.

\bibitem[{\citenamefont{{Hellings} and {Downs}}(1983)}]{1983ApJ...265L..39H}
\bibinfo{author}{\bibfnamefont{R.~W.} \bibnamefont{{Hellings}}}
  \bibnamefont{and} \bibinfo{author}{\bibfnamefont{G.~S.}
  \bibnamefont{{Downs}}}, \bibinfo{journal}{Astrophysical Journal Letters}
  \textbf{\bibinfo{volume}{265}}, \bibinfo{pages}{L39} (\bibinfo{year}{1983}).

\bibitem[{\citenamefont{{Jenet} et~al.}(2005)\citenamefont{{Jenet}, {Hobbs},
  {Lee}, and {Manchester}}}]{2005ApJ...625L.123J}
\bibinfo{author}{\bibfnamefont{F.~A.} \bibnamefont{{Jenet}}},
  \bibinfo{author}{\bibfnamefont{G.~B.} \bibnamefont{{Hobbs}}},
  \bibinfo{author}{\bibfnamefont{K.~J.} \bibnamefont{{Lee}}}, \bibnamefont{and}
  \bibinfo{author}{\bibfnamefont{R.~N.} \bibnamefont{{Manchester}}},
  \bibinfo{journal}{Astrophysical Journal Letters}
  \textbf{\bibinfo{volume}{625}}, \bibinfo{pages}{L123} (\bibinfo{year}{2005}),
  \eprint{astro-ph/0504458}.

\bibitem[{\citenamefont{{Furusawa} et~al.}(2025)\citenamefont{{Furusawa},
  {Kuroyanagi}, and {Ichiki}}}]{2025arXiv250510284F}
\bibinfo{author}{\bibfnamefont{K.}~\bibnamefont{{Furusawa}}},
  \bibinfo{author}{\bibfnamefont{S.}~\bibnamefont{{Kuroyanagi}}},
  \bibnamefont{and} \bibinfo{author}{\bibfnamefont{K.}~\bibnamefont{{Ichiki}}},
  \bibinfo{journal}{arXiv e-prints} \bibinfo{eid}{arXiv:2505.10284}
  (\bibinfo{year}{2025}), \eprint{2505.10284}.

\bibitem[{\citenamefont{{Bernardo} and {Ng}}(2023)}]{2023PhRvD.107d4007B}
\bibinfo{author}{\bibfnamefont{R.~C.} \bibnamefont{{Bernardo}}}
  \bibnamefont{and} \bibinfo{author}{\bibfnamefont{K.-W.} \bibnamefont{{Ng}}},
  \bibinfo{journal}{\prd} \textbf{\bibinfo{volume}{107}}, \bibinfo{eid}{044007}
  (\bibinfo{year}{2023}), \eprint{2208.12538}.

\bibitem[{\citenamefont{{Karniadakis} et~al.}(2021)\citenamefont{{Karniadakis},
  {Kevrekidis}, {Lu}, {Perdikaris}, {Wang}, and {Yang}}}]{2021NatRP...3..422K}
\bibinfo{author}{\bibfnamefont{G.~E.} \bibnamefont{{Karniadakis}}},
  \bibinfo{author}{\bibfnamefont{I.~G.} \bibnamefont{{Kevrekidis}}},
  \bibinfo{author}{\bibfnamefont{L.}~\bibnamefont{{Lu}}},
  \bibinfo{author}{\bibfnamefont{P.}~\bibnamefont{{Perdikaris}}},
  \bibinfo{author}{\bibfnamefont{S.}~\bibnamefont{{Wang}}}, \bibnamefont{and}
  \bibinfo{author}{\bibfnamefont{L.}~\bibnamefont{{Yang}}},
  \bibinfo{journal}{Nature Reviews Physics} \textbf{\bibinfo{volume}{3}},
  \bibinfo{pages}{422} (\bibinfo{year}{2021}).

\bibitem[{\citenamefont{{Hao} et~al.}(2022)\citenamefont{{Hao}, {Liu}, {Zhang},
  {Ying}, {Feng}, {Su}, and {Zhu}}}]{2022arXiv221108064H}
\bibinfo{author}{\bibfnamefont{Z.}~\bibnamefont{{Hao}}},
  \bibinfo{author}{\bibfnamefont{S.}~\bibnamefont{{Liu}}},
  \bibinfo{author}{\bibfnamefont{Y.}~\bibnamefont{{Zhang}}},
  \bibinfo{author}{\bibfnamefont{C.}~\bibnamefont{{Ying}}},
  \bibinfo{author}{\bibfnamefont{Y.}~\bibnamefont{{Feng}}},
  \bibinfo{author}{\bibfnamefont{H.}~\bibnamefont{{Su}}}, \bibnamefont{and}
  \bibinfo{author}{\bibfnamefont{J.}~\bibnamefont{{Zhu}}},
  \bibinfo{journal}{arXiv e-prints} \bibinfo{eid}{arXiv:2211.08064}
  (\bibinfo{year}{2022}), \eprint{2211.08064}.

\bibitem[{\citenamefont{{George} and {Huerta}}(2018)}]{2018PhLB..778...64G}
\bibinfo{author}{\bibfnamefont{D.}~\bibnamefont{{George}}} \bibnamefont{and}
  \bibinfo{author}{\bibfnamefont{E.~A.} \bibnamefont{{Huerta}}},
  \bibinfo{journal}{Physics Letters B} \textbf{\bibinfo{volume}{778}},
  \bibinfo{pages}{64} (\bibinfo{year}{2018}), \eprint{1711.03121}.

\bibitem[{\citenamefont{{Cuoco} et~al.}(2021)\citenamefont{{Cuoco}, {Powell},
  {Cavagli{\`a}}, {Ackley}, {Bejger}, {Chatterjee}, {Coughlin}, {Coughlin},
  {Easter}, {Essick} et~al.}}]{2021MLS&T...2a1002C}
\bibinfo{author}{\bibfnamefont{E.}~\bibnamefont{{Cuoco}}},
  \bibinfo{author}{\bibfnamefont{J.}~\bibnamefont{{Powell}}},
  \bibinfo{author}{\bibfnamefont{M.}~\bibnamefont{{Cavagli{\`a}}}},
  \bibinfo{author}{\bibfnamefont{K.}~\bibnamefont{{Ackley}}},
  \bibinfo{author}{\bibfnamefont{M.}~\bibnamefont{{Bejger}}},
  \bibinfo{author}{\bibfnamefont{C.}~\bibnamefont{{Chatterjee}}},
  \bibinfo{author}{\bibfnamefont{M.}~\bibnamefont{{Coughlin}}},
  \bibinfo{author}{\bibfnamefont{S.}~\bibnamefont{{Coughlin}}},
  \bibinfo{author}{\bibfnamefont{P.}~\bibnamefont{{Easter}}},
  \bibinfo{author}{\bibfnamefont{R.}~\bibnamefont{{Essick}}},
  \bibnamefont{et~al.}, \bibinfo{journal}{Machine Learning: Science and
  Technology} \textbf{\bibinfo{volume}{2}}, \bibinfo{eid}{011002}
  (\bibinfo{year}{2021}), \eprint{2005.03745}.

\bibitem[{\citenamefont{{Schmidt} et~al.}(2021)\citenamefont{{Schmidt},
  {Breschi}, {Gamba}, {Pagano}, {Rettegno}, {Riemenschneider}, {Bernuzzi},
  {Nagar}, and {Del Pozzo}}}]{2021PhRvD.103d3020S}
\bibinfo{author}{\bibfnamefont{S.}~\bibnamefont{{Schmidt}}},
  \bibinfo{author}{\bibfnamefont{M.}~\bibnamefont{{Breschi}}},
  \bibinfo{author}{\bibfnamefont{R.}~\bibnamefont{{Gamba}}},
  \bibinfo{author}{\bibfnamefont{G.}~\bibnamefont{{Pagano}}},
  \bibinfo{author}{\bibfnamefont{P.}~\bibnamefont{{Rettegno}}},
  \bibinfo{author}{\bibfnamefont{G.}~\bibnamefont{{Riemenschneider}}},
  \bibinfo{author}{\bibfnamefont{S.}~\bibnamefont{{Bernuzzi}}},
  \bibinfo{author}{\bibfnamefont{A.}~\bibnamefont{{Nagar}}}, \bibnamefont{and}
  \bibinfo{author}{\bibfnamefont{W.}~\bibnamefont{{Del Pozzo}}},
  \bibinfo{journal}{\prd} \textbf{\bibinfo{volume}{103}}, \bibinfo{eid}{043020}
  (\bibinfo{year}{2021}), \eprint{2011.01958}.

\bibitem[{\citenamefont{{Ormiston} et~al.}(2020)\citenamefont{{Ormiston},
  {Nguyen}, {Coughlin}, {Adhikari}, and {Katsavounidis}}}]{2020PhRvR...2c3066O}
\bibinfo{author}{\bibfnamefont{R.}~\bibnamefont{{Ormiston}}},
  \bibinfo{author}{\bibfnamefont{T.}~\bibnamefont{{Nguyen}}},
  \bibinfo{author}{\bibfnamefont{M.}~\bibnamefont{{Coughlin}}},
  \bibinfo{author}{\bibfnamefont{R.~X.} \bibnamefont{{Adhikari}}},
  \bibnamefont{and}
  \bibinfo{author}{\bibfnamefont{E.}~\bibnamefont{{Katsavounidis}}},
  \bibinfo{journal}{Physical Review Research} \textbf{\bibinfo{volume}{2}},
  \bibinfo{eid}{033066} (\bibinfo{year}{2020}), \eprint{2005.06534}.

\bibitem[{\citenamefont{{Vajente} et~al.}(2020)\citenamefont{{Vajente},
  {Huang}, {Isi}, {Driggers}, {Kissel}, {Szczepa{\'n}czyk}, and
  {Vitale}}}]{2020PhRvD.101d2003V}
\bibinfo{author}{\bibfnamefont{G.}~\bibnamefont{{Vajente}}},
  \bibinfo{author}{\bibfnamefont{Y.}~\bibnamefont{{Huang}}},
  \bibinfo{author}{\bibfnamefont{M.}~\bibnamefont{{Isi}}},
  \bibinfo{author}{\bibfnamefont{J.~C.} \bibnamefont{{Driggers}}},
  \bibinfo{author}{\bibfnamefont{J.~S.} \bibnamefont{{Kissel}}},
  \bibinfo{author}{\bibfnamefont{M.~J.} \bibnamefont{{Szczepa{\'n}czyk}}},
  \bibnamefont{and} \bibinfo{author}{\bibfnamefont{S.}~\bibnamefont{{Vitale}}},
  \bibinfo{journal}{\prd} \textbf{\bibinfo{volume}{101}}, \bibinfo{eid}{042003}
  (\bibinfo{year}{2020}), \eprint{1911.09083}.

\bibitem[{\citenamefont{{Sasaoka} et~al.}(2022)\citenamefont{{Sasaoka}, {Hou},
  {Somiya}, and {Takahashi}}}]{2022PhRvD.105j3030S}
\bibinfo{author}{\bibfnamefont{S.}~\bibnamefont{{Sasaoka}}},
  \bibinfo{author}{\bibfnamefont{Y.}~\bibnamefont{{Hou}}},
  \bibinfo{author}{\bibfnamefont{K.}~\bibnamefont{{Somiya}}}, \bibnamefont{and}
  \bibinfo{author}{\bibfnamefont{H.}~\bibnamefont{{Takahashi}}},
  \bibinfo{journal}{\prd} \textbf{\bibinfo{volume}{105}}, \bibinfo{eid}{103030}
  (\bibinfo{year}{2022}), \eprint{2202.12784}.

\bibitem[{\citenamefont{{Kolmus} et~al.}(2022)\citenamefont{{Kolmus}, {Baltus},
  {Janquart}, {van Laarhoven}, {Caudill}, and {Heskes}}}]{2022PhRvD.106b3032K}
\bibinfo{author}{\bibfnamefont{A.}~\bibnamefont{{Kolmus}}},
  \bibinfo{author}{\bibfnamefont{G.}~\bibnamefont{{Baltus}}},
  \bibinfo{author}{\bibfnamefont{J.}~\bibnamefont{{Janquart}}},
  \bibinfo{author}{\bibfnamefont{T.}~\bibnamefont{{van Laarhoven}}},
  \bibinfo{author}{\bibfnamefont{S.}~\bibnamefont{{Caudill}}},
  \bibnamefont{and} \bibinfo{author}{\bibfnamefont{T.}~\bibnamefont{{Heskes}}},
  \bibinfo{journal}{\prd} \textbf{\bibinfo{volume}{106}}, \bibinfo{eid}{023032}
  (\bibinfo{year}{2022}), \eprint{2111.00833}.

\bibitem[{\citenamefont{{Chatterjee} et~al.}(2022)\citenamefont{{Chatterjee},
  {Wen}, {Beveridge}, {Linqing Wen Team}, and {Damon Beveridge
  Team}}}]{2022APS..APRE17006C}
\bibinfo{author}{\bibfnamefont{C.}~\bibnamefont{{Chatterjee}}},
  \bibinfo{author}{\bibfnamefont{L.}~\bibnamefont{{Wen}}},
  \bibinfo{author}{\bibfnamefont{D.}~\bibnamefont{{Beveridge}}},
  \bibinfo{author}{\bibnamefont{{Linqing Wen Team}}}, \bibnamefont{and}
  \bibinfo{author}{\bibnamefont{{Damon Beveridge Team}}}, in
  \emph{\bibinfo{booktitle}{APS April Meeting Abstracts}}
  (\bibinfo{year}{2022}), vol. \bibinfo{volume}{2022} of
  \emph{\bibinfo{series}{APS Meeting Abstracts}}, p. \bibinfo{pages}{E17.006}.

\bibitem[{\citenamefont{{Bonetti} et~al.}(2024)\citenamefont{{Bonetti},
  {Franchini}, {Galuzzi}, and {Sesana}}}]{2024A&A...687A..42B}
\bibinfo{author}{\bibfnamefont{M.}~\bibnamefont{{Bonetti}}},
  \bibinfo{author}{\bibfnamefont{A.}~\bibnamefont{{Franchini}}},
  \bibinfo{author}{\bibfnamefont{B.~G.} \bibnamefont{{Galuzzi}}},
  \bibnamefont{and} \bibinfo{author}{\bibfnamefont{A.}~\bibnamefont{{Sesana}}},
  \bibinfo{journal}{Astronomy and Astrophysics} \textbf{\bibinfo{volume}{687}},
  \bibinfo{eid}{A42} (\bibinfo{year}{2024}), \eprint{2311.04276}.

\bibitem[{\citenamefont{{Laal} et~al.}(2025)\citenamefont{{Laal}, {Taylor},
  {Kelley}, {Simon}, {G{\"u}ltekin}, {Wright}, {B{\'e}csy}, {Casey-Clyde},
  {Chen}, {Cingoranelli} et~al.}}]{2025ApJ...982...55L}
\bibinfo{author}{\bibfnamefont{N.}~\bibnamefont{{Laal}}},
  \bibinfo{author}{\bibfnamefont{S.~R.} \bibnamefont{{Taylor}}},
  \bibinfo{author}{\bibfnamefont{L.~Z.} \bibnamefont{{Kelley}}},
  \bibinfo{author}{\bibfnamefont{J.}~\bibnamefont{{Simon}}},
  \bibinfo{author}{\bibfnamefont{K.}~\bibnamefont{{G{\"u}ltekin}}},
  \bibinfo{author}{\bibfnamefont{D.}~\bibnamefont{{Wright}}},
  \bibinfo{author}{\bibfnamefont{B.}~\bibnamefont{{B{\'e}csy}}},
  \bibinfo{author}{\bibfnamefont{J.~A.} \bibnamefont{{Casey-Clyde}}},
  \bibinfo{author}{\bibfnamefont{S.}~\bibnamefont{{Chen}}},
  \bibinfo{author}{\bibfnamefont{A.}~\bibnamefont{{Cingoranelli}}},
  \bibnamefont{et~al.}, \bibinfo{journal}{\apj} \textbf{\bibinfo{volume}{982}},
  \bibinfo{eid}{55} (\bibinfo{year}{2025}), \eprint{2411.10519}.

\bibitem[{\citenamefont{{Tejero-Cantero}
  et~al.}(2020)\citenamefont{{Tejero-Cantero}, {Boelts}, {Deistler},
  {Lueckmann}, {Durkan}, {Gon{\c{c}}alves}, {Greenberg}, and
  {Macke}}}]{tejero2020sbi}
\bibinfo{author}{\bibfnamefont{A.}~\bibnamefont{{Tejero-Cantero}}},
  \bibinfo{author}{\bibfnamefont{J.}~\bibnamefont{{Boelts}}},
  \bibinfo{author}{\bibfnamefont{M.}~\bibnamefont{{Deistler}}},
  \bibinfo{author}{\bibfnamefont{J.-M.} \bibnamefont{{Lueckmann}}},
  \bibinfo{author}{\bibfnamefont{C.}~\bibnamefont{{Durkan}}},
  \bibinfo{author}{\bibfnamefont{P.}~\bibnamefont{{Gon{\c{c}}alves}}},
  \bibinfo{author}{\bibfnamefont{D.}~\bibnamefont{{Greenberg}}},
  \bibnamefont{and} \bibinfo{author}{\bibfnamefont{J.}~\bibnamefont{{Macke}}},
  \bibinfo{journal}{The Journal of Open Source Software}
  \textbf{\bibinfo{volume}{5}}, \bibinfo{eid}{2505} (\bibinfo{year}{2020}),
  \eprint{2007.09114}.

\bibitem[{\citenamefont{{Papamakarios} and
  {Murray}}(2016)}]{2016arXiv160506376P}
\bibinfo{author}{\bibfnamefont{G.}~\bibnamefont{{Papamakarios}}}
  \bibnamefont{and} \bibinfo{author}{\bibfnamefont{I.}~\bibnamefont{{Murray}}},
  \bibinfo{journal}{arXiv e-prints} \bibinfo{eid}{arXiv:1605.06376}
  (\bibinfo{year}{2016}), \eprint{1605.06376}.

\bibitem[{\citenamefont{{Alsing} et~al.}(2019)\citenamefont{{Alsing},
  {Charnock}, {Feeney}, and {Wandelt}}}]{2019MNRAS.488.4440A}
\bibinfo{author}{\bibfnamefont{J.}~\bibnamefont{{Alsing}}},
  \bibinfo{author}{\bibfnamefont{T.}~\bibnamefont{{Charnock}}},
  \bibinfo{author}{\bibfnamefont{S.}~\bibnamefont{{Feeney}}}, \bibnamefont{and}
  \bibinfo{author}{\bibfnamefont{B.}~\bibnamefont{{Wandelt}}},
  \bibinfo{journal}{Monthly Notices of the RAS} \textbf{\bibinfo{volume}{488}},
  \bibinfo{pages}{4440} (\bibinfo{year}{2019}), \eprint{1903.00007}.

\bibitem[{\citenamefont{{Cranmer} et~al.}(2020)\citenamefont{{Cranmer},
  {Brehmer}, and {Louppe}}}]{2020PNAS..11730055C}
\bibinfo{author}{\bibfnamefont{K.}~\bibnamefont{{Cranmer}}},
  \bibinfo{author}{\bibfnamefont{J.}~\bibnamefont{{Brehmer}}},
  \bibnamefont{and} \bibinfo{author}{\bibfnamefont{G.}~\bibnamefont{{Louppe}}},
  \bibinfo{journal}{Proceedings of the National Academy of Science}
  \textbf{\bibinfo{volume}{117}}, \bibinfo{pages}{30055}
  (\bibinfo{year}{2020}), \eprint{1911.01429}.

\bibitem[{\citenamefont{{Hahn} et~al.}(2022)\citenamefont{{Hahn}, {Abidi},
  {Eickenberg}, {Ho}, {Lemos}, {Massara}, {Moradinezhad Dizgah}, and
  {R{\'e}galdo-Saint Blancard}}}]{2022mla..confE..24H}
\bibinfo{author}{\bibfnamefont{C.}~\bibnamefont{{Hahn}}},
  \bibinfo{author}{\bibfnamefont{M.}~\bibnamefont{{Abidi}}},
  \bibinfo{author}{\bibfnamefont{M.}~\bibnamefont{{Eickenberg}}},
  \bibinfo{author}{\bibfnamefont{S.}~\bibnamefont{{Ho}}},
  \bibinfo{author}{\bibfnamefont{P.}~\bibnamefont{{Lemos}}},
  \bibinfo{author}{\bibfnamefont{E.}~\bibnamefont{{Massara}}},
  \bibinfo{author}{\bibfnamefont{A.}~\bibnamefont{{Moradinezhad Dizgah}}},
  \bibnamefont{and}
  \bibinfo{author}{\bibfnamefont{B.}~\bibnamefont{{R{\'e}galdo-Saint
  Blancard}}}, in \emph{\bibinfo{booktitle}{Machine Learning for Astrophysics}}
  (\bibinfo{year}{2022}), p.~\bibinfo{pages}{24}.

\bibitem[{\citenamefont{{Alvey} et~al.}(2024)\citenamefont{{Alvey}, {Bhardwaj},
  {Domcke}, {Pieroni}, and {Weniger}}}]{2024PhRvD.109h3008A}
\bibinfo{author}{\bibfnamefont{J.}~\bibnamefont{{Alvey}}},
  \bibinfo{author}{\bibfnamefont{U.}~\bibnamefont{{Bhardwaj}}},
  \bibinfo{author}{\bibfnamefont{V.}~\bibnamefont{{Domcke}}},
  \bibinfo{author}{\bibfnamefont{M.}~\bibnamefont{{Pieroni}}},
  \bibnamefont{and}
  \bibinfo{author}{\bibfnamefont{C.}~\bibnamefont{{Weniger}}},
  \bibinfo{journal}{\prd} \textbf{\bibinfo{volume}{109}}, \bibinfo{eid}{083008}
  (\bibinfo{year}{2024}), \eprint{2309.07954}.

\bibitem[{\citenamefont{{Abedi} et~al.}(2025)\citenamefont{{Abedi}, {Jalali
  Kanafi}, and {Movahed}}}]{abedi2024}
\bibinfo{author}{\bibfnamefont{F.}~\bibnamefont{{Abedi}}},
  \bibinfo{author}{\bibfnamefont{M.~H.} \bibnamefont{{Jalali Kanafi}}},
  \bibnamefont{and} \bibinfo{author}{\bibfnamefont{S.~M.~S.}
  \bibnamefont{{Movahed}}}, \bibinfo{journal}{International Journal of
  Geometric Methods in Modern Physics} \textbf{\bibinfo{volume}{22}},
  \bibinfo{eid}{2540006-143} (\bibinfo{year}{2025}), \eprint{2410.01751}.

\bibitem[{\citenamefont{{Eckmann} et~al.}(1987)\citenamefont{{Eckmann},
  {Oliffson Kamphorst}, and {Ruelle}}}]{1987EL......4..973E}
\bibinfo{author}{\bibfnamefont{J.~P.} \bibnamefont{{Eckmann}}},
  \bibinfo{author}{\bibfnamefont{S.}~\bibnamefont{{Oliffson Kamphorst}}},
  \bibnamefont{and} \bibinfo{author}{\bibfnamefont{D.}~\bibnamefont{{Ruelle}}},
  \bibinfo{journal}{EPL (Europhysics Letters)} \textbf{\bibinfo{volume}{4}},
  \bibinfo{pages}{973} (\bibinfo{year}{1987}).

\bibitem[{\citenamefont{{Marwan} et~al.}(2007)\citenamefont{{Marwan}, {Carmen
  Romano}, {Thiel}, and {Kurths}}}]{2007PhR...438..237M}
\bibinfo{author}{\bibfnamefont{N.}~\bibnamefont{{Marwan}}},
  \bibinfo{author}{\bibfnamefont{M.}~\bibnamefont{{Carmen Romano}}},
  \bibinfo{author}{\bibfnamefont{M.}~\bibnamefont{{Thiel}}}, \bibnamefont{and}
  \bibinfo{author}{\bibfnamefont{J.}~\bibnamefont{{Kurths}}},
  \bibinfo{journal}{Physics Reports} \textbf{\bibinfo{volume}{438}},
  \bibinfo{pages}{237} (\bibinfo{year}{2007}), \eprint{2501.13933}.

\bibitem[{\citenamefont{{Zhan}}(2011)}]{2011SSPMA}
\bibinfo{author}{\bibfnamefont{H.}~\bibnamefont{{Zhan}}},
  \bibinfo{journal}{Scientia Sinica Physica, Mechanica \& Astronomica}
  \textbf{\bibinfo{volume}{41}}, \bibinfo{pages}{1441} (\bibinfo{year}{2011}).

\bibitem[{\citenamefont{Chen et~al.}(2018)\citenamefont{Chen, Yang, and
  Kumara}}]{chen2018recurrence}
\bibinfo{author}{\bibfnamefont{C.-B.} \bibnamefont{Chen}},
  \bibinfo{author}{\bibfnamefont{H.}~\bibnamefont{Yang}}, \bibnamefont{and}
  \bibinfo{author}{\bibfnamefont{S.}~\bibnamefont{Kumara}},
  \bibinfo{journal}{Chaos: An Interdisciplinary Journal of Nonlinear Science}
  \textbf{\bibinfo{volume}{28}} (\bibinfo{year}{2018}).

\bibitem[{\citenamefont{Takens}(1981)}]{takens1981detecting}
\bibinfo{author}{\bibfnamefont{F.}~\bibnamefont{Takens}}, in
  \emph{\bibinfo{booktitle}{Dynamical systems and turbulence, Warwick 1980}}
  (\bibinfo{publisher}{Springer}, \bibinfo{year}{1981}), pp.
  \bibinfo{pages}{366--381}.

\bibitem[{\citenamefont{Packard et~al.}(1980)\citenamefont{Packard,
  Crutchfield, Farmer, and Shaw}}]{packard1980geometry}
\bibinfo{author}{\bibfnamefont{N.~H.} \bibnamefont{Packard}},
  \bibinfo{author}{\bibfnamefont{J.~P.} \bibnamefont{Crutchfield}},
  \bibinfo{author}{\bibfnamefont{J.~D.} \bibnamefont{Farmer}},
  \bibnamefont{and} \bibinfo{author}{\bibfnamefont{R.~S.} \bibnamefont{Shaw}},
  \bibinfo{journal}{Physical review letters} \textbf{\bibinfo{volume}{45}},
  \bibinfo{pages}{712} (\bibinfo{year}{1980}).

\bibitem[{\citenamefont{{Myers} et~al.}(2019)\citenamefont{{Myers}, {Munch},
  and {Khasawneh}}}]{2019PhRvE.100b2314M}
\bibinfo{author}{\bibfnamefont{A.}~\bibnamefont{{Myers}}},
  \bibinfo{author}{\bibfnamefont{E.}~\bibnamefont{{Munch}}}, \bibnamefont{and}
  \bibinfo{author}{\bibfnamefont{F.~A.} \bibnamefont{{Khasawneh}}},
  \bibinfo{journal}{\pre} \textbf{\bibinfo{volume}{100}}, \bibinfo{eid}{022314}
  (\bibinfo{year}{2019}), \eprint{1904.07403}.

\bibitem[{\citenamefont{Yesilli et~al.}(2022)\citenamefont{Yesilli, Khasawneh,
  and Otto}}]{yesilli2022topological}
\bibinfo{author}{\bibfnamefont{M.~C.} \bibnamefont{Yesilli}},
  \bibinfo{author}{\bibfnamefont{F.~A.} \bibnamefont{Khasawneh}},
  \bibnamefont{and} \bibinfo{author}{\bibfnamefont{A.}~\bibnamefont{Otto}},
  \bibinfo{journal}{The International Journal of Advanced Manufacturing
  Technology} \textbf{\bibinfo{volume}{119}}, \bibinfo{pages}{5687}
  (\bibinfo{year}{2022}).

\bibitem[{\citenamefont{{Myers} et~al.}(2023)\citenamefont{{Myers}, {Chumley},
  {Khasawneh}, and {Munch}}}]{2023PhRvE.107c4303M}
\bibinfo{author}{\bibfnamefont{A.~D.} \bibnamefont{{Myers}}},
  \bibinfo{author}{\bibfnamefont{M.~M.} \bibnamefont{{Chumley}}},
  \bibinfo{author}{\bibfnamefont{F.~A.} \bibnamefont{{Khasawneh}}},
  \bibnamefont{and} \bibinfo{author}{\bibfnamefont{E.}~\bibnamefont{{Munch}}},
  \bibinfo{journal}{\pre} \textbf{\bibinfo{volume}{107}}, \bibinfo{eid}{034303}
  (\bibinfo{year}{2023}), \eprint{2206.02530}.

\bibitem[{\citenamefont{Lacasa et~al.}(2012)\citenamefont{Lacasa, Nunez,
  Rold{\'a}n, Parrondo, and Luque}}]{lacasa2012time}
\bibinfo{author}{\bibfnamefont{L.}~\bibnamefont{Lacasa}},
  \bibinfo{author}{\bibfnamefont{A.}~\bibnamefont{Nunez}},
  \bibinfo{author}{\bibfnamefont{{\'E}.}~\bibnamefont{Rold{\'a}n}},
  \bibinfo{author}{\bibfnamefont{J.~M.} \bibnamefont{Parrondo}},
  \bibnamefont{and} \bibinfo{author}{\bibfnamefont{B.}~\bibnamefont{Luque}},
  \bibinfo{journal}{The European Physical Journal B}
  \textbf{\bibinfo{volume}{85}}, \bibinfo{pages}{217} (\bibinfo{year}{2012}).

\bibitem[{\citenamefont{{Yang} et~al.}(2009)\citenamefont{{Yang}, {Wang},
  {Yang}, and {Mang}}}]{yang2009visibility}
\bibinfo{author}{\bibfnamefont{Y.}~\bibnamefont{{Yang}}},
  \bibinfo{author}{\bibfnamefont{J.}~\bibnamefont{{Wang}}},
  \bibinfo{author}{\bibfnamefont{H.}~\bibnamefont{{Yang}}}, \bibnamefont{and}
  \bibinfo{author}{\bibfnamefont{J.}~\bibnamefont{{Mang}}},
  \bibinfo{journal}{Physica A Statistical Mechanics and its Applications}
  \textbf{\bibinfo{volume}{388}}, \bibinfo{pages}{4431} (\bibinfo{year}{2009}).

\bibitem[{\citenamefont{Small}(2013)}]{small2013complex}
\bibinfo{author}{\bibfnamefont{M.}~\bibnamefont{Small}}, in
  \emph{\bibinfo{booktitle}{2013 IEEE International Symposium on Circuits and
  Systems (ISCAS)}} (\bibinfo{organization}{IEEE}, \bibinfo{year}{2013}), pp.
  \bibinfo{pages}{2509--2512}.

\bibitem[{\citenamefont{{McCullough} et~al.}(2015)\citenamefont{{McCullough},
  {Small}, {Stemler}, and {Iu}}}]{2015Chaos..25e3101M}
\bibinfo{author}{\bibfnamefont{M.}~\bibnamefont{{McCullough}}},
  \bibinfo{author}{\bibfnamefont{M.}~\bibnamefont{{Small}}},
  \bibinfo{author}{\bibfnamefont{T.}~\bibnamefont{{Stemler}}},
  \bibnamefont{and} \bibinfo{author}{\bibfnamefont{H.~H.-C.}
  \bibnamefont{{Iu}}}, \bibinfo{journal}{Chaos} \textbf{\bibinfo{volume}{25}},
  \bibinfo{eid}{053101} (\bibinfo{year}{2015}), \eprint{1501.06656}.

\bibitem[{\citenamefont{Campanharo et~al.}(2011)\citenamefont{Campanharo,
  Sirer, Malmgren, Ramos, and Amaral}}]{campanharo2011duality}
\bibinfo{author}{\bibfnamefont{A.~S. L.~O.} \bibnamefont{Campanharo}},
  \bibinfo{author}{\bibfnamefont{M.~I.} \bibnamefont{Sirer}},
  \bibinfo{author}{\bibfnamefont{R.~D.} \bibnamefont{Malmgren}},
  \bibinfo{author}{\bibfnamefont{F.~M.} \bibnamefont{Ramos}}, \bibnamefont{and}
  \bibinfo{author}{\bibfnamefont{L.~A.~N.} \bibnamefont{Amaral}},
  \bibinfo{journal}{PLOS ONE} \textbf{\bibinfo{volume}{6}}, \bibinfo{pages}{1}
  (\bibinfo{year}{2011}),
  \urlprefix\url{https://doi.org/10.1371/journal.pone.0023378}.

\bibitem[{\citenamefont{Silva et~al.}(2021)\citenamefont{Silva, Silva, Ribeiro,
  and Silva}}]{silva2021time}
\bibinfo{author}{\bibfnamefont{V.~F.} \bibnamefont{Silva}},
  \bibinfo{author}{\bibfnamefont{M.~E.} \bibnamefont{Silva}},
  \bibinfo{author}{\bibfnamefont{P.}~\bibnamefont{Ribeiro}}, \bibnamefont{and}
  \bibinfo{author}{\bibfnamefont{F.}~\bibnamefont{Silva}},
  \bibinfo{journal}{Wiley Interdisciplinary Reviews: Data Mining and Knowledge
  Discovery} \textbf{\bibinfo{volume}{11}}, \bibinfo{pages}{e1404}
  (\bibinfo{year}{2021}).

\bibitem[{\citenamefont{{Zou} et~al.}(2019)\citenamefont{{Zou}, {Donner},
  {Marwan}, {Donges}, and {Kurths}}}]{2019PhR...787....1Z}
\bibinfo{author}{\bibfnamefont{Y.}~\bibnamefont{{Zou}}},
  \bibinfo{author}{\bibfnamefont{R.~V.} \bibnamefont{{Donner}}},
  \bibinfo{author}{\bibfnamefont{N.}~\bibnamefont{{Marwan}}},
  \bibinfo{author}{\bibfnamefont{J.~F.} \bibnamefont{{Donges}}},
  \bibnamefont{and} \bibinfo{author}{\bibfnamefont{J.}~\bibnamefont{{Kurths}}},
  \bibinfo{journal}{Physics Reports} \textbf{\bibinfo{volume}{787}},
  \bibinfo{pages}{1} (\bibinfo{year}{2019}), \eprint{2501.18737}.

\bibitem[{\citenamefont{{Albert} and
  {Barab{\'a}si}}(2002)}]{2002RvMP...74...47A}
\bibinfo{author}{\bibfnamefont{R.}~\bibnamefont{{Albert}}} \bibnamefont{and}
  \bibinfo{author}{\bibfnamefont{A.-L.} \bibnamefont{{Barab{\'a}si}}},
  \bibinfo{journal}{Reviews of Modern Physics} \textbf{\bibinfo{volume}{74}},
  \bibinfo{pages}{47} (\bibinfo{year}{2002}), \eprint{cond-mat/0106096}.

\bibitem[{\citenamefont{{Yu} et~al.}(2006)\citenamefont{{Yu}, {Righero}, and
  {Kocarev}}}]{2006PhRvL..97r8701Y}
\bibinfo{author}{\bibfnamefont{D.}~\bibnamefont{{Yu}}},
  \bibinfo{author}{\bibfnamefont{M.}~\bibnamefont{{Righero}}},
  \bibnamefont{and}
  \bibinfo{author}{\bibfnamefont{L.}~\bibnamefont{{Kocarev}}},
  \bibinfo{journal}{\prl} \textbf{\bibinfo{volume}{97}}, \bibinfo{eid}{188701}
  (\bibinfo{year}{2006}).

\bibitem[{\citenamefont{{Ranjan} and {Zhang}}(2013)}]{2013PhyA..392.3833R}
\bibinfo{author}{\bibfnamefont{G.}~\bibnamefont{{Ranjan}}} \bibnamefont{and}
  \bibinfo{author}{\bibfnamefont{Z.-L.} \bibnamefont{{Zhang}}},
  \bibinfo{journal}{Physica A Statistical Mechanics and its Applications}
  \textbf{\bibinfo{volume}{392}}, \bibinfo{pages}{3833} (\bibinfo{year}{2013}).

\bibitem[{\citenamefont{Kantz and Schreiber}(2003)}]{kantz2003nonlinear}
\bibinfo{author}{\bibfnamefont{H.}~\bibnamefont{Kantz}} \bibnamefont{and}
  \bibinfo{author}{\bibfnamefont{T.}~\bibnamefont{Schreiber}},
  \emph{\bibinfo{title}{Nonlinear time series analysis}}
  (\bibinfo{publisher}{Cambridge university press}, \bibinfo{year}{2003}).

\bibitem[{\citenamefont{Newman}(2018)}]{network_book2018}
\bibinfo{author}{\bibfnamefont{M.}~\bibnamefont{Newman}},
  \emph{\bibinfo{title}{Networks}} (\bibinfo{publisher}{Oxford University
  Press}, \bibinfo{year}{2018}),
  \urlprefix\url{https://doi.org/10.1093/oso/9780198805090.001.0001}.

\bibitem[{\citenamefont{{Barab{\'a}si}}(2016)}]{2016nesc.book.....B}
\bibinfo{author}{\bibfnamefont{A.-L.} \bibnamefont{{Barab{\'a}si}}},
  \emph{\bibinfo{title}{{Network Science}}} (\bibinfo{year}{2016}).

\bibitem[{\citenamefont{{Ueda} and {Itoh}}(1999)}]{1999ApJ...526..560U}
\bibinfo{author}{\bibfnamefont{H.}~\bibnamefont{{Ueda}}} \bibnamefont{and}
  \bibinfo{author}{\bibfnamefont{M.}~\bibnamefont{{Itoh}}},
  \bibinfo{journal}{\apj} \textbf{\bibinfo{volume}{526}}, \bibinfo{pages}{560}
  (\bibinfo{year}{1999}).

\bibitem[{\citenamefont{{Krioukov} et~al.}(2012)\citenamefont{{Krioukov},
  {Kitsak}, {Sinkovits}, {Rideout}, {Meyer}, and
  {Bogu{\~n}{\'a}}}}]{2012NatSR...2..793K}
\bibinfo{author}{\bibfnamefont{D.}~\bibnamefont{{Krioukov}}},
  \bibinfo{author}{\bibfnamefont{M.}~\bibnamefont{{Kitsak}}},
  \bibinfo{author}{\bibfnamefont{R.~S.} \bibnamefont{{Sinkovits}}},
  \bibinfo{author}{\bibfnamefont{D.}~\bibnamefont{{Rideout}}},
  \bibinfo{author}{\bibfnamefont{D.}~\bibnamefont{{Meyer}}}, \bibnamefont{and}
  \bibinfo{author}{\bibfnamefont{M.}~\bibnamefont{{Bogu{\~n}{\'a}}}},
  \bibinfo{journal}{Scientific Reports} \textbf{\bibinfo{volume}{2}},
  \bibinfo{eid}{793} (\bibinfo{year}{2012}), \eprint{1203.2109}.

\bibitem[{\citenamefont{{Coutinho} et~al.}(2016)\citenamefont{{Coutinho},
  {Hong}, {Albrecht}, {Dey}, {Barab{\'a}si}, {Torrey}, {Vogelsberger}, and
  {Hernquist}}}]{2016arXiv160403236C}
\bibinfo{author}{\bibfnamefont{B.~C.} \bibnamefont{{Coutinho}}},
  \bibinfo{author}{\bibfnamefont{S.}~\bibnamefont{{Hong}}},
  \bibinfo{author}{\bibfnamefont{K.}~\bibnamefont{{Albrecht}}},
  \bibinfo{author}{\bibfnamefont{A.}~\bibnamefont{{Dey}}},
  \bibinfo{author}{\bibfnamefont{A.-L.} \bibnamefont{{Barab{\'a}si}}},
  \bibinfo{author}{\bibfnamefont{P.}~\bibnamefont{{Torrey}}},
  \bibinfo{author}{\bibfnamefont{M.}~\bibnamefont{{Vogelsberger}}},
  \bibnamefont{and}
  \bibinfo{author}{\bibfnamefont{L.}~\bibnamefont{{Hernquist}}},
  \bibinfo{journal}{arXiv e-prints} \bibinfo{eid}{arXiv:1604.03236}
  (\bibinfo{year}{2016}), \eprint{1604.03236}.

\bibitem[{\citenamefont{{Hong} et~al.}(2016)\citenamefont{{Hong}, {Coutinho},
  {Dey}, {Barab{\'a}si}, {Vogelsberger}, {Hernquist}, and
  {Gebhardt}}}]{2016MNRAS.459.2690H}
\bibinfo{author}{\bibfnamefont{S.}~\bibnamefont{{Hong}}},
  \bibinfo{author}{\bibfnamefont{B.~C.} \bibnamefont{{Coutinho}}},
  \bibinfo{author}{\bibfnamefont{A.}~\bibnamefont{{Dey}}},
  \bibinfo{author}{\bibfnamefont{A.-L.} \bibnamefont{{Barab{\'a}si}}},
  \bibinfo{author}{\bibfnamefont{M.}~\bibnamefont{{Vogelsberger}}},
  \bibinfo{author}{\bibfnamefont{L.}~\bibnamefont{{Hernquist}}},
  \bibnamefont{and}
  \bibinfo{author}{\bibfnamefont{K.}~\bibnamefont{{Gebhardt}}},
  \bibinfo{journal}{Monthly Notices of the RAS} \textbf{\bibinfo{volume}{459}},
  \bibinfo{pages}{2690} (\bibinfo{year}{2016}), \eprint{1603.02285}.

\bibitem[{\citenamefont{Kololgi et~al.}(2025)\citenamefont{Kololgi, Naidoo,
  Saintonge, and Lahav}}]{kololgi2025learning}
\bibinfo{author}{\bibfnamefont{D.}~\bibnamefont{Kololgi}},
  \bibinfo{author}{\bibfnamefont{K.}~\bibnamefont{Naidoo}},
  \bibinfo{author}{\bibfnamefont{A.}~\bibnamefont{Saintonge}},
  \bibnamefont{and} \bibinfo{author}{\bibfnamefont{O.}~\bibnamefont{Lahav}},
  \bibinfo{journal}{arXiv preprint arXiv:2512.05909}  (\bibinfo{year}{2025}),
  \eprint{2512.05909}.

\bibitem[{\citenamefont{{Garc{\'i}a} et~al.}(2024)\citenamefont{{Garc{\'i}a},
  {Illiano}, {Torres}, {Papitto}, {Coti Zelati}, {de Martino}, and
  {Patruno}}}]{2024A&A...692A.187G}
\bibinfo{author}{\bibfnamefont{C.~R.} \bibnamefont{{Garc{\'i}a}}},
  \bibinfo{author}{\bibfnamefont{G.}~\bibnamefont{{Illiano}}},
  \bibinfo{author}{\bibfnamefont{D.~F.} \bibnamefont{{Torres}}},
  \bibinfo{author}{\bibfnamefont{A.}~\bibnamefont{{Papitto}}},
  \bibinfo{author}{\bibfnamefont{F.}~\bibnamefont{{Coti Zelati}}},
  \bibinfo{author}{\bibfnamefont{D.}~\bibnamefont{{de Martino}}},
  \bibnamefont{and}
  \bibinfo{author}{\bibfnamefont{A.}~\bibnamefont{{Patruno}}},
  \bibinfo{journal}{Astronomy and Astrophysics} \textbf{\bibinfo{volume}{692}},
  \bibinfo{eid}{A187} (\bibinfo{year}{2024}), \eprint{2410.13650}.

\bibitem[{\citenamefont{{Garc{\'i}a} et~al.}(2022)\citenamefont{{Garc{\'i}a},
  {Torres}, and {Patruno}}}]{2022MNRAS.515.3883G}
\bibinfo{author}{\bibfnamefont{C.~R.} \bibnamefont{{Garc{\'i}a}}},
  \bibinfo{author}{\bibfnamefont{D.~F.} \bibnamefont{{Torres}}},
  \bibnamefont{and}
  \bibinfo{author}{\bibfnamefont{A.}~\bibnamefont{{Patruno}}},
  \bibinfo{journal}{Monthly Notices of the RAS} \textbf{\bibinfo{volume}{515}},
  \bibinfo{pages}{3883} (\bibinfo{year}{2022}), \eprint{2207.06311}.

\bibitem[{\citenamefont{{Vohl} et~al.}(2024)\citenamefont{{Vohl}, {van
  Leeuwen}, and {Maan}}}]{2024A&A...687A.113V}
\bibinfo{author}{\bibfnamefont{D.}~\bibnamefont{{Vohl}}},
  \bibinfo{author}{\bibfnamefont{J.}~\bibnamefont{{van Leeuwen}}},
  \bibnamefont{and} \bibinfo{author}{\bibfnamefont{Y.}~\bibnamefont{{Maan}}},
  \bibinfo{journal}{Astronomy and Astrophysics} \textbf{\bibinfo{volume}{687}},
  \bibinfo{eid}{A113} (\bibinfo{year}{2024}), \eprint{2311.09201}.

\bibitem[{\citenamefont{{Yip} et~al.}(2024)\citenamefont{{Yip}, {Biagetti},
  {Cole}, {Viswanathan}, and {Shiu}}}]{2024JCAP...09..034Y}
\bibinfo{author}{\bibfnamefont{J.~H.~T.} \bibnamefont{{Yip}}},
  \bibinfo{author}{\bibfnamefont{M.}~\bibnamefont{{Biagetti}}},
  \bibinfo{author}{\bibfnamefont{A.}~\bibnamefont{{Cole}}},
  \bibinfo{author}{\bibfnamefont{K.}~\bibnamefont{{Viswanathan}}},
  \bibnamefont{and} \bibinfo{author}{\bibfnamefont{G.}~\bibnamefont{{Shiu}}},
  \bibinfo{journal}{Journal of Cosmology and Astroparticle Physics}
  \textbf{\bibinfo{volume}{2024}}, \bibinfo{eid}{034} (\bibinfo{year}{2024}),
  \eprint{2403.13985}.

\bibitem[{\citenamefont{{Jalali Kanafi} and {Movahed}}(2025)}]{Jalali2025}
\bibinfo{author}{\bibfnamefont{M.~H.} \bibnamefont{{Jalali Kanafi}}}
  \bibnamefont{and} \bibinfo{author}{\bibfnamefont{S.~M.~S.}
  \bibnamefont{{Movahed}}}, \bibinfo{journal}{arXiv preprint arXiv:2511.03636}
  (\bibinfo{year}{2025}).

\bibitem[{\citenamefont{{Nguyen} et~al.}(2024)\citenamefont{{Nguyen},
  {Schmidt}, {Tucci}, {Reinecke}, and {Kosti{\'c}}}}]{2024PhRvL.133v1006N}
\bibinfo{author}{\bibfnamefont{N.-M.} \bibnamefont{{Nguyen}}},
  \bibinfo{author}{\bibfnamefont{F.}~\bibnamefont{{Schmidt}}},
  \bibinfo{author}{\bibfnamefont{B.}~\bibnamefont{{Tucci}}},
  \bibinfo{author}{\bibfnamefont{M.}~\bibnamefont{{Reinecke}}},
  \bibnamefont{and}
  \bibinfo{author}{\bibfnamefont{A.}~\bibnamefont{{Kosti{\'c}}}},
  \bibinfo{journal}{\prl} \textbf{\bibinfo{volume}{133}}, \bibinfo{eid}{221006}
  (\bibinfo{year}{2024}), \eprint{2403.03220}.

\bibitem[{\citenamefont{{Demorest} et~al.}(2013)\citenamefont{{Demorest},
  {Ferdman}, {Gonzalez}, {Nice}, {Ransom}, {Stairs}, {Arzoumanian}, {Brazier},
  {Burke-Spolaor}, {Chamberlin} et~al.}}]{2013ApJ...762...94D}
\bibinfo{author}{\bibfnamefont{P.~B.} \bibnamefont{{Demorest}}},
  \bibinfo{author}{\bibfnamefont{R.~D.} \bibnamefont{{Ferdman}}},
  \bibinfo{author}{\bibfnamefont{M.~E.} \bibnamefont{{Gonzalez}}},
  \bibinfo{author}{\bibfnamefont{D.}~\bibnamefont{{Nice}}},
  \bibinfo{author}{\bibfnamefont{S.}~\bibnamefont{{Ransom}}},
  \bibinfo{author}{\bibfnamefont{I.~H.} \bibnamefont{{Stairs}}},
  \bibinfo{author}{\bibfnamefont{Z.}~\bibnamefont{{Arzoumanian}}},
  \bibinfo{author}{\bibfnamefont{A.}~\bibnamefont{{Brazier}}},
  \bibinfo{author}{\bibfnamefont{S.}~\bibnamefont{{Burke-Spolaor}}},
  \bibinfo{author}{\bibfnamefont{S.~J.} \bibnamefont{{Chamberlin}}},
  \bibnamefont{et~al.}, \bibinfo{journal}{\apj} \textbf{\bibinfo{volume}{762}},
  \bibinfo{eid}{94} (\bibinfo{year}{2013}), \eprint{1201.6641}.

\bibitem[{\citenamefont{{Anholm} et~al.}(2009)\citenamefont{{Anholm},
  {Ballmer}, {Creighton}, {Price}, and {Siemens}}}]{2009PhRvD..79h4030A}
\bibinfo{author}{\bibfnamefont{M.}~\bibnamefont{{Anholm}}},
  \bibinfo{author}{\bibfnamefont{S.}~\bibnamefont{{Ballmer}}},
  \bibinfo{author}{\bibfnamefont{J.~D.~E.} \bibnamefont{{Creighton}}},
  \bibinfo{author}{\bibfnamefont{L.~R.} \bibnamefont{{Price}}},
  \bibnamefont{and}
  \bibinfo{author}{\bibfnamefont{X.}~\bibnamefont{{Siemens}}},
  \bibinfo{journal}{\prd} \textbf{\bibinfo{volume}{79}}, \bibinfo{eid}{084030}
  (\bibinfo{year}{2009}), \eprint{0809.0701}.

\bibitem[{\citenamefont{{Agazie}
  et~al.}(2023{\natexlab{a}})\citenamefont{{Agazie}, {Alam}, {Anumarlapudi},
  {Archibald}, {Arzoumanian}, {Baker}, {Blecha}, {Bonidie}, {Brazier}, {Brook}
  et~al.}}]{2023ApJ...951L...9A}
\bibinfo{author}{\bibfnamefont{G.}~\bibnamefont{{Agazie}}},
  \bibinfo{author}{\bibfnamefont{M.~F.} \bibnamefont{{Alam}}},
  \bibinfo{author}{\bibfnamefont{A.}~\bibnamefont{{Anumarlapudi}}},
  \bibinfo{author}{\bibfnamefont{A.~M.} \bibnamefont{{Archibald}}},
  \bibinfo{author}{\bibfnamefont{Z.}~\bibnamefont{{Arzoumanian}}},
  \bibinfo{author}{\bibfnamefont{P.~T.} \bibnamefont{{Baker}}},
  \bibinfo{author}{\bibfnamefont{L.}~\bibnamefont{{Blecha}}},
  \bibinfo{author}{\bibfnamefont{V.}~\bibnamefont{{Bonidie}}},
  \bibinfo{author}{\bibfnamefont{A.}~\bibnamefont{{Brazier}}},
  \bibinfo{author}{\bibfnamefont{P.~R.} \bibnamefont{{Brook}}},
  \bibnamefont{et~al.}, \bibinfo{journal}{Astrophysical Journal Letters}
  \textbf{\bibinfo{volume}{951}}, \bibinfo{eid}{L9}
  (\bibinfo{year}{2023}{\natexlab{a}}), \eprint{2306.16217}.

\bibitem[{\citenamefont{{Agazie}
  et~al.}(2023{\natexlab{b}})\citenamefont{{Agazie}, {Anumarlapudi},
  {Archibald}, {Arzoumanian}, {Baker}, {B{\'e}csy}, {Blecha}, {Brazier},
  {Brook}, {Burke-Spolaor} et~al.}}]{2023ApJ...951L...8A}
\bibinfo{author}{\bibfnamefont{G.}~\bibnamefont{{Agazie}}},
  \bibinfo{author}{\bibfnamefont{A.}~\bibnamefont{{Anumarlapudi}}},
  \bibinfo{author}{\bibfnamefont{A.~M.} \bibnamefont{{Archibald}}},
  \bibinfo{author}{\bibfnamefont{Z.}~\bibnamefont{{Arzoumanian}}},
  \bibinfo{author}{\bibfnamefont{P.~T.} \bibnamefont{{Baker}}},
  \bibinfo{author}{\bibfnamefont{B.}~\bibnamefont{{B{\'e}csy}}},
  \bibinfo{author}{\bibfnamefont{L.}~\bibnamefont{{Blecha}}},
  \bibinfo{author}{\bibfnamefont{A.}~\bibnamefont{{Brazier}}},
  \bibinfo{author}{\bibfnamefont{P.~R.} \bibnamefont{{Brook}}},
  \bibinfo{author}{\bibfnamefont{S.}~\bibnamefont{{Burke-Spolaor}}},
  \bibnamefont{et~al.}, \bibinfo{journal}{Astrophysical Journal Letters}
  \textbf{\bibinfo{volume}{951}}, \bibinfo{eid}{L8}
  (\bibinfo{year}{2023}{\natexlab{b}}), \eprint{2306.16213}.

\bibitem[{\citenamefont{{Chamberlin} et~al.}(2015)\citenamefont{{Chamberlin},
  {Creighton}, {Siemens}, {Demorest}, {Ellis}, {Price}, and
  {Romano}}}]{2015PhRvD..91d4048C}
\bibinfo{author}{\bibfnamefont{S.~J.} \bibnamefont{{Chamberlin}}},
  \bibinfo{author}{\bibfnamefont{J.~D.~E.} \bibnamefont{{Creighton}}},
  \bibinfo{author}{\bibfnamefont{X.}~\bibnamefont{{Siemens}}},
  \bibinfo{author}{\bibfnamefont{P.}~\bibnamefont{{Demorest}}},
  \bibinfo{author}{\bibfnamefont{J.}~\bibnamefont{{Ellis}}},
  \bibinfo{author}{\bibfnamefont{L.~R.} \bibnamefont{{Price}}},
  \bibnamefont{and} \bibinfo{author}{\bibfnamefont{J.~D.}
  \bibnamefont{{Romano}}}, \bibinfo{journal}{\prd}
  \textbf{\bibinfo{volume}{91}}, \bibinfo{eid}{044048} (\bibinfo{year}{2015}),
  \eprint{1410.8256}.

\bibitem[{\citenamefont{Euler}(1741)}]{euler1741}
\bibinfo{author}{\bibfnamefont{L.}~\bibnamefont{Euler}},
  \bibinfo{journal}{Commentarii Academiae Scientiarum Imperialis
  Petropolitanae} \textbf{\bibinfo{volume}{8}}, \bibinfo{pages}{128}
  (\bibinfo{year}{1741}).

\bibitem[{\citenamefont{{Edelson} and {Krolik}}(1988)}]{1988ApJ...333..646E}
\bibinfo{author}{\bibfnamefont{R.~A.} \bibnamefont{{Edelson}}}
  \bibnamefont{and} \bibinfo{author}{\bibfnamefont{J.~H.}
  \bibnamefont{{Krolik}}}, \bibinfo{journal}{\apj}
  \textbf{\bibinfo{volume}{333}}, \bibinfo{pages}{646} (\bibinfo{year}{1988}).

\bibitem[{\citenamefont{{Max-Moerbeck}
  et~al.}(2014)\citenamefont{{Max-Moerbeck}, {Richards}, {Hovatta}, {Pavlidou},
  {Pearson}, and {Readhead}}}]{2014MNRAS.445..437M}
\bibinfo{author}{\bibfnamefont{W.}~\bibnamefont{{Max-Moerbeck}}},
  \bibinfo{author}{\bibfnamefont{J.~L.} \bibnamefont{{Richards}}},
  \bibinfo{author}{\bibfnamefont{T.}~\bibnamefont{{Hovatta}}},
  \bibinfo{author}{\bibfnamefont{V.}~\bibnamefont{{Pavlidou}}},
  \bibinfo{author}{\bibfnamefont{T.~J.} \bibnamefont{{Pearson}}},
  \bibnamefont{and} \bibinfo{author}{\bibfnamefont{A.~C.~S.}
  \bibnamefont{{Readhead}}}, \bibinfo{journal}{Monthly Notices of the RAS}
  \textbf{\bibinfo{volume}{445}}, \bibinfo{pages}{437} (\bibinfo{year}{2014}),
  \eprint{1408.6265}.

\bibitem[{\citenamefont{{Watts} and {Strogatz}}(1998)}]{1998Natur.393..440W}
\bibinfo{author}{\bibfnamefont{D.~J.} \bibnamefont{{Watts}}} \bibnamefont{and}
  \bibinfo{author}{\bibfnamefont{S.~H.} \bibnamefont{{Strogatz}}},
  \bibinfo{journal}{\nat} \textbf{\bibinfo{volume}{393}}, \bibinfo{pages}{440}
  (\bibinfo{year}{1998}).

\bibitem[{\citenamefont{Albert and Barab\'asi}(2002)}]{RevModPhys.74.47}
\bibinfo{author}{\bibfnamefont{R.}~\bibnamefont{Albert}} \bibnamefont{and}
  \bibinfo{author}{\bibfnamefont{A.-L.} \bibnamefont{Barab\'asi}},
  \bibinfo{journal}{Rev. Mod. Phys.} \textbf{\bibinfo{volume}{74}},
  \bibinfo{pages}{47} (\bibinfo{year}{2002}),
  \urlprefix\url{https://link.aps.org/doi/10.1103/RevModPhys.74.47}.

\bibitem[{\citenamefont{{Onnela} et~al.}(2005)\citenamefont{{Onnela},
  {Saram{\"a}ki}, {Kert{\'e}sz}, and {Kaski}}}]{Onnela2005PhRvE}
\bibinfo{author}{\bibfnamefont{J.-P.} \bibnamefont{{Onnela}}},
  \bibinfo{author}{\bibfnamefont{J.}~\bibnamefont{{Saram{\"a}ki}}},
  \bibinfo{author}{\bibfnamefont{J.}~\bibnamefont{{Kert{\'e}sz}}},
  \bibnamefont{and} \bibinfo{author}{\bibfnamefont{K.}~\bibnamefont{{Kaski}}},
  \bibinfo{journal}{\pre} \textbf{\bibinfo{volume}{71}}, \bibinfo{eid}{065103}
  (\bibinfo{year}{2005}), \eprint{cond-mat/0408629}.

\bibitem[{\citenamefont{Newman}(2010)}]{Newman10.1093}
\bibinfo{author}{\bibfnamefont{M.}~\bibnamefont{Newman}},
  \emph{\bibinfo{title}{Networks: An Introduction}} (\bibinfo{publisher}{Oxford
  University Press}, \bibinfo{year}{2010}),
  \urlprefix\url{https://doi.org/10.1093/acprof:oso/9780199206650.001.0001}.

\bibitem[{\citenamefont{{Gunderson} and
  {Bravo-Hermsdorff}}(2020)}]{2020arXiv200203959G}
\bibinfo{author}{\bibfnamefont{L.~M.} \bibnamefont{{Gunderson}}}
  \bibnamefont{and}
  \bibinfo{author}{\bibfnamefont{G.}~\bibnamefont{{Bravo-Hermsdorff}}},
  \bibinfo{journal}{arXiv e-prints} \bibinfo{eid}{arXiv:2002.03959}
  (\bibinfo{year}{2020}), \eprint{2002.03959}.

\bibitem[{\citenamefont{Hanley}(1982)}]{Hanley1982}
\bibinfo{author}{\bibfnamefont{J.}~\bibnamefont{Hanley}},
  \bibinfo{journal}{Radiology} \textbf{\bibinfo{volume}{143}},
  \bibinfo{pages}{29} (\bibinfo{year}{1982}).

\bibitem[{\citenamefont{Welch}(1947)}]{a967ba42-d0b9-3e65-976f-80cc6086b406}
\bibinfo{author}{\bibfnamefont{B.~L.} \bibnamefont{Welch}},
  \bibinfo{journal}{Biometrika} \textbf{\bibinfo{volume}{34}},
  \bibinfo{pages}{28} (\bibinfo{year}{1947}), ISSN \bibinfo{issn}{00063444},
  \urlprefix\url{http://www.jstor.org/stable/2332510}.

\bibitem[{\citenamefont{Cohen}(1988)}]{cohen1988spa}
\bibinfo{author}{\bibfnamefont{J.}~\bibnamefont{Cohen}},
  \emph{\bibinfo{title}{{Statistical Power Analysis for the Behavioral
  Sciences}}} (\bibinfo{publisher}{Lawrence Erlbaum Associates},
  \bibinfo{year}{1988}), \bibinfo{edition}{2nd} ed.

\bibitem[{\citenamefont{{Gelman} et~al.}(2014)\citenamefont{{Gelman}, {Carlin},
  {Stern}, {Dunson}, {Vehtari}, and {Rubin}}}]{2014bda..book.....G}
\bibinfo{author}{\bibfnamefont{A.}~\bibnamefont{{Gelman}}},
  \bibinfo{author}{\bibfnamefont{J.~B.} \bibnamefont{{Carlin}}},
  \bibinfo{author}{\bibfnamefont{H.~S.} \bibnamefont{{Stern}}},
  \bibinfo{author}{\bibfnamefont{D.~B.} \bibnamefont{{Dunson}}},
  \bibinfo{author}{\bibfnamefont{A.}~\bibnamefont{{Vehtari}}},
  \bibnamefont{and} \bibinfo{author}{\bibfnamefont{D.~B.}
  \bibnamefont{{Rubin}}}, \emph{\bibinfo{title}{{Bayesian Data Analysis}}}
  (\bibinfo{publisher}{Chapman and Hall/CRC}, \bibinfo{year}{2014}),
  \bibinfo{edition}{3rd} ed.

\bibitem[{\citenamefont{{Heydenreich} et~al.}(2021)\citenamefont{{Heydenreich},
  {Br{\"u}ck}, and {Harnois-D{\'e}raps}}}]{2021A&A...648A..74H}
\bibinfo{author}{\bibfnamefont{S.}~\bibnamefont{{Heydenreich}}},
  \bibinfo{author}{\bibfnamefont{B.}~\bibnamefont{{Br{\"u}ck}}},
  \bibnamefont{and}
  \bibinfo{author}{\bibfnamefont{J.}~\bibnamefont{{Harnois-D{\'e}raps}}},
  \bibinfo{journal}{Astronomy and Astrophysics} \textbf{\bibinfo{volume}{648}},
  \bibinfo{eid}{A74} (\bibinfo{year}{2021}), \eprint{2007.13724}.

\bibitem[{\citenamefont{{Grand{\'o}n}}(2023)}]{2023PhDT........23G}
\bibinfo{author}{\bibfnamefont{D.}~\bibnamefont{{Grand{\'o}n}}}, Ph.D. thesis,
  \bibinfo{school}{University of Chile} (\bibinfo{year}{2023}).

\bibitem[{\citenamefont{{Vel{\'a}zquez}
  et~al.}(2024)\citenamefont{{Vel{\'a}zquez}, {Escamilla}, {Mukherjee}, and
  {V{\'a}zquez}}}]{2024Univ...10..464V}
\bibinfo{author}{\bibfnamefont{J.~d.~J.} \bibnamefont{{Vel{\'a}zquez}}},
  \bibinfo{author}{\bibfnamefont{L.~A.} \bibnamefont{{Escamilla}}},
  \bibinfo{author}{\bibfnamefont{P.}~\bibnamefont{{Mukherjee}}},
  \bibnamefont{and} \bibinfo{author}{\bibfnamefont{J.~A.}
  \bibnamefont{{V{\'a}zquez}}}, \bibinfo{journal}{Universe}
  \textbf{\bibinfo{volume}{10}}, \bibinfo{eid}{464} (\bibinfo{year}{2024}),
  \eprint{2410.02061}.

\bibitem[{\citenamefont{Abdi and Williams}(2010)}]{abdi2010principal}
\bibinfo{author}{\bibfnamefont{H.}~\bibnamefont{Abdi}} \bibnamefont{and}
  \bibinfo{author}{\bibfnamefont{L.~J.} \bibnamefont{Williams}},
  \bibinfo{journal}{WIREs Computational Statistics}
  \textbf{\bibinfo{volume}{2}}, \bibinfo{pages}{433} (\bibinfo{year}{2010}),
  \eprint{https://wires.onlinelibrary.wiley.com/doi/pdf/10.1002/wics.101},
  \urlprefix\url{https://wires.onlinelibrary.wiley.com/doi/abs/10.1002/wics.101}.

\bibitem[{\citenamefont{{Hartlap} et~al.}(2007)\citenamefont{{Hartlap},
  {Simon}, and {Schneider}}}]{2007A&A...464..399H}
\bibinfo{author}{\bibfnamefont{J.}~\bibnamefont{{Hartlap}}},
  \bibinfo{author}{\bibfnamefont{P.}~\bibnamefont{{Simon}}}, \bibnamefont{and}
  \bibinfo{author}{\bibfnamefont{P.}~\bibnamefont{{Schneider}}},
  \bibinfo{journal}{Astronomy and Astrophysics} \textbf{\bibinfo{volume}{464}},
  \bibinfo{pages}{399} (\bibinfo{year}{2007}), \eprint{astro-ph/0608064}.

\bibitem[{\citenamefont{{Agazie}
  et~al.}(2023{\natexlab{c}})\citenamefont{{Agazie}, {Anumarlapudi},
  {Archibald}, {Arzoumanian}, {Baker}, {B{\'e}csy}, {Blecha}, {Brazier},
  {Brook}, {Burke-Spolaor} et~al.}}]{2023ApJ...951L..10A}
\bibinfo{author}{\bibfnamefont{G.}~\bibnamefont{{Agazie}}},
  \bibinfo{author}{\bibfnamefont{A.}~\bibnamefont{{Anumarlapudi}}},
  \bibinfo{author}{\bibfnamefont{A.~M.} \bibnamefont{{Archibald}}},
  \bibinfo{author}{\bibfnamefont{Z.}~\bibnamefont{{Arzoumanian}}},
  \bibinfo{author}{\bibfnamefont{P.~T.} \bibnamefont{{Baker}}},
  \bibinfo{author}{\bibfnamefont{B.}~\bibnamefont{{B{\'e}csy}}},
  \bibinfo{author}{\bibfnamefont{L.}~\bibnamefont{{Blecha}}},
  \bibinfo{author}{\bibfnamefont{A.}~\bibnamefont{{Brazier}}},
  \bibinfo{author}{\bibfnamefont{P.~R.} \bibnamefont{{Brook}}},
  \bibinfo{author}{\bibfnamefont{S.}~\bibnamefont{{Burke-Spolaor}}},
  \bibnamefont{et~al.}, \bibinfo{journal}{Astrophysical Journal Letters}
  \textbf{\bibinfo{volume}{951}}, \bibinfo{eid}{L10}
  (\bibinfo{year}{2023}{\natexlab{c}}), \eprint{2306.16218}.

\bibitem[{\citenamefont{Zomorodian}(2005)}]{zomorodian2005topology11}
\bibinfo{author}{\bibfnamefont{A.~J.} \bibnamefont{Zomorodian}},
  \emph{\bibinfo{title}{Topology for computing}} (\bibinfo{publisher}{Cambridge
  university press}, \bibinfo{year}{2005}).

\bibitem[{\citenamefont{Wasserman}(2018)}]{wasserman2018topological}
\bibinfo{author}{\bibfnamefont{L.}~\bibnamefont{Wasserman}},
  \bibinfo{journal}{Annual Review of Statistics and Its Application}
  \textbf{\bibinfo{volume}{5}}, \bibinfo{pages}{501} (\bibinfo{year}{2018}),
  ISSN \bibinfo{issn}{2326-831X},
  \urlprefix\url{https://www.annualreviews.org/content/journals/10.1146/annurev-statistics-031017-100045}.

\bibitem[{\citenamefont{Dey and Wang}(2022)}]{dey2022computational}
\bibinfo{author}{\bibfnamefont{T.~K.} \bibnamefont{Dey}} \bibnamefont{and}
  \bibinfo{author}{\bibfnamefont{Y.}~\bibnamefont{Wang}},
  \emph{\bibinfo{title}{Computational Topology for Data Analysis}}
  (\bibinfo{publisher}{Cambridge University Press}, \bibinfo{year}{2022}).

\bibitem[{\citenamefont{Edelsbrunner and
  Harer}(2010)}]{edelsbrunner2022computational}
\bibinfo{author}{\bibfnamefont{H.}~\bibnamefont{Edelsbrunner}}
  \bibnamefont{and} \bibinfo{author}{\bibfnamefont{J.}~\bibnamefont{Harer}},
  \emph{\bibinfo{title}{Computational Topology: an Introduction}}
  (\bibinfo{publisher}{American Mathematical Society},
  \bibinfo{address}{Providence, R.I.}, \bibinfo{year}{2010}).

\bibitem[{\citenamefont{{Jalali Kanafi} et~al.}(2024)\citenamefont{{Jalali
  Kanafi}, {Ansarifard}, and {Movahed}}}]{2024MNRAS.535..657J}
\bibinfo{author}{\bibfnamefont{M.~H.} \bibnamefont{{Jalali Kanafi}}},
  \bibinfo{author}{\bibfnamefont{S.}~\bibnamefont{{Ansarifard}}},
  \bibnamefont{and} \bibinfo{author}{\bibfnamefont{S.~M.~S.}
  \bibnamefont{{Movahed}}}, \bibinfo{journal}{Monthly Notices of the RAS}
  \textbf{\bibinfo{volume}{535}}, \bibinfo{pages}{657} (\bibinfo{year}{2024}),
  \eprint{2311.13520}.

\bibitem[{\citenamefont{{Hoy}}(2022)}]{2022PhRvD.106h3003H}
\bibinfo{author}{\bibfnamefont{C.}~\bibnamefont{{Hoy}}},
  \bibinfo{journal}{\prd} \textbf{\bibinfo{volume}{106}}, \bibinfo{eid}{083003}
  (\bibinfo{year}{2022}), \eprint{2208.00106}.

\bibitem[{\citenamefont{{Paradiso} et~al.}(2024)\citenamefont{{Paradiso},
  {DiMarco}, {Chen}, {McGee}, and {Percival}}}]{2024MNRAS.528.1531P}
\bibinfo{author}{\bibfnamefont{S.}~\bibnamefont{{Paradiso}}},
  \bibinfo{author}{\bibfnamefont{M.}~\bibnamefont{{DiMarco}}},
  \bibinfo{author}{\bibfnamefont{M.}~\bibnamefont{{Chen}}},
  \bibinfo{author}{\bibfnamefont{G.}~\bibnamefont{{McGee}}}, \bibnamefont{and}
  \bibinfo{author}{\bibfnamefont{W.~J.} \bibnamefont{{Percival}}},
  \bibinfo{journal}{Monthly Notices of the RAS} \textbf{\bibinfo{volume}{528}},
  \bibinfo{pages}{1531} (\bibinfo{year}{2024}), \eprint{2310.06747}.

\bibitem[{\citenamefont{{Adityan} and {Raj}}(2025)}]{Adityan2025}
\bibinfo{author}{\bibfnamefont{S.}~\bibnamefont{{Adityan}}} \bibnamefont{and}
  \bibinfo{author}{\bibfnamefont{A.~S.} \bibnamefont{{Raj}}},
  \bibinfo{journal}{Astrophysics and Space Science}
  \textbf{\bibinfo{volume}{370}}, \bibinfo{eid}{35} (\bibinfo{year}{2025}).

\bibitem[{\citenamefont{Hagberg et~al.}(2008)\citenamefont{Hagberg, Schult, and
  Swart}}]{hagberg2008}
\bibinfo{author}{\bibfnamefont{A.~A.} \bibnamefont{Hagberg}},
  \bibinfo{author}{\bibfnamefont{D.~A.} \bibnamefont{Schult}},
  \bibnamefont{and} \bibinfo{author}{\bibfnamefont{P.~J.} \bibnamefont{Swart}},
  in \emph{\bibinfo{booktitle}{Proceedings of the 7th Python in Science
  Conference}}, edited by
  \bibinfo{editor}{\bibfnamefont{G.}~\bibnamefont{Varoquaux}},
  \bibinfo{editor}{\bibfnamefont{T.}~\bibnamefont{Vaught}}, \bibnamefont{and}
  \bibinfo{editor}{\bibfnamefont{J.}~\bibnamefont{Millman}}
  (\bibinfo{year}{2008}), pp. \bibinfo{pages}{11 -- 15},
  \urlprefix\url{http://conference.scipy.org/proceedings/SciPy2008/paper_2/}.

\bibitem[{\citenamefont{Lewis}(2025)}]{Lewis2019xzd}
\bibinfo{author}{\bibfnamefont{A.}~\bibnamefont{Lewis}},
  \bibinfo{journal}{JCAP} \textbf{\bibinfo{volume}{08}}, \bibinfo{pages}{025}
  (\bibinfo{year}{2025}), \eprint{1910.13970}.

\bibitem[{\citenamefont{{Allen} and {Ottewill}}(1997)}]{1997PhRvD..56..545A}
\bibinfo{author}{\bibfnamefont{B.}~\bibnamefont{{Allen}}} \bibnamefont{and}
  \bibinfo{author}{\bibfnamefont{A.~C.} \bibnamefont{{Ottewill}}},
  \bibinfo{journal}{\prd} \textbf{\bibinfo{volume}{56}}, \bibinfo{pages}{545}
  (\bibinfo{year}{1997}), \eprint{gr-qc/9607068}.

\end{thebibliography}



\end{document}